\documentclass[a4paper,11pt]{article}
\usepackage{jheppub}
\usepackage[utf8]{inputenc}

\usepackage{todonotes}

\usepackage{lscape}

\usepackage{comment}

\usepackage{dsfont}
\usepackage{simplewick}

\usepackage{mathtools,amsmath,amssymb}

\DeclareMathSizes{1}{1}{1}{1}
\usepackage{exscale}

\usepackage{graphicx}

\usepackage{lmodern}
\usepackage[T1]{fontenc}
\usepackage{microtype}

\usepackage{hyperref}

\usepackage{xspace}

\usepackage{array}

\usepackage{subfigure}

\usepackage{pgf}
\usepackage{pgfplots}
\pgfplotsset{compat=1.15}
\usepackage{tikz}
\usetikzlibrary{shapes,arrows,automata,positioning,backgrounds,intersections}
\usepackage{tikz-3dplot}

\newcommand{\YM}{{\mathrm{\scriptscriptstyle YM}}}

\usepackage{xspace}

 \usepackage{breqn}
  
\usepackage{placeins}

\DeclareMathOperator{\phaneq}{\phantom{{}=}}

\newcommand{\THag}{T_{\rm H}}

\newcommand{\LL}{\mathcal{L}}

\DeclareMathOperator{\Li}{Li}
\DeclareMathOperator{\sech}{sech}
\DeclareMathOperator{\csch}{csch}

\newcommand{\Pb}{\mathbf{P}}
\newcommand{\Qb}{\mathbf{Q}}
\newcommand{\yy}{\tanh(\tfrac{1}{4\THag})}

\newcommand{\e}{\operatorname{e}}
\newcommand{\de}{\operatorname{d}\!}

\newcommand{\eqncom}{\, , }

\newcommand{\CA}{\mathcal{A}}
\newcommand{\CB}{\mathcal{B}}

\newcommand{\CN}{\mathcal{N}}
\newcommand{\CO}{\mathcal{O}}

\newcommand{\CY}{\mathcal{Y}}
\newcommand{\nn}{\nonumber}
\newcommand{\spa}{\ , \ \ }
\newcommand{\ds}{\displaystyle}

\newcommand{\C}{\mathbb{C}}
\newcommand{\R}{\mathbb{R}}

\DeclareMathOperator{\checkstar}{\check {\star}}
\DeclareMathOperator{\hatstar}{\hat {\star}}

\DeclareMathOperator{\tr}{tr}

\renewcommand{\tfrac}[2]{{\textstyle \frac{#1}{#2}}}

\makeatletter
\DeclareRobustCommand*{\bfseries}{%
  \not@math@alphabet\bfseries\mathbf
  \fontseries\bfdefault\selectfont
  \boldmath
}

\makeatother

\usepackage{lipsum}
\makeatletter
\if@titlepage
  \renewenvironment{abstract}{%
      \titlepage
      \null\vfil
      \@beginparpenalty\@lowpenalty
      \begin{center}%
        \bfseries \abstractname
        \@endparpenalty\@M
      \end{center}}%
     {\par\vfil\null\endtitlepage}
\else
  \renewenvironment{abstract}{%
      \if@twocolumn
        \section*{\abstractname}%
      \else
        \small
        \begin{center}%
          {\bfseries \abstractname\vspace{-.5em}\vspace{\z@}}%
        \end{center}%
        \quotation
      \fi}
      {\if@twocolumn\else\endquotation\fi}
\fi
\makeatother

\title{Solving the Hagedorn temperature of \texorpdfstring{AdS$_5$/CFT$_4$}{AdS5/CFT4} via the Quantum Spectral Curve: Chemical potentials and deformations}
\author{Troels Harmark and Matthias Wilhelm}

\begin{document}

\begingroup\parindent0pt
\begin{flushright}\footnotesize
\end{flushright}
\vspace*{4em}
\centering
\begingroup\LARGE
\bf
Solving the Hagedorn temperature of AdS$_5$/CFT$_4$ via the Quantum Spectral Curve:\\ Chemical potentials and deformations
\par\endgroup
\vspace{2.5em}
\begingroup\large{\bf Troels Harmark and Matthias Wilhelm}
\par\endgroup
\vspace{1em}
\begingroup\itshape
Niels Bohr Institute, Copenhagen University,\\
Blegdamsvej 17, 2100 Copenhagen \O{}, Denmark\\

\par\endgroup
\vspace{1em}
\begingroup\ttfamily
harmark@nbi.ku.dk, matthias.wilhelm@nbi.ku.dk \\
\par\endgroup
\vspace{2.5em}
\endgroup

\begin{abstract}
\noindent
 We describe how to calculate the Hagedorn temperature of $\mathcal{N}=4$ SYM theory and type IIB superstring theory on $AdS_5\times S^5$ via the Quantum Spectral Curve (QSC) -- providing further details on our previous letters \cite{Harmark:2017yrv} and \cite{Harmark:2018red}.
 We solve the QSC equations perturbatively at weak 't Hooft coupling $\lambda$ up to seven-loop order and numerically at finite coupling, finding that the perturbative results can be expressed in terms of single-valued harmonic polylogarithms.
 Moreover, we generalize the QSC to describe the Hagedorn temperature in the presence of chemical potentials.
 Finally, we show that the Hagedorn temperature in certain deformations of $\mathcal{N}=4$ SYM theory (real-$\beta$ and $\gamma_i$ deformation) agrees with the one in $\mathcal{N}=4$ SYM theory at any value of $\lambda$. 
\end{abstract}


\thispagestyle{empty}

\newpage
\tableofcontents

\section{Introduction}
\label{sec: Introduction}

Integrability in the context of the AdS/CFT correspondence has opened a unique window of non-perturbative understanding of the properties of strongly coupled gauge theories and string theories in strongly curved backgrounds, see e.g.\ Refs.\ \cite{Beisert:2010jr,Bombardelli:2016rwb} for reviews.
It has been developed furthest for the so-called spectral problem, the problem of finding the scaling dimensions $\Delta$ of composite operators in planar $\mathcal{N}=4$ SYM theory and thus the energies of the corresponding strings in type IIB superstring theory on $AdS_5\times S^5$.
The finite-coupling solution to the spectral problem is given by the thermodynamic Bethe ansatz (TBA) \cite{Arutyunov:2009zu,Bombardelli:2009ns,Gromov:2009bc,Arutyunov:2009ur,Gromov:2009tv,Cavaglia:2010nm}, an infinite set of integral equations.
These equations have been subsequently brought into the form of finite-difference equations, the so-called Quantum Spectral Curve (QSC) equations \cite{Gromov:2013pga,Gromov:2014caa}, which can be efficiently solved perturbatively at weak coupling \cite{Marboe:2014gma,Marboe:2017dmb,Marboe:2018ugv} and numerically at finite coupling \cite{Gromov:2015wca,Hegedus:2016eop}.
Further applications of the QSC include the pomeron and BFKL regime \cite{Alfimov:2014bwa,Gromov:2015vua,Alfimov:2020obh},
cusped Wilson lines \cite{Gromov:2015dfa,Cavaglia:2018lxi,Grabner:2020nis,Gromov:2021ahm,Cavaglia:2021bnz}, the quark-antiquark potential \cite{Gromov:2016rrp}, color-twist operators \cite{Cavaglia:2020hdb} and
integrable deformations of $\mathcal{N}=4$ SYM theory \cite{Kazakov:2015efa,Klabbers:2017vtw,Gromov:2017cja,Marboe:2019wyc,Levkovich-Maslyuk:2020rlp}.

An important aspect of understanding gauge theories and string theories, which has received less attention in the context of integrability, concerns their thermodynamic properties.
One such property that occurs on both sides of the AdS/CFT correspondence is Hagedorn behavior, an exponential growth of the density of states with the energy 
that leads to a pole in the planar partition function at the so-called Hagedorn temperature $\THag$.
For string theory in flat space, the Hagedorn temperature could be calculated long time ago \cite{Sundborg:1984uk}. In $AdS_5\times S^5$, an analogous calculation has not been possible, due to the problems with quantizing string theory in curved spacetime.
In the planar gauge theory, the Hagedorn temperature has been calculated in the free theory \cite{Sundborg:1999ue} and to the first order at weak coupling \cite{Spradlin:2004pp}. 
The physical interpretation of the Hagedorn temperature is that of a limiting temperature; it signals the breakdown of the low-energy description and a confinement-deconfinement-like transition on the gauge-theory side, which corresponds to the Hawking-Page transition between a gas of closed strings and a black hole on the string-theory side \cite{Atick:1988si, Witten:1998zw,Sundborg:1999ue,Aharony:2003sx}.

In our letters \cite{Harmark:2017yrv,Harmark:2018red}, we have recently shown how to calculate the Hagedorn temperature of planar $\mathcal{N}=4$ SYM theory and type IIB superstring theory on $AdS_5\times S^5$ via integrability.
Concretely, we have derived TBA equations for the Hagedorn temperature \cite{Harmark:2017yrv} and recast them into the form of the QSC \cite{Harmark:2018red}.
Solving these equations perturbatively at weak coupling, we calculated the Hagedorn temperature up to three loops. Moreover, we solved the equations numerically at finite coupling, finding in particular that Hagedorn temperature of type IIB superstring theory on $AdS_5\times S^5$ asymptotes to the one on flat 10D Minkowski space (calculated in Ref.\ \cite{Sundborg:1984uk}) in the limit of strong coupling, i.e.\ vanishing curvature.

The first aim of the present paper is to provide several details on the Hagedorn QSC and its derivation from the TBA, which we deferred in our letters \cite{Harmark:2017yrv,Harmark:2018red}. Moreover, we extend the weak-coupling solution from three loops to seven loops, finding interesting number-theoretic properties.
In particular, we find that the perturbative results can be expressed in terms of so-called single-valued harmonic polylogarithms \cite{Brown:2004ugm}, which have previously occurred in scattering amplitudes and generalize the single-valued multiple zeta values that occur in the spectral problem \cite{Leurent:2013mr,Marboe:2014gma,Marboe:2018ugv}.
Finally, we also provide numerical values for the first three corrections to the leading strong-coupling behavior, finding perfect agreement with the analytic calculation \cite{MaldacenaPrivateCommunication,Urbach:2022xzw} of the first correction.\footnote{We thank Juan Maldacena for sharing his calculation with us and for making us aware of the independent calculation \cite{Urbach:2022xzw} that appeared since.}

The second aim of this paper is to extend said results for the Hagedorn temperature to a class of integrable deformations of the maximally supersymmetric Yang-Mills theory and to include chemical potentials.
The Hagedorn temperature for integrable deformations has been previously studied in Ref.~\cite{Fokken:2014moa}.
The Hagedorn temperature of $\CN=4$ SYM theory with chemical potentials has been studied previously in Refs.~\cite{Yamada:2006rx,Harmark:2006di,Harmark:2006ie,Harmark:2007px,Harmark:2014mpa,Suzuki:2017ipd}.
We show in this paper how to use the QSC to determine the Hagedorn temperature for a class of integrable deformations and in the presence of arbitrary chemical potentials. 

\begin{figure}[t]
\centering

\begin{tikzpicture}
 
\node[] (torus) at (0,0) 
{
 \begin{tikzpicture}[line width=.7pt,scale=1]
 \begin{axis}[axis equal image,
     clip=false,
     xlabel=\empty, ylabel=\empty, zlabel=\empty,
     axis lines=middle,
      hide axis,
     ]
             \addplot3[domain=0:360,y domain=0:360, samples=30,
             surf,z buffer=sort,
     colormap/redyellow, 
             ]
             ({15+(10+4*cos(x))*cos(y)} ,
             {15+(10+4*cos(x))*sin(y)},
             {4+4*sin(x)});
\addplot3 [domain=-38:142, samples y=1,
            name path=xline,
            black, thin,
        ] ({15+(10+4*cos(x))*cos(0)} ,
             {15+(10+4*cos(x))*sin(0)},
             {4+4*sin(x)});
\addplot3 [domain=0:360, samples=30,samples y=1,
            name path=xline,
            black, thin,
        ] ({15+(10+4*cos(84))*cos(x)} ,
             {15+(10+4*cos(84))*sin(x)},
             {4+4*sin(84)});   
\node[label={90:{$\frac{1}{T}$}}] at ({15+(10+4*cos(84))*cos(90)} ,
             {15+(10+4*cos(84))*sin(90)},
             {4+4*sin(84)}) {};
\node[label={0:{$L$}}] at ({15+(10+4*cos(4))*cos(0)} ,
             {15+(10+4*cos(4))*sin(0)},
             {4+4*sin(4)}) {};             
 \end{axis}
 \end{tikzpicture} 
 };
 \node[] (cylinder 1) at (-0.35\textwidth,-0.15\textheight) {
 \begin{tikzpicture}[line width=.7pt,scale=1]
 \begin{axis}[axis equal image,
     clip=false,
     xlabel=\empty, ylabel=\empty, zlabel=\empty,
     axis lines=middle,
      hide axis
     ]
             \addplot3[domain=0:30,y domain=0:360, samples=20,
             surf,z buffer=sort,colormap/redyellow,
             ]
             (
             {15+(4)*cos(y)} ,{x},
             {15+(4)*sin(y)});
             \addplot3 [domain=-30:150, samples=20,samples y=1,
            name path=xline,
            black, thin,
        ] ({15+(4*cos(x))} ,{15},
             {15+(4*sin(x))});   
\node[label={270:{$\quad L$}}] at (
	      {15+(4*cos(4)))} ,{15},
             {15+(4*sin(4))}) {};  
 \end{axis}
 \end{tikzpicture} 
 }; 
 \node[] (cylinder 2) at (0.35\textwidth,-0.15\textheight) {
  \begin{tikzpicture}[line width=.7pt,scale=1]
 \begin{axis}[axis equal image,
     clip=false,
     xlabel=\empty, ylabel=\empty, zlabel=\empty,
     axis lines=middle,
      hide axis
     ]
             \addplot3[domain=0:16,y domain=0:360, samples=20,
             surf,z buffer=sort,colormap/redyellow,
             ]
             (
             {15+(4)*cos(y)} ,
             {15+(4)*sin(y)},{x});
             \addplot3 [domain=200:380, samples=20,samples y=1,
            name path=xline,
            black, thin,
        ] ({15+(4*cos(x))} ,
             {15+(4*sin(x))},{8});   
\node[label={0:{$\frac{1}{T}$}}] at (
	      {15+(4*cos(4)))} ,
             {15+(4*sin(4))},{8}) {};  
 \end{axis}
 \end{tikzpicture} 
 };
 \draw[->,thick] (torus) -- (cylinder 1) node[midway, right] {$\scriptstyle \,\,\, T\to0$};
 \draw[->,thick] (torus) -- (cylinder 2) node[midway, left] {$\scriptstyle L\to\infty\,\,$};
 \draw[<->,thick] (cylinder 1) -- (cylinder 2) node[midway, above] {{\scriptsize double Wick rotation}};
 \node[below=of cylinder 1.south,node distance=-\baselineskip] (text 1) {spectral problem};
 \node[below=of cylinder 2.south,node distance=-\baselineskip] (text 2) {Hagedorn temperature};
\end{tikzpicture} 
\label{fig: torus and cylinders} 
\caption{Calculating the spin-chain free energy at a finite temperature $T$ for a chain of finite length $L$ requires to consider the integrable model on a torus with circumferences $1/T$ and $L$ (top). In the spectral problem, the temperature is taken to be zero, resulting in a cylinder with circumference $L$ (left). In the case of the Hagedorn temperature, we have to take the limit $L\to \infty$, resulting in a cylinder with circumference $1/T$ (right). The former and the latter situation are related by a double Wick rotation.} 
\end{figure}
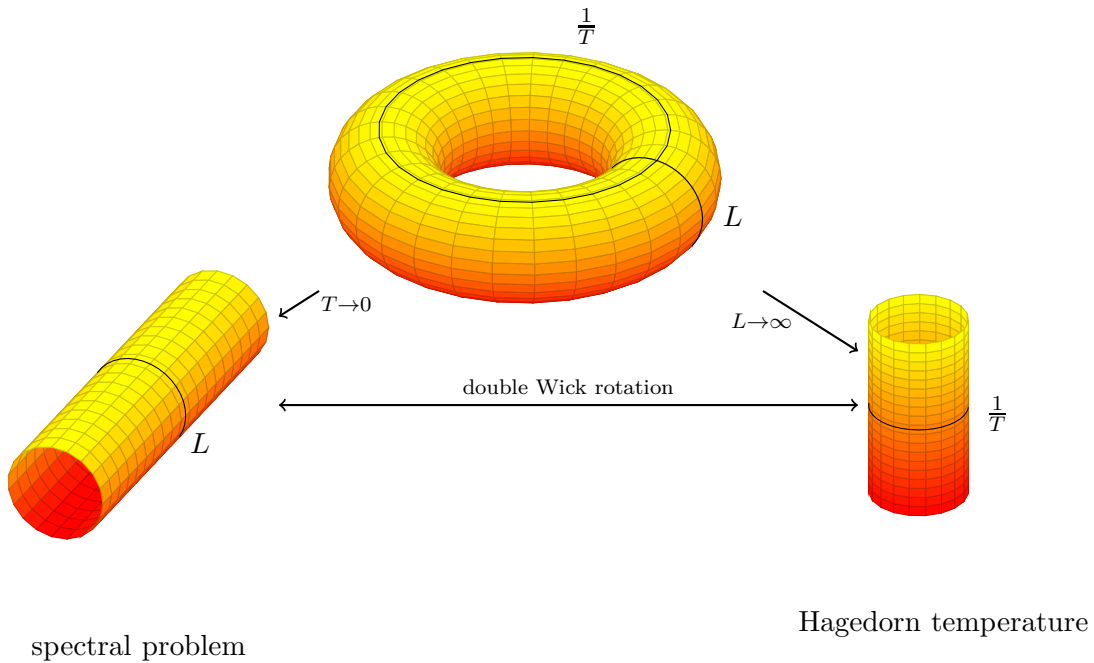

Let us briefly stress some salient features of the TBA and QSC for the Hagedorn temperature \cite{Harmark:2017yrv,Harmark:2018red}, contrasting this case to the one in the spectral problem; see also Fig.\ \ref{fig: torus and cylinders}.
In the spectral problem, the integrable model is solved at zero temperature on a cylinder with a finite circumference $L$ that takes into account the finite length of the operator, i.e.\ the finite number of fields the operator contains. This is related to solving the integral model on a cylinder with circumference $1/T$ by a double Wick rotation, such that the TBA can be used. While the case with finite temperature is known as the physical theory, its double-Wick-rotated version with finite length that is relevant in the spectral problem is called mirror theory.
In order to calculate the Hagedorn temperature, we are interested in the physical theory at finite $1/T$. The other dimension accounting for the finite length of an operator is sent to infinity as the Hagedorn singularity is governed by the high-energy limit of the density of states, where only operators with large length -- or rather large classical scaling dimension -- contribute; see Ref.\ \cite{Harmark:2017yrv} for details.

One effect of the double Wick rotation concerns the analytic structure on the QSC.
The planar coupling $g^2=\frac{g_\YM^2N}{16\pi^2}$ enters the QSC via branch cuts in its fundamental functions. The effect of the double Wick rotation is to exchange so-called short branch cuts on the interval $(-2g,+2g)$ with so-called long branch cuts on $(-\infty,-2g)\cup (+2g,+\infty)$, and vice versa.
The QSC for the Hagedorn temperature thus exhibits the opposite branch-cut structure compared to the one for the spectral problem.

A second effect of the double Wick rotation concerns the boundary conditions. In a certain class of integrable deformations of $\mathcal{N}=4$ SYM theory, characterized by diagonal twists depending on the Cartan charges of $\mathfrak{psu}(2,2|4)$, the deformation parameters enter the TBA by imposing twisted boundary conditions along the direction with finite circumference $L$ \cite{Arutyunov:2010gu}.
The twists result in exponential asymptotics of the fundamental functions of the QSC \cite{Kazakov:2015efa,Gromov:2015dfa}.
Similarly, the Hagedorn temperature and chemical potentials for the different Cartan charges of $\mathfrak{psu}(2,2|4)$ enter by imposing twisted boundary conditions along the direction with finite circumference $1/T$.
Again, we see that the situation for the Hagedorn temperature is related to the one for the spectral problem by a double Wick rotation, which exchanges these two circumferences.
In particular, the Hagedorn temperature formally enters the QSC as a twist, with the consequence that the fundamental functions of the QSC exhibit exponential asymptotics as well. 

We can already see at this heuristic level that introducing twisted boundary conditions along the direction with finite circumference $L$ in the calculation of the Hagedorn temperature has no effect, as we take the limit where $L$ goes to infinity. Thus, as we will discuss in more detail in Sec.\ \ref{sec: deformations}, the Hagedorn temperature of the corresponding integrable deformations of $\mathcal{N}=4$ SYM theory coincides with the Hagedorn temperature in undeformed $\mathcal{N}=4$ SYM theory.

In the twisted spectral problem, the twists (as well as the coupling constants and the Cartan charges $S_1,S_2,J_1,J_2,J_3$) are fixed and considered as the input while the scaling dimension $\Delta$ is kept free and becomes the output. 
The Hagedorn temperature is defined as the temperature for which the free energy per unit classical scaling dimension equals $-1$ \cite{Harmark:2017yrv}.
In the spectral problem, the free energy is connected to the scaling dimension $\Delta$. Thus, for the Hagedorn temperature, we are keeping the scaling dimension $\Delta$ (and thus the free energy) fixed while the single twist encoding the Hagedorn temperature is kept free and constitutes the output.

Finally, it is worth noting that the TBA and QSC in the spectral problem strictly speaking compute the Witten index, while we are interested in the spin-chain partition function or free energy. This difference can be accounted for by a (fermionic) sign in the twists.  

The remainder of this paper is structured as follows: In Sec.\ \ref{sec: QSC}, we describe in detail the QSC equations determining the Hagedorn temperature.
We proceed in Sec.\ \ref{sec: solving} with a description of how to solve these equations, first perturbatively at weak coupling and then numerically at finite coupling.
Sec.\ \ref{sec: chemical potentials} is devoted to the inclusion of chemical potentials, followed by a discussion of the Hagedorn temperature for integrable deformations of $\mathcal{N}=4$ SYM theory in Sec.\ \ref{sec: deformations}.
We include several appendices on technical details of the perturbative solution at weak coupling (App.\ \ref{app: eta functions}) as well as on the TBA equations for the Hagedorn temperature and their relation to the Y-system (App.\ \ref{app: TBA and Y}), T-system (App.\ \ref{app: asymptotic T}--\ref{app: zeroth order}) and QSC (App.\ \ref{app: asymptotic Q}--\ref{app: asymptotic Qai}).

\section{Quantum Spectral Curve for the Hagedorn temperature}
\label{sec: QSC}

In this section, we will give a brief introduction to the Quantum Spectral Curve (QSC) and describe how it can be applied to the problem of determining the Hagedorn temperature $\THag$. In particular, we provide several details which we had to omit in our letter \cite{Harmark:2018red}. 

\subsection{Generalities}
\label{subsec: QSC generalities}

Some of the features of the QSC are universal, in such as they are only reflecting the $\mathfrak{psu}(2,2|4)$ symmetry of $\mathcal{N}=4$ SYM theory and occur for all observables in it to which the QSC has up to now been applied.
Other features are specific to a particular observable.
In this subsection, we will give an overview of the former, while discussing the latter in the subsequent subsections.
For a more general review of the QSC, see e.g.\ Refs.\ \cite{Gromov:2017blm,Kazakov:2018ugh,Levkovich-Maslyuk:2019awk}.

The QSC is also known as analytic Q-system. This Q-system is an equivalent formulation of the T-system, which in turn is an equivalent formulation of the Y-system and the TBA. For more details on the transitions between these formulations, see App.\ \ref{app: TBA and Y}, \ref{app: asymptotic T} and \ref{app: asymptotic Q}.
The Q-system consists of $2^8=256$ functions $Q_{A|I}(u)$ labeled by subsets $A,B\subset \{1,2,3,4\}$. They are (multivariate) functions of the spectral parameter $u\in \mathbb{C}$.
The Q-functions satisfy the finite-difference equations
\begin{align}
Q_{A|I}Q_{Aab|I}&=Q^+_{Aa|I}Q^-_{Ab|I}-Q^-_{Aa|I}Q^+_{Ab|I}\,,
\label{eq: Q-system general 1}\\
Q_{Aa|I}Q_{A|Ii}&=Q^+_{Aa|Ii}Q^-_{A|I}-Q^+_{A|I}Q^-_{Aa|Ii}\,,
\label{eq: Q-system general 2}
\\
Q_{A|I}Q_{A|Iij}&=Q^+_{A|Ii}Q^-_{A|Ij}-Q^-_{A|Ii}Q^+_{A|Ij}\,,
\label{eq: Q-system general 3}
\end{align}
where throughout this paper we are using the notation $f^\pm (u) = f(u\pm \tfrac{i}{2})$.

Note that the Q-functions themselves are gauge dependent. The corresponding Y-system is left invariant under the following gauge transformations:
\begin{equation}
\label{eq: gauge transformations}
 Q_{A|I}\to\frac{g_1^{[+(|A|-|I|+1)]}}{g_1^{[-(|A|-|I|+1)]}}Q_{A|I}\,,\qquad 
 Q_{A|I}\to\frac{g_2^{[+|A|-|I|]}}{g_2^{[-|A|+|I|]}}Q_{A|I}\,,
\end{equation}
where $g_1$ and $g_2$ are arbitrary functions.
Moreover, the Q-system has a $GL(4)\times GL(4)$ symmetry, called $H$ symmetry:
\begin{equation}
\label{eq: H symmetry}
 Q_{A|I}\to \sum_{|B|=|A|,|J|=|I|}(H_1^{[|A|-|I|]})_A{}^B(H_2^{[|A|-|I|]})_I{}^JQ_{B|J}\,,
\end{equation}
where $H_1$ and $H_2$ are $i$-periodic matrices, and $H_A{}^B\equiv H_{a_1}{}^{b_1}H_{a_2}{}^{b_2}\dots H_{a_{|A|}}{}^{b_{|A|}}$.

Using the relations \eqref{eq: Q-system general 1}--\eqref{eq: Q-system general 3}, all functions $Q_{A|I}$ can be written in terms of the functions $\Pb_a\equiv Q_{a|\varnothing}$, $\Qb_i\equiv Q_{\varnothing|i}$, $Q_{a|i}$ and $Q_{\varnothing|\varnothing}$; see e.g.\ Ref.\ \cite{Kazakov:2015efa} for the explicit construction. Moreover, we can use the first gauge transformation \eqref{eq: gauge transformations} to set $Q_{\varnothing|\varnothing}=1$, which we will do throughout.
In terms of these functions, the finite-difference equation \eqref{eq: Q-system general 2} reads 
\begin{equation}
\label{eq: QSC equation 1}
Q_{a|i}^+ - Q_{a|i}^- = \Pb_a \Qb_i \,. 
\end{equation}

It will be convenient to define Hodge-dual Q  functions $Q^{A|I}$ via
\begin{equation}
\label{eq: Hodge dual Q}
 Q^{A|I}\equiv (-1)^{|A||\bar{I}|}\epsilon^{\bar{A}A}\epsilon^{\bar{I}I} Q_{\bar{A}|\bar{I}}\,,
\end{equation}
where $\bar{A}$ ($\bar{I}$) denotes the complement of $A$ ($I$), and $\epsilon$ is the completely anti-symmetric tensor with four indices. In particular, we use the notation $\Pb^a\equiv Q^{a|\varnothing}$ and $\Qb^i\equiv Q^{\varnothing|i}$.
The Q-functions and their Hodge duals satisfy the relations 
\begin{equation}
\label{eq: QSC equation 2}
\Pb_a = - \Qb^i Q^+_{a|i} \,, \quad \Qb_i = - \Pb^a Q^+_{a|i}
\end{equation}
and 
\begin{equation}
\label{eq: QSC orthonormality}
Q_{a|i}Q^{b|i}=-\delta_a^b\,,\qquad Q_{a|i}Q^{a|j}=-\delta_i^j \,.
\end{equation}

For many cases of interest, the QSC possesses an additional symmetry -- called left-right symmetry due to its manifestation in the T-system and Y-system.  
In the left-right-symmetric case, the Q-functions with upper and lower indices satisfy another relation on top of Hodge duality. In particular,
\begin{equation}
\label{eq: QSC upper indices}
\Pb^a = \chi^{ab} \Pb_b \,, \qquad  \Qb^i = \chi^{ij} \Qb_j \,, \qquad Q^{a|i} = \chi^{ab} \chi^{ij} Q_{b|j} \,,
\end{equation}
where 
\begin{equation}
\label{eq: chi matrix}
 \chi=\begin{pmatrix}
       0&0&0&-1\\
       0&0&+1&0\\
       0&-1&0&0\\
       +1&0&0&0
      \end{pmatrix}
      \,.
\end{equation}
The left-right-symmetric case occurs for example in the spectral problem in well-studied closed subsectors and for the Hagedorn temperature in the absence of chemical potentials, or if these chemical potentials satisfy certain constraints to be discussed in Sec.\ \ref{sec: chemical potentials}.

\subsection{Asymptotic solution}
\label{subsec: asymptotic solution}

The asymptotic behavior of the different Q-functions, i.e.\ their behavior for large values of the spectral parameter $u$,  depends on the particular observable that is to be determined by the QSC; and by specifying the asymptotics, one can specify the observable.

Our starting point for determining the asymptotic behavior of the QSC is the T-system we derived in Ref.\ \cite{Harmark:2017yrv}. For large spectral parameter $u$, we have shown that it asymptotes to a constant T-system, which is quoted in Eqs.\ \eqref{consTbfgen1} and \eqref{consTbfgen2}.
Using the relation between the Q-system and the T-system discussed in App.\ \ref{app: asymptotic Q}, this constant T-system can be reproduced via the following Q-functions:
\begin{equation}
\label{eq: asymptotic Q-system: Ps after H}
 \begin{aligned}
  \Pb_{1}(u) &= A_1\left(-\e^{-\frac{1}{2\THag}}\right)^{-i u}\,,\\
  \Pb_{2}(u) &= A_2\left(u+i\frac{1-3 \yy^2}{4 \yy}\right)\left(-\e^{-\frac{1}{2\THag}}\right)^{-i u} \,,\\
  \Pb_{3}(u) &= A_3\left(-\e^{-\frac{1}{2\THag}}\right)^{i u}\,,\\
  \Pb_{4}(u) &= A_4\left(u-i\frac{1-3 \yy^2}{4 \yy}\right)\left(-\e^{-\frac{1}{2\THag}}\right)^{i u}  \,,
 \end{aligned}
\end{equation}
and
\begin{equation}
\label{eq: asymptotic Q-system: Qs after H}
 \begin{aligned}
\Qb_{1}(u)&=B_1\,,\\
\Qb_{2}(u)&=B_2 u\,,\\
\Qb_{3}(u)&=B_3 u^2 \,,\\
\Qb_{4}(u)&=B_4 u \left(u^2+3 \yy^2-2\right) \,,
 \end{aligned}
\end{equation}
where
\begin{equation}
\label{eq: A and B}
A_1A_4=A_2A_3=\frac{i}{\tanh^2 \frac{1}{4\THag}}
\,,\qquad
3B_1B_4=B_2B_3=-8 i\cosh^4 \frac{1}{4\THag}
\,.
\end{equation}
The corresponding $Q_{a|i}(u)$ are given in App.\ \ref{app: asymptotic Qai}.
Note that we have already used some of the $H$ symmetry \eqref{eq: H symmetry} to bring the asymptotic solution into this form. 

It is further convenient to use the remaining $H$ symmetry to set 
\begin{equation}
\label{eq: gauge A and B}
A_1=iA_2=-A_3=-iA_4=\left(\tanh\frac{1}{4\THag}\right)^{-1}\,\qquad
 B_1=B_2 = 1\,.
\end{equation}
In this gauge the asymptotic $\Pb_{a}$ functions for $a=1,2$ and $a=3,4$ transform into each other under $u\rightarrow -u$, while the asymptotic $\Qb_{i}(u)$ functions are either even or odd.
 
The exponential asymptotics of the $\Pb_a$ contain a sign, which is a consequence of the fact that the Hagedorn QSC is based of the partition function instead of the Witten index. This sign is understood to contain a small imaginary part to resolve the branch-cut ambiguity:
\begin{equation}
\label{eq: sign}
 \left(-\e^{-\frac{1}{2\THag}}\right)^{\mp i u}\equiv\left(-\e^{-\frac{1}{2\THag}}+i0\right)^{\mp i u}=\e^{\pm \pi u}\left(\e^{-\frac{1}{2\THag}}\right)^{\mp i u}\,.
\end{equation}

We remark that the asymptotic behavior of the Hagedorn QSC is a special case of the one occurring for the spectral problem in twisted $\mathcal{N}=4$ SYM theory discussed in Ref.~\cite{Kazakov:2015efa}; concretely, the asymptotics in Ref.\ \cite{Kazakov:2015efa} reduce to the ones above upon interchanging, in their notation, $\Pb_a \leftrightarrow \Qb_i$, $x_a \leftrightarrow y_i$ and $A_a \leftrightarrow B_i$, as well as setting,   $x_1=x_2=1/x_3=1/x_4=-\e^{-\frac{1}{2\THag}}$, $y_1=y_2=y_3=y_4=1$ and $\lambda_a=\nu_i=0$.
We will discuss this observation in more detail when discussing chemical potentials in Sec.\ \ref{sec: chemical potentials}.

At tree level, i.e.\ in the free theory, the T-system is constant and obtained by setting $\THag=\THag^{(0)}=\frac{1}{2\log(2+\sqrt{3})}$ in the asymptotic solution. Thus, the tree-level QSC is obtained by setting  $\THag=\THag^{(0)}=\frac{1}{2\log(2+\sqrt{3})}$ in Eqs.\ \eqref{eq: asymptotic Q-system: Ps after H}--\eqref{eq: A and B} and \eqref{eq: assymptotic Qai 1}--\eqref{eq: assymptotic Qai 2}.
It is worth noting that the tree-level QSC is given by polynomials times exponential factors, while the tree-level QSC in the spectral problem typically contains also negative powers of $u$.

\subsection{Branch cuts}
\label{subsec: branch cuts}

\begin{figure}[t]
\centering
 \begin{subfigure}{
  \begin{tikzpicture}[scale=0.6,framed,draw=black!55,background rectangle/.style={draw=black!25,fill=white,
  rounded corners=1ex, ultra thick}, show background rectangle]
  \draw[thick,color=black] (-2,0) -- (2,0);
  \draw[thick,dashed,color=gray] (-2,1) -- (2,1);
  \draw[thick,dashed,color=gray] (-2,2) -- (2,2);
  \draw[thick,dashed,color=gray] (-2,-1) -- (2,-1);
  \draw[thick,dashed,color=gray] (-2,-2) -- (2,-2);
  \node[right] at (2,0) {$\scriptstyle 2g$};
  \node[thick,right,color=gray] at (2,1) {$\scriptstyle 2g+i$};
  \node[thick,right,color=gray] at (2,2) {$\scriptstyle 2g+2i$};
  \node[thick,right,color=gray] at (2,-1) {$\scriptstyle 2g-i$};
  \node[right,color=gray] at (2,-2) {$\scriptstyle 2g-2i$};
  \node[above] at (-3,0) {$ \Qb_i$};
  \node[below,color=gray] at (-3,0) {$ \tilde\Qb_i$};
  \draw[thick,->,color=black]    (-3,0) to[out=+45,in=+90] (0,0);
  \draw[thick,<-,color=gray]    (-3,0) to[out=-45,in=-90] (0,0);
 \end{tikzpicture}}
  \label{subfig: branch cuts of Q}
  \end{subfigure}
 \begin{subfigure}{
  \begin{tikzpicture}[scale=0.6,framed,inner frame xsep=0,draw=black!55,background rectangle/.style={draw=black!25,fill=white,
  rounded corners=1ex, ultra thick}, show background rectangle]
  \draw[thick,color=black] (-4,0) -- (-2,0);
  \draw[thick,dashed,color=gray] (-4,1) -- (-2,1);
  \draw[thick,dashed,color=gray] (-4,2) -- (-2,2);
  \draw[thick,dashed,color=gray] (-4,-1) -- (-2,-1);
  \draw[thick,dashed,color=gray] (-4,-2) -- (-2,-2);
    \draw[thick,color=black] (4,0) -- (2,0);
  \draw[thick,dashed,color=gray] (4,1) -- (2,1);
  \draw[thick,dashed,color=gray] (4,2) -- (2,2);
  \draw[thick,dashed,color=gray] (4,-1) -- (2,-1);
  \draw[thick,dashed,color=gray] (4,-2) -- (2,-2);
  \node[left] at (2,0) {$\scriptstyle 2g$};
  \node[thick,left,color=gray] at (2,1) {$\scriptstyle 2g+i$};
  \node[thick,left,color=gray] at (2,2) {$\scriptstyle 2g+2i$};
  \node[thick,left,color=gray] at (2,-1) {$\scriptstyle 2g-i$};
  \node[left,color=gray] at (2,-2) {$\scriptstyle 2g-2i$};
  \node[above] at (-1,0) {$ \Pb_a$};
  \node[below,color=gray] at (-1,0) {$ \tilde\Pb_a$};
  \draw[thick,->,color=black]    (-1,0) to[out=135,in=+90] (-3,0);
  \draw[thick,<-,color=gray]    (-1,0) to[out=-135,in=-90] (-3,0);
 \end{tikzpicture}}
  \label{subfig: branch cuts of P}
  \end{subfigure}
\caption{Branch cut structure of the QSC for the Hagedorn temperature. There exists a Riemann sheet for which $ \Qb_i$ has a single `short' cut and $ \Pb_a$ has a single `long' cut. (Note that the branch-cut structure is the opposite of the one occurring for the spectral problem.)}
\label{fig: branch cut structure on sheet with single cuts}
\end{figure}
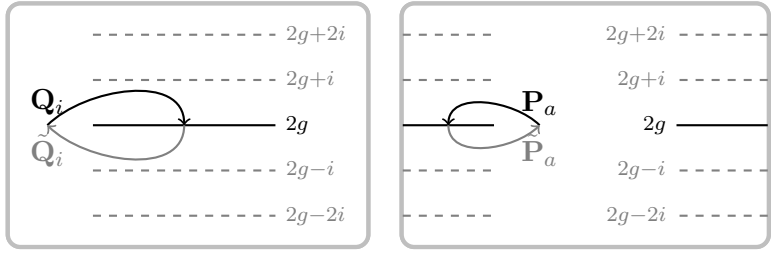

Beyond the asymptotic limit, the analytic structure of the QSC is characterized by the structure of its branch cuts.
Recall that the Hagedorn QSC is based on the physical TBA, while the QSC for the spectral problem is based on the mirror TBA. Since the physical and mirror theory are related by a double Wick rotation, the branch-cut structure of the Hagedorn QSC is exactly opposite to that for the spectral problem.
In particular, there exists a Riemann sheet on which the $\Qb_i$ have a single `short' branch cut on the interval $(-2g,+2g)$, while the $\Pb_a$ have a single `long' cut on $(-\infty,-2g)\cup (+2g,\infty)$; see Fig.\ \ref{fig: branch cut structure on sheet with single cuts}.

\begin{figure}[t]
\centering
   \begin{subfigure}{
  \begin{tikzpicture}[scale=0.6,framed,draw=black!55,background rectangle/.style={draw=black!25,fill=white,
  rounded corners=1ex, ultra thick}, show background rectangle]
  \draw[thick,color=black] (-2,0) -- (2,0);
  \draw[thick,color=black] (-2,-1) -- (2,-1);
  \draw[thick,color=black] (-2,-2) -- (2,-2);
  \draw[thick,dashed,color=gray] (-2,1) -- (2,1);
  \draw[thick,dashed,color=gray] (-2,2) -- (2,2);
  \node[right] at (2,0) {$\scriptstyle 2g$};
  \node[thick,right,color=gray] at (2,1) {$\scriptstyle 2g+i$};
  \node[thick,right,color=gray] at (2,2) {$\scriptstyle 2g+2i$};
  \node[thick,right] at (2,-1) {$\scriptstyle 2g-i$};
  \node[right] at (2,-2) {$\scriptstyle 2g-2i$};
  \node[above] at (-3,0) {$  \Pb_a$};
  \node[below,color=gray] at (-3,0) {$\tilde\Pb_a$};
  \draw[thick,->,color=black]    (-3,0) to[out=+45,in=+90] (0,0);
  \draw[thick,<-,color=gray]    (-3,0) to[out=-45,in=-90] (0,0);
 \end{tikzpicture}}
  \label{subfig: branch cuts of Ptilde UHPA}
  \end{subfigure}
   \begin{subfigure}{
  \begin{tikzpicture}[scale=0.6,framed,draw=black!55,background rectangle/.style={draw=black!25,fill=white,
  rounded corners=1ex, ultra thick}, show background rectangle]
  \draw[thick,color=black] (-2,0) -- (2,0);
  \draw[thick,color=black] (-2,1) -- (2,1);
  \draw[thick,color=black] (-2,2) -- (2,2);
  \draw[thick,dashed,color=gray] (-2,-1) -- (2,-1);
  \draw[thick,dashed,color=gray] (-2,-2) -- (2,-2);
  \node[right] at (2,0) {$\scriptstyle 2g$};
  \node[thick,right] at (2,1) {$\scriptstyle 2g+i$};
  \node[thick,right] at (2,2) {$\scriptstyle 2g+2i$};
  \node[thick,right,color=gray] at (2,-1) {$\scriptstyle 2g-i$};
  \node[right,color=gray] at (2,-2) {$\scriptstyle 2g-2i$};
  \node[above] at (-3,0) {$ \tilde\Pb_a$};
 \end{tikzpicture}}
  \label{subfig: branch cuts of P LHPA}
  \end{subfigure}
 \caption{For the purpose of solving the QSC for the Hagedorn temperature, a different Riemann sheet for $\Pb_a$ is advantageous, obtained from the one depicted in Fig.\ \ref{fig: branch cut structure on sheet with single cuts} via analytic continuation. On this sheet, $\Pb_a$ is an UHPA function with short branch cuts $(-2g-in,+2g-in)$ for $n\in\mathbb{N}_0$, while $\tilde\Pb_a$ is a LHPA function with short branch cuts $(-2g+in,+2g+in)$.
 }
\label{fig: branch cut structure on sheet with short cuts}
\end{figure}
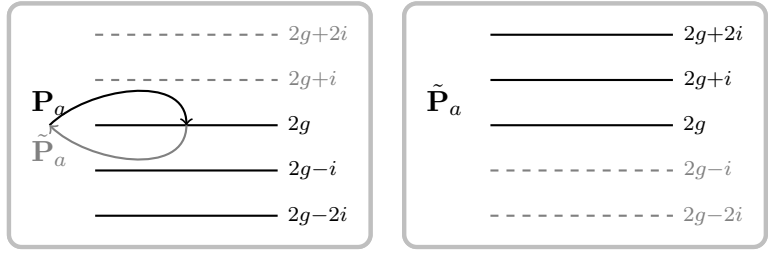

Upon analytic continuation of $\Pb_a$ across the single long cut, one arrives on a Riemann sheet on which $\Pb_a$ is analytic in the upper half plane (UHPA) but possesses an infinite series of shorts cuts in the lower half plane at $(-2g-in,+2g-in)$ for $n\in\mathbb{N}_0$. Similarly, the analytic continuation $\tilde{\Pb}_a$ of $\Pb_a$ across the first short cut is analytic in the lower half plane (LHPA) but possesses an infinite series of shorts cuts in the upper half plane at $(-2g+in,+2g+in)$; see Fig.\ \ref{fig: branch cut structure on sheet with short cuts}.

The sheet with the short branch cut corresponds to the following relation between the spectral plane and the rapidity plane: 
\begin{equation}
x(u) = \frac{u}{2g} \left( 1 + \sqrt{1-\frac{4g^2}{u^2}} \right) ,
\end{equation} 
known as the Zhukowski variable.
In order to determine the QSC at weak or finite $g$, we thus make the ansatz
\begin{equation}
\label{eq: ansatz}
 \begin{aligned}
\Qb_{1}(u)&=1+\sum_{n=1}^\infty \frac{c_{1,2n}(g) g^{2n}}{x(u)^{2n}}\,,\\
\Qb_{2}(u)&=g x(u)\left(1+\sum_{n=1}^\infty \frac{c_{2,2n-1}(g) g^{2(n-1)}}{x(u)^{2n}}\right)\,,\\
\Qb_{3}(u)&= -8 i \cosh(\tfrac{1}{4\THag})^4 (g x(u))^2\left(1+\sum_{n=2}^\infty \frac{c_{3,2n-2}(g) g^{2(n-2)}}{x(u)^{2n}}\right)\,,\\
\Qb_{4}(u)&=-\frac{8 i }{3}\cosh(\tfrac{1}{4\THag})^4(g x(u))^3\left(1+\frac{c_{4,-1}(g) g^{-2}}{x(u)^{2}}+\sum_{n=2}^\infty \frac{c_{4,2n-3}(g) g^{2(n-3)}}{x(u)^{2n}}\right)\,,
 \end{aligned}
\end{equation}
where $c_{i,n}=\mathcal{O}(g^0)$ and starts to contribute at order $1/u^n$.%
\footnote{Note that we have imposed that all functions are either even or odd also at negative powers of $u$; for positive powers, this is a consequence of $H$ symmetry. Moreover, we have already imposed the gauge choice \eqref{eq: gauge A and B} and made the gauge choice that the sum in $\Qb_{3}$ starts with $n=2$.}
In particular, we have $c_{4,-1}=-1+\mathcal{O}(g^2)$. 
The coefficients in this ansatz can be fixed by the gluing conditions (and asymptotic conditions) discussed in the next subsection, up to some remaining gauge symmetry discussed in the subsection thereafter. 

Note that this procedure is exactly opposite to the procedure for the QSC in the spectral problem, where an ansatz for $\Pb_a$ is made.

\subsection{Closing the equations}
\label{subsec: closing eqs}

In the original formulation of the QSC for the spectral problem, the discontinuities of $\Pb_a(u)$ and $\Qb_i(u)$ are determined via certain $i$-periodic functions $\mu_{ab}(u)$ and $\omega_{ij}(u)$ \cite{Gromov:2013pga,Gromov:2014caa}. Our generalization to the Hagedorn temperature will instead be based on an alternative formulation \cite{Gromov:2015vua,Gromov:2015dfa,Gromov:2016rrp,Gromov:2017blm}, in which the analytic continuation across the cut is realized via complex conjugation. The corresponding equations are known as gluing conditions.

The gluing conditions connect $\tilde\Pb_a(u)$ and $\Pb_a(-u)$ on the Riemann sheet with short cuts depicted in Fig.\ \ref{fig: branch cut structure on sheet with short cuts}.
It turns out that for $u\in (-2g,+2g)$,
\begin{equation}
\label{eq: gluing conditions}
 \tilde\Pb_a(u)=(-1)^{1+a}\overline{\Pb_a(u)}=(-1)^{1+a}\Pb_a(-u)\times \begin{cases}
         e^{+2\pi u}& a=1,2,\\                                                                
         e^{-2\pi u}& a=3,4,                                                               \end{cases}
\end{equation}
where complex conjugation is understood to change the sign of the $i0$ in Eq.\ \eqref{eq: sign}.

However, in contrast to the case of the spectral problem, the gluing conditions are not sufficient to fix all coefficients.
In addition we should impose further restrictions on the large-$u$ asymptotics of the $\Pb_a$ functions that are not guaranteed by the ansatz \eqref{eq: ansatz} and the asymptotics of the $Q_{a|i}$ functions. 
In particular, for the perturbative analysis in Subsec.~\ref{subsec: perturbative solution} we should impose the following behavior at $u\to \infty$:
\begin{equation}
\label{eq: asymptotics condition}
 \frac{\Pb_2(u)}{\Pb_1(u)}=-i u +\mathcal{O}(u^0)\,,\qquad
 \frac{\Pb_4(u)}{\Pb_3(u)}=-i u +\mathcal{O}(u^0)\,.
\end{equation}
Note that this follows from  Eqs.\ \eqref{eq: asymptotic Q-system: Ps after H} and \eqref{eq: A and B} when imposing the gauge choice \eqref{eq: gauge A and B}. One can easily modify this for other gauge choices. Similarly, one should also impose the correct asymptotics of the $\Pb_a$ functions as an additional requirement in the numerical solution of Subsec.\ \ref{subsec: numeric solution}.

\subsection{Gauge fixing}
\label{subsec: gauge fixing}

In the particular gauge \eqref{eq: gauge A and B} we chose, we have 
\begin{equation}
\label{eq: gauge fixing}
 \Pb_1(u)\sim -\Pb_3(-u)\,,\qquad \Pb_2(u)\sim +\Pb_4(-u)\,,
\end{equation}
where $\sim$ denotes that the functions have the same coefficients in a large-$u$ expansion while differing in there analytic structure. In particular, $\Pb_1(u)$ and $\Pb_2(u)$ are UPHA while $\Pb_3(-u)$ and  $\Pb_4(-u)$ are LHPA.
The above relations imply similar relations for $Q_{a|i}$:
\begin{equation}
\label{eq: asymptotics condition Qai}
 Q_{a|i}(u)\sim (-1)^{a+i} Q_{a+2|i}(-u)
\end{equation}
for $a=1,2$.
While this property is manifest for the leading coefficients via Eq.\ \eqref{eq: gauge fixing}, it is a gauge choice that we can also impose on the higher-order coefficients to eliminate the gauge freedom at higher loop orders in the weak-coupling expansion as well as in the numeric approach at finite coupling. 

\section{Solving the Quantum Spectral Curve}
\label{sec: solving}

Having derived the QSC equations for the Hagedorn temperature in the previous section, we now set out to solve it -- first perturbatively at weak coupling in Subsec.\ \ref{subsec: perturbative solution} and then numerically at finite coupling in Subsec.\ \ref{subsec: numeric solution}.
While some of the discussion provides further details of the procedure used in our letter \cite{Harmark:2018red}, we also provide new perturbative results up to seven loops and discuss their rich number-theoretic structure, finding in particular that they can be expressed in terms of single-valued harmonic polylogarithms.

\subsection{Perturbative solution at weak coupling}
\label{subsec: perturbative solution}

\subsubsection*{Solution algorithm}

The starting point for the perturbative solution at weak coupling is the tree-level QSC presented in Subsec.\ \ref{subsec: asymptotic solution} and App.\ \ref{app: asymptotic Qai}. 
We can determine the perturbative corrections to the tree-level QSC order by order in the coupling constant $g$ using a slightly modified version of the general solution strategy of Ref.\ \cite{Gromov:2015vua}.

At each order $\ell$, we start with the ansatz \eqref{eq: ansatz} for the $\Qb_i$ functions, which truncates at a finite order in $1/u$ and contains the undetermined coefficients $\THag^{(\ell)}$ as well as $c_{i,n}^{(j)}$ with $\max(0,n)+j=\ell$, where we write $\THag(g^2)=\sum_{\ell=0}^\infty g^{2\ell}\THag^{(\ell)}$ and similarly for $c_{i,n}(g^2)$.
Combining Eqs.\ \eqref{eq: QSC equation 1} and \eqref{eq: QSC equation 2}, we know that the exact solution $Q_{a|i}(u)$ to the QSC satisfies
\begin{equation}
 0= Q_{a|i}(u+\tfrac{i}{2})-Q_{a|i}(u-\tfrac{i}{2})+\Qb_i\Qb^jQ_{a|j}(u+\tfrac{i}{2})\,.
\end{equation}
We define the mismatch in this equation when instead using the tree-level solution $Q_{a|i}^{(0)}$ as
\begin{equation}
 dS_{a|i}\equiv Q_{a|i}^{(0)}(u+\tfrac{i}{2})-Q_{a|i}^{(0)}(u-\tfrac{i}{2})+\Qb_i\Qb^jQ_{a|j}^{(0)}(u+\tfrac{i}{2})\,.
\end{equation}
We can write the exact solution in terms of the tree-level solution $Q_{a|i}^{(0)}$ as 
\begin{equation}
 Q_{a|i}(u)=Q_{a|i}^{(0)}(u)+b_a{}^c(u+\tfrac{i}{2})Q_{c|i}^{(0)}(u)\,.
\end{equation}
Consequently, the matrix $b_a{}^c$ satisfies the first-order difference equation
\begin{equation}
\label{eq: b equation}
 b_a{}^c(u+i)-b_a{}^c(u)=dS_{a|i}(u)Q^{(0)c|i}(u-\tfrac{i}{2})+b_a{}^b(u+i)dS_{b|i}(u)Q^{(0)c|i}(u-\tfrac{i}{2})\,.
\end{equation}

At any loop order of perturbation theory at weak coupling, 
\begin{equation}
 \left(-\e^{-\frac{1}{2\THag(g^2)}}\right)^{\pm iu}=\left(-\e^{-\frac{1}{2\THag(0)}}\right)^{\pm iu}(1+\mathcal{O}(g^2))\,.
\end{equation}
The exponential factors in the functions $\Pb_a$, $Q_{a|i}$ and $b_a{}^c$ are thus the same at any order in perturbation theory.
For convenience, we split off the exponential factor from $b_a{}^c$ as 
\begin{equation}
\label{eq: b underscore}
 b_a{}^c=\begin{cases}
	 \underline{b}_a{}^c & \text{if }(a\leq2 \text{ and } c\leq2)\text{ or }(a\geq3 \text{ and } c\geq3)\,,\\
          \underline{b}_a{}^c\left(-\e^{-\frac{1}{2\THag{(0)}}}\right)^{+2iu} &\text{if }a\geq3 \text{ and } c\leq2\,,\\ 
          \underline{b}_a{}^c\left(-\e^{-\frac{1}{2\THag{(0)}}}\right)^{-2iu} &\text{if }a\leq2 \text{ and } c\geq3\,.
         \end{cases}
\end{equation}

The finite-difference equation \eqref{eq: b equation} for $\underline{b}_a{}^c$ then takes the form 
\begin{equation}
\label{eq: stylized finite-difference equation}
 z f(u+i) -f(u)=h(u)\,,
 \end{equation}
                        with
\begin  {equation}z\in \{1,(2+\sqrt{3})^{-2},(2+\sqrt{3})^2\}
\, , \qquad
(2+\sqrt{3}) = \left(-\e^{-\frac{1}{2\THag{(0)}}}\right)^{-1} \,,
\end{equation}
and where $h$ is given while $f$ is the unknown.
Finite-difference equations of this type have previously been studied in the context of the twisted spectral problem \cite{Kazakov:2015efa} and of cusped Wilson lines \cite{Gromov:2015dfa}. In particular, the work \cite{Gromov:2015dfa} contains the \textsc{Mathematica} package \texttt{TwistTools.m} that implements some of the identities  for solving these finite-difference equations. We derive further identities that are required in our case in App.\ \ref{app: eta functions}.
Following the notation of Ref.\ \cite{Gromov:2015dfa}, denote by $\Sigma$ the operator generating $f$ from $h$ and $z$, i.e.\  $f(u)=\Sigma(h(u),z)$.%
\footnote{Strictly speaking, the authors of Ref.\ \cite{Gromov:2015dfa} denote by $f(u)=\Sigma(h(u))$ the solution to the finite-difference equation $f(u+i) -f(u)=h(u)$, where they allow for an exponential dependence in $f$ and $h$. However, we have found it advantageous to factor out the exponential factor $z^{-iu}$, cf.\ Eqs.\ \eqref{eq: b underscore}--\eqref{eq: stylized finite-difference equation}, indicating this by $\Sigma(\cdot,z)$.}
To have a unique solution, we define $\Sigma(0,z)=0$.
Note that the action of $\Sigma$ crucially depends on whether $z=1$ or $z\neq1$, e.g.\
\begin{equation}
 \begin{aligned}
  \Sigma(1,z)&=\frac{1}{z-1}\,,\qquad &&\text{if }z\neq1\,,\\
  \Sigma(1,1)&=-iu\,.
 \end{aligned}
\end{equation}
For $h(u)$ given by a polynomial in $u$, $\Sigma(h(u),z)$ is again a polynomial in $u$.
Moreover,
\begin{equation}
 \Sigma\left(\frac{1}{u^a},z\right)=-\sum_{n=0}^\infty \frac{z^n}{(u+in)^a}\equiv -\eta_a^z(u)\,.
\end{equation}
At subsequent loop orders, the single $\eta$ functions $\eta_a^z(u)$ can occur on the right-hand-side of the finite-difference equation \eqref{eq: stylized finite-difference equation}. 
However, at any loop order at weak coupling, a solution to the twisted finite-difference equation \eqref{eq: stylized finite-difference equation} can be found in the space of rational functions in $u$ multiplying the generalized multiple $\eta$ functions
\cite{Gromov:2015dfa}:
\begin{equation}
\label{eq: eta definition}
 \eta^{z_1,\dots,z_k}_{s_1,\dots,s_k}(u)\equiv \sum_{n_1>n_2>\dots >n_k\geq 0}\frac{z_1^{n_1}\dots z_k^{n_k}}{(u+i n_1)^{s_1}\dots(u+i n_k)^{s_k}}
 \,.
\end{equation}
These functions satisfy several important properties that we summaries in App.\ \ref{app: eta functions}; some of these have already been described in Ref.\ \cite{Gromov:2015dfa}, while others are new.

The finite-difference equation \eqref{eq: stylized finite-difference equation}, also admits homogeneous solutions.
For $z=1$, these are given by constants, which we have to add to $b$ with coefficients that have to be determined in the following steps.
For general $z$, there is also the homogeneous solution
\begin{equation}
 \mathcal{P}^z_a(u)=\sum_{n=-\infty}^\infty\frac{z^n}{(u+in)^a}=\eta^z_a(u)-\overline{\eta^z_a}(u-i)\,.
\end{equation}
It is removed by imposing $\Pb_a$ to be UHPA.

Note that in the twisted spectral problem \cite{Kazakov:2015efa}, the variables $z$ are pure phases, $|z|=1$, such that $z^*=1/z$. In our case, however, both $z=(2+\sqrt{3})^2>1$ and $z=1/(2+\sqrt{3})^2<1$ occur, and they are real. The functions above are understood to be defined by the given sum representations for values inside of their respective regions of convergence, and by analytic continuation outside of them; see also the discussions in Refs.\ \cite{Marboe:2014gma,Kazakov:2015efa}.

We are now in the position to fix the various coefficients we introduced.%
\footnote{As can be seen from the discussion in Subsec.\ \ref{subsec: asymptotic solution}, already the tree-level coefficients contain $\sqrt{3}$. As discussed in Ref.\ \cite{Marboe:2014gma} for the case of the spectral problem, it is thus more efficient to solve for the algebraic part of the higher-loop coefficients numerically with high accuracy, reconstructing the exact values using the PSLQ algorithm implemented e.g.\ in \textsc{Mathematica} as \texttt{FindIntegerNullVector}.}
Using the orthogonality conditions \eqref{eq: QSC orthonormality}, we find
\begin{equation}
 \begin{aligned}
  -\delta_a^b&=Q_{a|i}Q^{b|i}=(Q_{a|i}^{(0)}+b_a^+{}^cQ_{c|i}^{(0)})(Q^{b|i}{}^{(0)}+b^{+b}{}_d Q^{d|i(0)})=-\delta_a^b+b_a^+{}^b+b^{+b}{}_a+b_a^+{}^cb^{+b}{}_c\,.
 \end{aligned}
\end{equation}
Imposing this fixes half of the coefficients in $b$ stemming from the homogeneous solution.

Via Eq.\ \eqref{eq: QSC equation 2}, we then obtain $\Pb_a$ from $\Qb_i$ and $Q_{a|i}$.
This allows us to impose the gluing conditions \eqref{eq: gluing conditions}.
As in Ref.\ \cite{Gromov:2015vua}, they enter by imposing that $\Pb_a(u)+\tilde\Pb_a(u)$ and $(\Pb_a(u)-\tilde\Pb_a(u))/\sqrt{u^2-4g^2}$ are regular at $u=0$.%
\footnote{In practice, it suffices to impose these  conditions for $a=1,2$.}
In the process, the generalized $\eta$ functions are evaluated at $u=i$, where they are proportional to multiple polylogarithms (MPLs).
More precisely,
\begin{equation}
 \eta^{z_1,\dots,z_k}_{s_1,\dots,s_k}(i)=\frac{(-i)^{s_1+\dots+s_k}}{z_1 \dots z_k} \Li_{z_1,\dots,z_k}(s_1,\dots,s_k)\,,
\end{equation}
where%
\footnote{Note that -- compared to Ref.\ \cite{Duhr:2014woa} -- we are using the opposite order for the arguments of $\Li_{s_1,\dots,s_k}(z_1,\dots,z_k)$; cf.\ (27) of that paper.}
\begin{equation}
 \Li_{s_1,\dots,s_k}(z_1,\dots,z_k)\equiv \sum_{n_1>n_2>\dots >n_k> 0}\frac{z_1^{n_1}\dots z_k^{n_k}}{ n_1^{s_1}\dots n_k^{s_k}}
 \,.
\end{equation}

For all $z_i=1$, the multiple polylogarithms reduce to multiple zeta values (MZV):
\begin{equation}
 \Li_{s_1,\dots,s_k}(1,\dots,1)=\zeta_{s_1,\dots,s_k}\,.
\end{equation}
The weight of these functions and numbers is defined as $s_1+\dots+s_k$.
Note that MPLs have branch cuts; in particular, already classical polylogarithms $\Li_{s}(z)$ have branch cuts in $z$ on the interval $(1,\infty)$. When evaluating say $\Li_{s}(z)$ at $z=\left(-\e^{-\frac{1}{2\THag(0)}}\right)^{-2}=(2+\sqrt{3})^2>1$, we use the $+i0$ prescription \eqref{eq: sign}.
The occurrence of MPLs at this step already indicates that the Hagedorn temperature can be written in terms of these functions.
Knowledge about the identities between the MPLs is crucial when solving for the undetermined coefficients, as solutions to the gluing conditions only exist when taking these identities into account. 

MPLs satisfy the so-called shuffle and stuffle relations, see e.g.\ Ref.\ \cite{Duhr:2014woa} for a review.
They can be used to reduce MPLs of weight less than four to classical polylogarithms.
Moreover, the following inversion identity for classical polylogarithms  exists for $z\notin (0,1]$:
\begin{equation}
 \Li_n(z)+(-1)^n\Li_n(1/z)=-\frac{(2\pi i)^n}{n!}B_n\left(\frac{1}{2}+\frac{\ln(-z)}{2\pi i}\right)\,,
\end{equation}
where $B_n$ is the Bernoulli polynomial.
We can numerically evaluate MPLs to arbitrary precision using \texttt{GiNaC} \cite{Vollinga:2004sn}. This can be used to find identities among MPLs using the PSLQ algorithms implemented e.g.\ in \textsc{Mathematica} as \texttt{FindIntegerNullVector}.

As already indicated in Sec.\ \ref{sec: QSC}, the gluing conditions do not suffice to fix all undetermined coefficients.
In addition, we have to impose the correct asymptotics at large $u$ via Eq.\ \eqref{eq: asymptotics condition}.
This requires to expand $\eta$ functions at large $u$; see App.\ \ref{app: eta functions} for details of this expansion.
Similarly, an expansion is also required when imposing the final condition \eqref{eq: asymptotics condition Qai}.

\subsubsection*{Perturbative results and their number-theoretic properties}

Using the procedure described above, we have solved the Quantum Spectral Curve and determined the Hagedorn temperature up to and including seven-loop order. 
We now present these perturbative results and discuss their number-theoretic properties. We also attach the perturbative results in the ancillary file \texttt{PerturbativeResults.m}.

We write
\begin{equation}
 \THag(g^2)=\sum_{\ell=0}^\infty g^{2\ell}\THag^{(\ell)}\,.
\end{equation}
In general, we find that $\THag^{(\ell)}/\THag^{(0)}$ for $\ell=2,3,4,\dots$ is of mixed transcendentality, with a highest transcendental piece of transcendental degree $2\ell-3$.
This is similar to the spectrum of anomalous dimensions in $\mathcal{N}=4$ SYM theory, which is also of mixed transcendentality, while for example $\ell$-loop scattering amplitudes in $\mathcal{N}=4$ SYM theory are of uniform maximal transcendentality $2\ell$ (see e.g.\ Ref.\ \cite{Duhr:2014woa}).

As previously mentioned, the tree-level Hagedorn temperature is 
\begin{equation}
 \THag^{(0)}=\frac{1}{2\log(2+\sqrt{3})}\approx 0.3796628588\dots\,,
\end{equation}
in full agreement with Ref.\ \cite{Sundborg:1999ue}.
At one-loop order, we find
\begin{equation}
 \THag^{(1)}=\frac{1}{\log(2+\sqrt{3})} \approx 0.7593257175\dots\,,
\end{equation}
in full agreement with Ref.\ \cite{Spradlin:2004pp}.
At two-loop order, we have 
\begin{equation}
 \THag^{(2)}=48 -\frac{86}{ \sqrt{3}} - \frac{48 \Li_1\left(\tfrac{1}{(2+\sqrt{3})^2}\right)}{\log(2+\sqrt{3})}\approx -4.367638556\dots\,,
\end{equation}
in full agreement with Ref.\ \cite{Harmark:2017yrv}.
At three-loop order, we find 
\begin{equation}
\begin{aligned}
 \THag^{(3)}&=-\frac{20}{\sqrt{3}}+\left(\frac{1900}{3}-384 \sqrt{3}\right) \log (2+\sqrt{3})
+\left(384 \sqrt{3}-864\right) \Li_1\left(\tfrac{1}{(2+\sqrt{3})^2}\right)
 \\
 &\phaneq+\frac{432 \Li_1\left(\tfrac{1}{(2+\sqrt{3})^2}\right)^2}{\log (2+\sqrt{3})}+
 624 \Li_2\left(\tfrac{1}{(2+\sqrt{3})^2}\right)
+416 \log (2+\sqrt{3})\Li_1\left(\tfrac{1}{(2+\sqrt{3})^2}\right)\\&\phaneq+\frac{312 \Li_3\left(\tfrac{1}{(2+\sqrt{3})^2}\right)}{\log (2+\sqrt{3})}
 \\
 &
\approx 37.22529358\dots\,,
\end{aligned}
\end{equation}
in full agreement with Ref.\ \cite{Harmark:2018red}.
At four-loop order, we find 
\begin{dmath}
\label{eq: T four MPL}
   \THag^{(4)}=
   +\frac{40}{\sqrt{3}}
   +\left(1272-704 \sqrt{3} \right)\log (2+\sqrt{3})
   +704 \sqrt{3}\Li_1\left(\tfrac{1}{(2+\sqrt{3})^2}\right)
   +\left(8448-\frac{43906}{3 \sqrt{3}}\right) \log ^2(2+\sqrt{3})
   +\left(-18816 +11904 \sqrt{3}\right) \log (2+\sqrt{3})\Li_1\left(\tfrac{1}{(2+\sqrt{3})^2}\right)
   +\left(15552-5952 \sqrt{3}\right)\Li_1\left(\tfrac{1}{(2+\sqrt{3})^2}\right)^2
   -\frac{5184 \Li_1\left(\tfrac{1}{(2+\sqrt{3})^2}\right)^3}{\log (2+\sqrt{3})}
   -6048 \log (2+\sqrt{3})\Li_1\left(\tfrac{1}{(2+\sqrt{3})^2}\right)^2
   +\left(8928 -3840 \sqrt{3}\right) \log (2+\sqrt{3})\Li_2\left(\tfrac{1}{(2+\sqrt{3})^2}\right)
   -288 \Li_{2,1}\left(\tfrac{1}{(2+\sqrt{3})^2},(2+\sqrt{3})^2\right)
   -288 \Li_1\left(\tfrac{1}{(2+\sqrt{3})^2}\right) \zeta_2
   -8928 \Li_2\left(\tfrac{1}{(2+\sqrt{3})^2}\right) \Li_1\left(\tfrac{1}{(2+\sqrt{3})^2}\right)
   +\left(5952 -2560 \sqrt{3}\right) \log ^2(2+\sqrt{3})\Li_1\left(\tfrac{1}{(2+\sqrt{3})^2}\right) 
   +\left(5040-1920 \sqrt{3}\right)\Li_3\left(\tfrac{1}{(2+\sqrt{3})^2}\right)
   -\frac{4608 \Li_3\left(\tfrac{1}{(2+\sqrt{3})^2}\right) \Li_1\left(\tfrac{1}{(2+\sqrt{3})^2}\right)}{\log (2+\sqrt{3})}
   -\frac{144 \Li_1\left(\tfrac{1}{(2+\sqrt{3})^2}\right) \zeta_3}{\log (2+\sqrt{3})}
   -1440 \log ^2(2+\sqrt{3})\Li_2\left(\tfrac{1}{(2+\sqrt{3})^2}\right)
   -4320 \log (2+\sqrt{3})\Li_3\left(\tfrac{1}{(2+\sqrt{3})^2}\right)
   -5400 \Li_4\left(\tfrac{1}{(2+\sqrt{3})^2}\right)
   -\frac{2700 \Li_5\left(\tfrac{1}{(2+\sqrt{3})^2}\right)}{\log (2+\sqrt{3})} \\
   \approx-372.0410892
\end{dmath}

While the Hagedorn temperature up to and including three-loop order is written in terms of classical polylogarithms, the four-loop Hagedorn temperature \eqref{eq: T four MPL} contains the multiple polylogarithm $\Li_{1,2}\left(\tfrac{1}{(2+\sqrt{3})^2},(2+\sqrt{3})^2\right)$. While $\Li_{1,2}\left(\tfrac{1}{(2+\sqrt{3})^2},(2+\sqrt{3})^2\right)$ is of weight three and can thus be expressed in terms of classical polylogarithms, this would introduce $1-\tfrac{1}{(2+\sqrt{3})^2}$ as argument, thus obscuring the origin of this number in the expansion of a generalized $\eta$ function.
Beyond four-loop order, however, it is not possible to write the Hagedorn temperature in terms of classical polylogarithms, as can be seen via the Lie cobracket test in Ref.\ \cite{Goncharov:2010jf}.\footnote{We thank Andrew McLeod for this comment.}

The higher-loop results for $\THag$ become increasingly lengthy when written in terms of MPLs, so we refrain from giving the full expressions here. 
Instead, we will now study the number-theoretic properties of these results, yielding much more compact expressions.

It follows from the discussion above that the arguments of the MPLs occurring in the Hagedorn temperature  are necessarily $z=(2+\sqrt{3})^{-2}$ to some power. Promoting the MPLs to functions of $z$ in this way, we can use that MPLs depending only on a single variable can be expressed in terms of a smaller class of functions called harmonic polylogarithms \cite{Remiddi:1999ew}, as discussed in the following.
MPLs are equivalent to Goncharov polylogarithms \cite{Chen,G91b,Goncharov:1998kja} via\footnote{Recall that -- compared to Ref.\ \cite{Duhr:2014woa} -- we are using the opposite order for the arguments of $\Li_{s_1,\dots,s_k}(z_1,\dots,z_k)$; cf.\ (27) of that paper. However, we are using the same order as Ref.\ \cite{Duhr:2014woa} for $G$; cf.\ (20) of that paper.}
\begin{equation}
\label{eq: Li to G}
 \Li_{s_1,\dots,s_k}(z_1,\dots,z_k)=(-1)^kG\big(\underbrace{0,\dots,0}_{s_1-1},\tfrac{1}{z_1},\dots,\underbrace{0,\dots,0}_{s_k-1},\tfrac{1}{z_1\dots z_k}\big),
\end{equation}
where
\begin{equation}
\begin{aligned}
 G(a_1,\dots, a_n;z) &= \int_0^z \frac{dt}{t-a_1} G(a_2,\dots, a_n,t)\,, \qquad G(;z)=1\,,
 \\
 G(\underbrace{0,\dots,0}_{p};z) &= \frac{\log^p z}{p!}\,.
\end{aligned} 
\end{equation}
Harmonic polylogarithms (HPLs) \cite{Remiddi:1999ew} are defined as 
\begin{equation}
\label{eq: harmonic polylogarithm}
 H(\vec{a};x)=(-1)^{p}G(\vec{a};x),\qquad \text{with }a_i\in \{0,1\}\,, 
\end{equation}
where $p$ counts the number of $a_i$ that are equal to $1$.
Converting the Goncharov polylogarithms resulting from Eq.\ \eqref{eq: Li to G} to Goncharov polylogarithms of the type \eqref{eq: harmonic polylogarithm} can be achieved via a so-called fibration with respect to $\tfrac{1}{(2+\sqrt{3})^2}$, as implemented e.g.\ in \texttt{HyperInt} \cite{Panzer:2014caa}.%
\footnote{Note that compared to \texttt{HyperInt} we are using the opposite order for the arguments of $\Li$, but the same order for the argument in $G$.}

We observe that the weak-coupling results can be expressed in terms of an even smaller class of functions, namely single-valued harmonic polylogarithms (SVHPLs).
SVHPLs $\LL$ were introduced by Francis Brown \cite{Brown:2004ugm} as particular combinations of HPLs of complex conjugated arguments $z$, $z^*$ by requiring the branch cuts in the harmonic polylogarithms to cancel such that one obtains a single-valued function in the $(z,z^*)$ plane.
For example,
\begin{equation}
 \LL(0;z)=H(0;z)+H(0;z^*)=\log(z)+\log(z^*)\,.
\end{equation}
Up to weight 6, they are explicitly given in auxiliary files of Ref.\ \cite{Dixon:2012yy}.
For higher weight, they can be generated using \texttt{HyperLogProcedures} \cite{hyperlogprocedures}, and we attach the relevant expressions up to weight 11 in the auxiliary file \texttt{SVHPLreplacementsUpTo11.m}.
In order to express our result in terms of SVHPLs, we make an ansatz in terms of single-valued harmonic polylogarithms, reexpress them in terms of Goncharov polylogarithms and fix the coefficients by going to a fibration basis as implemented e.g.\ in \texttt{HyperInt} \cite{Panzer:2014caa}.\footnote{We thank Andrew McLeod for his help in the conversion.}
We moreover define the shorthand notation  
\begin{equation}
 \LL_{\vec{a}}\equiv\LL\left(\vec{a};\frac{1}{(2+\sqrt{3})^{2}}\right).
\end{equation}

In terms of SVHPLs, the perturbative results take a much more compact form. The first few orders read
\begin{equation}
 \THag^{(0)}=-\frac{2}{\LL_0}\,,\qquad
 \THag^{(1)}=-\frac{4}{\LL_0}\,,\qquad
 \THag^{(2)}=\frac{1}{\LL_0}\left[96 \LL_1+\left(48-\frac{86}{\sqrt{3}}\right)\LL_0\right]\,,
\end{equation}
\begin{dmath}
 \THag^{(3)}=\frac{1}{\LL_0}\left[-\frac{20}{\sqrt{3}}\LL_0+\left(192 \sqrt{3}-\frac{950}{3}\right) \LL_{0,0}+\left(192 \sqrt{3}-432\right) \left(\LL_{0,1}+\LL_{1,0}\right)-864
   \LL_{1,1}-104 \left(\LL_{0,0,1}-2 \LL_{0,1,0}+\LL_{1,0,0}\right)\right]\,,
\end{dmath}
\begin{dmath}
 \THag^{(4)}=\frac{1}{\LL_0}\left[\frac{40 \LL_0}{\sqrt{3}}+\left(352 \sqrt{3}-636\right) \LL_{0,0}+352 \sqrt{3} \left(\LL_{0,1}+\LL_{1,0}\right)+\left(3168-\frac{21953}{4
   \sqrt{3}}\right) \LL_{0,0,0}+\left(4704-2976 \sqrt{3}\right) \left(\LL_{0,0,1}+\LL_{0,1,0}+\LL_{1,0,0}\right)+\left(7776-2976
   \sqrt{3}\right) \left(\LL_{0,1,1}+\LL_{1,0,1}+\LL_{1,1,0}\right)+15552 \LL_{1,1,1}+288 \zeta_3
   \LL_1-4464 \LL_{0,1,1,0}+\left(1116-480 \sqrt{3}\right)
   \left(\LL_{0,0,0,1}-\LL_{0,0,1,0}-\LL_{0,1,0,0}+\LL_{1,0,0,0}\right)-792 \left(\LL_{0,1,0,1}+\LL_{1,0,1,0}\right)+3024
   \left(\LL_{0,0,1,1}+\LL_{1,1,0,0}\right)-270 \left(\LL_{0,0,0,1,0}-2 \LL_{0,0,1,0,0}+\LL_{0,1,0,0,0}\right)\right]\,,
\end{dmath}
\begin{dmath}
 \THag^{(5)}=\frac{1}{\LL_0}\left[-\frac{80 \LL_0}{\sqrt{3}}+\left(1728-1024 \sqrt{3}\right) \LL_{0,0}-1024 \sqrt{3}
   \left(\LL_{0,1}+\LL_{1,0}\right)+\left(39696-\frac{136369}{2 \sqrt{3}}\right) \LL_{0,0,0}+\left(26464-15072 \sqrt{3}\right)
   \left(\LL_{0,0,1}+\LL_{0,1,0}+\LL_{1,0,0}\right)-15072 \sqrt{3} \left(\LL_{0,1,1}+\LL_{1,0,1}+\LL_{1,1,0}\right)+\left(288-192 \sqrt{3}\right) \zeta_3 \LL_0+\left(4432
   \sqrt{3}-\frac{40026}{5}\right) \LL_{0,0,0,0}+\left(37008 \sqrt{3}-60552\right) \left(\LL_{0,0,1,0}+\LL_{0,1,0,0}\right)-288
   \LL_{0,1,1,0}+\left(34816 \sqrt{3}-60408\right) \left(\LL_{0,0,0,1}+\LL_{1,0,0,0}\right)+\left(67392 \sqrt{3}-102480\right)
   \left(2 \LL_{0,0,1,1}+\LL_{0,1,0,1}+\LL_{1,0,1,0}+2 \LL_{1,1,0,0}\right)+\left(67392 \sqrt{3}-186624\right)
   \left(\LL_{0,1,1,1}+\LL_{1,0,1,1}+\LL_{1,1,0,1}+\LL_{1,1,1,0}\right)-373248 \LL_{1,1,1,1}+\left(4512 \sqrt{3}-7776\right) \zeta_3 \left(\LL_{0,1}+\LL_{1,0}\right)-17280 \zeta_3 \LL_{1,1}+\left(2256 \sqrt{3}-432\right)
   \left(\LL_{0,0,1,1,0}+\LL_{0,1,1,0,0}\right)+\left(9556-6192 \sqrt{3}\right) \left(-2 \LL_{0,0,0,0,1}+\LL_{0,0,0,1,0}+2
   \LL_{0,0,1,0,0}+\LL_{0,1,0,0,0}-2 \LL_{1,0,0,0,0}\right)+\left(12384 \sqrt{3}-41472\right)
   \left(\LL_{0,1,0,0,1}+\LL_{1,0,0,1,0}\right)+\left(41904-14640 \sqrt{3}\right) \left(\LL_{0,0,1,0,1}+\LL_{1,0,1,0,0}\right)+41472
   \left(\LL_{0,1,1,0,1}+\LL_{1,0,1,1,0}\right)+\left(83376-27024 \sqrt{3}\right) \left(\LL_{0,0,0,1,1}-2
   \LL_{1,0,0,0,1}+\LL_{1,1,0,0,0}\right)-42336 \left(\LL_{1,0,0,1,1}+\LL_{1,1,0,0,1}\right)+864
   \left(\LL_{0,1,0,1,1}+\LL_{1,1,0,1,0}\right)-40608 \left(\LL_{0,0,1,1,1}-2 \LL_{0,1,1,1,0}+\LL_{1,1,1,0,0}\right)-2880
   \zeta_5 \LL_1-720 \zeta_3
   \left(\LL_{0,0,1}-\LL_{0,1,0}+\LL_{1,0,0}\right)-24384
   \LL_{0,0,1,1,0,0}+\left(3348-1416 \sqrt{3}\right)
   \left(\LL_{0,0,0,0,1,0}-\LL_{0,0,0,1,0,0}-\LL_{0,0,1,0,0,0}+\LL_{0,1,0,0,0,0}\right)-3072 \LL_{0,1,0,0,1,0}-4992
   \left(\LL_{0,0,1,0,1,0}+\LL_{0,1,0,1,0,0}\right)+14760 \left(\LL_{0,0,0,1,1,0}+\LL_{0,1,1,0,0,0}\right)-7920 \LL_{1,0,0,0,0,1}+360
   \left(\LL_{0,0,1,0,0,1}+\LL_{1,0,0,1,0,0}\right)+7560 \left(\LL_{0,0,0,1,0,1}+\LL_{1,0,1,0,0,0}\right)-\frac{3864}{5}
   \left(\LL_{0,0,0,0,1,0,0}-2 \LL_{0,0,0,1,0,0,0}+\LL_{0,0,1,0,0,0,0}\right)\right]
   \\
   \approx +4132.973342\,.
\end{dmath}

For space reasons, we refrain from showing the full six-loop and seven-loop results here; they can be found in the attached ancillary file \texttt{PerturbativeResults.m}.
Their numeric values are
\begin{equation}
\THag^{(6)}\approx -49510.01767\,,\qquad\THag^{(7)}\approx +625284.5652\,.
\end{equation}
While we have stopped at seven-loop order, there is no technical or conceptual obstacle to going to higher loop orders.

It was previously observed for the Konishi anomalous dimension that the weak-coupling expansion can be expressed in terms of single-valued multiple zeta values (SVMZV) \cite{Leurent:2013mr,Marboe:2014gma,Marboe:2018ugv}, which are SVHPLs evaluated at argument $1$.
Note that for cusped Wilson loops \cite{Gromov:2015dfa}, weak-coupling results have only been obtained in the limit of small twist such that only classical polylogarithms occur; for these, the promotion to single-valued functions is trivial.

We observe that the perturbative results at weak coupling are palindromic in the arguments of the SVHPLs, i.e.\ they stay the same when sending $\LL_{a_1,a_2,\dots,a_n}\to\LL_{a_n,\dots,a_2,a_1}$. It would be interesting to understand the reason for this.

Moreover, the piece of $\THag^{(\ell)}$ with the highest transcendental degree seems to follow a simple structure. In the Konishi anomalous dimension, a corresponding piece could be determined in closed form for any loop order \cite{Leurent:2013mr}, and it would be interesting to develop a similar understanding here.

\subsection{Numeric solution at finite coupling}
\label{subsec: numeric solution}

In this section, we describe how to find the Hagedorn temperature $\THag$ numerically for finite values of the planar coupling $g$. We employ a modified version of the method of Ref.\ \cite{Gromov:2015wca}; see also Refs.\  \cite{Gromov:2017blm,Alfimov:2018cms}.
Moreover, we discuss the strong-coupling behaviour of the Hagedorn temperature.

\subsubsection*{Numeric algorithm}

The key to solving the QSC numerically is to use the ansatz \eqref{eq: ansatz} for the four $\Qb_{i}(u)$ functions, but with the sum truncated so that $n$ has a maximal value that we call $K$.
This enables us to determine the four $\Qb_{i}(u)$ functions in terms of a finite number of parameters $\THag$ and $c_{i,n}$ with $i=1,2,3,4$ and $n \leq K$. To find the values of these parameters for a given coupling $g$, we numerically and thus approximately solve the various conditions discussed in Sec.\ \ref{sec: QSC}, similar to what we previously did at weak coupling. 

For the 16 functions $Q_{a|i}(u)$, we make now the approximate ansatz 
\begin{equation}
\label{eq: Qai numerical ansatz}
Q_{a|i}(u) =  \big(-\e^{-\frac{1}{2T_{\rm H}}}\big)^{-s_a i u} u^{p_{a|i}}  \sum_{n=0}^{N} \frac{B_{a|i,n}}{u^n}\,,
\end{equation}
which is truncated at a finite order depending on the value of $N$ and where 
\begin{equation}
s_a = \left\{
\begin{array}{rl}
1 &\ \mbox{for} \ a=1,2 \,, \\
-1 &\ \mbox{for} \ a=3,4  \,,
\end{array}
\right.
\end{equation}
\begin{equation}
 p_{a|i}=s_a+a+i-3 \,.
\end{equation}
Notice that $Q_{a|i}(u)$ is exponentially convergent for $N \rightarrow \infty$ when the imaginary part of $s_a u$ is large. Thus, this determines the region in which we can use Eq.\ \eqref{eq: Qai numerical ansatz} as a good approximation for sufficiently large $N$. The next step is to find the coefficients $B_{a|i,n}$ given the parameters $\THag$ and $c_{i,n}$. 

A first step is to consider the leading large-$u$ behavior of Eq.~\eqref{eq: QSC equation 1}. On the LHS we use the asymptotic behavior of Eq.~\eqref{eq: Qai numerical ansatz}. On the RHS we use Eqs.~\eqref{eq: asymptotic Q-system: Ps after H} and \eqref{eq: asymptotic Q-system: Qs after H} for the asymptotic behaviors of $\Pb_a$ and $\Qb_i$. Comparing the LHS and RHS, we find
\begin{equation}
B_{a|i,0} = - i s_a \frac{ \e^{-\frac{1}{4T_{\rm H}}}}{ 1+  \e^{-\frac{1}{2T_{\rm H}}}} A_a B_i \,.
\end{equation}
However, one cannot continue in this fashion since one needs to eliminate the $\Pb_a(u)$ functions to find an approximate solution at large $u$. 
To this end, we notice that one can combine Eqs.\ \eqref{eq: QSC equation 1}, \eqref{eq: QSC equation 2} and \eqref{eq: QSC upper indices} to find
\begin{equation}
\label{eq: QSC equation combo}
Q_{a|i}^+ - Q_{a|i}^- = -  Q^+_{a|j} \chi^{jk}   \Qb_i  \Qb_k  \,. 
\end{equation}
This gives a direct relation between the $\Qb_i(u)$ functions and the $Q_{a|i}(u)$ functions, enabling us to use the approximate truncated version of Eq.\ \eqref{eq: ansatz} as well as the approximate truncated expression \eqref{eq: Qai numerical ansatz}. 
To solve Eq.~\eqref{eq: QSC equation combo} at large $u$, we expand the expression
\begin{equation}
\label{eq: QSC equation combo 2}
  \big(-\e^{-\frac{1}{2T_{\rm H}}}\big)^{s_a i u} u^{-p_{a|i}}    \left( Q_{a|i}^+ - Q_{a|i}^- +  Q^+_{a|j} \chi^{jk}   \Qb_i  \Qb_k \right) = \sum_{n=1}^\infty u^{3-n} V_{a|i,n}   \,, 
\end{equation}
in powers of $1/u$ for large $u$, giving the coefficients $V_{a|i,n}$. In this way one finds equations
\begin{equation}
\label{Veqs}
V_{a|i,n} = 0 \,,
\end{equation}
that determine the $B_{a|i,n}$ coefficients.
However, the procedure turns out to be slightly more complicated than one might naively expect. 
Expanding the LHS of Eq.~\eqref{eq: QSC equation combo 2}, one finds that it goes like $u^2$ for large $u$, which is why we started the sum on the RHS with $n=1$. The origin of this behavior is the last term on the LHS. This could seem surprising since we used above that in Eq.~\eqref{eq: QSC equation 1} all three terms starts at the same order. However, this is because in that case we imposed the asymptotics \eqref{eq: asymptotic Q-system: Ps after H} on the $\Pb_a(u)$ functions. Now, instead, we have eliminated the $\Pb_a(u)$ functions and this means one cannot assume anymore  that their asymptotics are of the form \eqref{eq: asymptotic Q-system: Ps after H}. Indeed, it is not difficult to show that with generic behavior of the $B_{a|i,n}$ and $c_{i,n}$ coefficients, the leading behavior of the $\Pb_a(u)$ functions is 
\begin{equation}
\Pb_a (u) \sim u^{a+s_a} \left( - \e^{\frac{1}{2T_{\rm H}}} \right)^{-s_a i u} \,,
\end{equation}
for large $u$.
Therefore, requiring the asymptotic behavior  \eqref{eq: asymptotic Q-system: Ps after H} imposes additional conditions on the $B_{a|i,n}$ and $c_{i,n}$ coefficients.%
\footnote{We note that this is different from the QSC case studied numerically in Ref.\ \cite{Gromov:2015wca} since the last term in Eq.\ \eqref{eq: QSC equation combo} is further suppressed in their case, and hence this issue did not arise there.}

With this in mind, we now describe the procedure to solve the equations \eqref{Veqs} at large $u$. 
One starts by choosing a numerical value for the coupling $g$, $T_{\rm H}$ and the parameters $c_{i,n}$, except for $c_{4,-1}$ and $c_{4,1}$ that are left as free variables.

For $n=1,2,3$, one finds for each $n$ that $V_{a|4,n}=0$ give 4 equations that one can solve linearly for the 4 variables $B_{a|4,n}$, with $a=1,2,3,4$. 
For $n=4,...,9$, one finds for each $n$ that $V_{a|i,n}=0$ with $a=1,2,3,4$ and $i=2,3,4$ give 12  equations that one can solve linearly for the 12 variables $B_{a|2,n-3}$, $B_{a|3,n-3}$ and $B_{a|4,n}$ with $a=1,2,3,4$. 

At this point, all the equations $V_{a|i,n}=0$ are solved for $n=1,...,5$. For $n=6$, one gets a single independent equation which one can solve linearly for $c_{4,-1}$. This in turn solves the equations for $n=7$. Instead for $n=8$ one finds a single independent equation which one can solve linearly for $c_{4,1}$. Finally, for $n=9$ one finds two equations $V_{1|1,9}=0$ and $V_{3|1,9}=0$ that one can solve linearly for $B_{1|1,1}$ and $B_{3|1,1}$. With this, all equations $V_{a|i,n}=0$ are solved for $n=1,...,9$. 

At $n=10$ one has 16 equations $V_{a|i,10}=0$ that one can solve linearly for the 16 variables $B_{a|1,2}$, $B_{a|2,7}$, $B_{a|3,7}$ and $B_{a|4,10}$ with $a=1,2,3,4$. 
At this point, one can look at the orthogonality relations \eqref{eq: QSC orthonormality} to obtain the two parameters $B_{2|1,1}$ and $B_{4|1,1}$. Expanding 
\begin{equation}
\label{eq: orthoexp}
Q_{a|i} Q^{a|j} + \delta^j_i 
\end{equation}
for large $u$, and inserting the solutions of the variables found so far, one finds that its expansion starts at order $u^3$. One can now solve for $B_{2|1,1}$ and $B_{4|1,1}$ by demanding that the $u^3$ terms are zero. 

One proceeds now for $n \geq 11$ as follows. For each successive $n$ one has 16 equations $V_{a|i,n}=0$ that one can solve linearly for the 16 variables $B_{a|1,n-8}$, $B_{a|2,n-3}$, $B_{a|3,n-3}$ and $B_{a|4,n}$ with $a=1,2,3,4$. In this way one can proceed order by order in $n$. To determine the approximate expression \eqref{eq: Qai numerical ansatz} for a given $N$, one needs to solve $V_{a|i,n}=0$ up to and including the order $n=N+8$.
Given this, one can check that the large-$u$ expansion of Eq.~\eqref{eq: orthoexp} is zero to order $u^{5-N}$. 

To recap, given numerical values for the coupling $g$, $T_{\rm H}$ and the parameters $c_{i,n}$ (except for $c_{4,-1}$ and $c_{4,1}$) we have now found the approximate large-$u$ $Q_{a|i}$  \eqref{eq: Qai numerical ansatz} to the desired accuracy given by $N$, and we have in addition obtained $c_{4,-1}$ and $c_{4,1}$. 
One can now use this to determine the functions $\Pb_a$ and its analytic continuation $\tilde{\Pb}_a$ for $a=1,2$ on the real axis. Indeed, 
starting with a large and positive imaginary $u$, we can use Eq.\ \eqref{eq: QSC equation combo} iteratively to find $Q_{a|i}$ closer and closer to the real axis. We choose the starting imaginary part of $u$, written here as $U=\mbox{Im}(u)$, to be a sufficiently large and positive odd integer.  We choose the real part of $u$ such that after the iterative procedure one can find $Q^+_{a|i}(u)$ for $u \in I_P$, where $I_P$ is a suitably chosen set of $P$ points in the real-valued interval $(-2g,2g)$.%
\footnote{We use the function $-2g \cos[\pi (n-\frac{1}{2})/|I_P|]$ for the distribution of points in $I_P$, where $|I_P|$ is the number of points.}
This is possible for $a=1,2$ thanks to the above-mentioned exponential convergence. Building on this, one obtains $\Pb_a(u)$ for $u \in I_P$ from the first line of Eq.\ \eqref{eq: QSC equation 2} where $\Qb_{i}(u)$ is obtained from the truncated version of  Eq.\ \eqref{eq: ansatz} as described above. One can furthermore find the analytic continuation $\tilde{\Pb}_a(u)$ for $u \in I_P$ from 
\begin{equation}
\tilde{\Pb}_a = - \tilde{\Qb}^i Q^+_{a|i} \,,
\end{equation}
where $\tilde{\Qb}_i$ is obtained from the truncated version of  Eq.\ \eqref{eq: ansatz} by replacing $x \rightarrow \tilde{x}= 1/x$. 

Having established a numerical procedure to compute $\Pb_a$ and $\tilde{\Pb}_a$ for $a=1,2$ and $u \in (-2g,2g)$, given values for the coupling $g$, $T_{\rm H}$ and the parameters $c_{i,n}$ (except for $c_{4,-1}$ and $c_{4,1}$), we can construct a function $F$ that parametrize how far we are from obeying the gluing conditions \eqref{eq: gluing conditions}:
\begin{equation}
F(T_{\rm H},\{c_{i,n}\})= \sum_{a=1}^2 \sum_{u \in I_p} \left| \frac{\overline{\Pb_a(u)}}{\tilde\Pb_a(u)} + (-1)^a \right|^2 \,.
\end{equation}
Given a value of the coupling $g$, one can now use a numerical minimization procedure based on $F(T_{\rm H},\{c_{i,n}\})$ that, given a suitable starting guess for $T_{\rm H}$ and the relevant $c_{i,n}$ coefficients, can approach values of these parameters that minimize $F(T_{\rm H},\{c_{i,n}\})$, preferably so that it gets very close to zero. We have implemented this using the Levenberg-Marquardt algorithm, an improved version of Newton's method. 

\subsubsection*{Results}
\begin{figure}
\centering
 \begin{tikzpicture}
  \begin{axis}[width=0.95\textwidth,height=0.5\textwidth,
  /pgf/number format/set thousands separator = {},
    xlabel = $g^2$,
    ylabel = $\THag$,
    xmin=0,
    xtick distance=0.02,
    xmax=0.1,
    xlabel near ticks,
    ylabel near ticks,
    legend pos = north west,
    ]
    \addplot [only marks, black,mark size=1] table[col sep=comma,x index=0,y index=1,header=false] {datagsquared.csv};
     \addplot [blue,domain=0:0.1,dash pattern=on 5pt off 5pt] {0.3796628587501034616094920614718360257603178667293620488696476738937120349688174722116960880927288187};
     \addplot [red,domain=0:0.1,dash pattern=on 3pt off 3pt on 5pt off 3pt] {0.3796628587501034616094920614718360257603178667293620488696476738937120349688174722116960880927288187+2*x*0.3796628587501034616094920614718360257603178667293620488696476738937120349688174722116960880927288187};
     \addplot [green,domain=0:0.1,dash pattern=on 3pt off 2pt on 1pt off 2pt] {0.3796628587501034616094920614718360257603178667293620488696476738937120349688174722116960880927288187
     +2*x*0.3796628587501034616094920614718360257603178667293620488696476738937120349688174722116960880927288187
     -4.367638555796405907668892892349321686325629017277959319603305534781703926169081640936457034108387146*x^2};
     \addplot [orange,domain=0:0.1,dash pattern=on 3pt off 3pt] {0.3796628587501034616094920614718360257603178667293620488696476738937120349688174722116960880927288187
     +2*x*0.3796628587501034616094920614718360257603178667293620488696476738937120349688174722116960880927288187
     -4.367638555796405907668892892349321686325629017277959319603305534781703926169081640936457034108387146*x^2
     +37.22529357787421243436491861519803356882481034413398637710896967279804321042713467179649224766014504*x^3};
      \addplot [pink,domain=0:0.1,dash pattern=on 3pt off 1pt] {0.3796628587501034616094920614718360257603178667293620488696476738937120349688174722116960880927288187
     +2*x*0.3796628587501034616094920614718360257603178667293620488696476738937120349688174722116960880927288187
     -4.367638555796405907668892892349321686325629017277959319603305534781703926169081640936457034108387146*x^2
     +37.22529357787421243436491861519803356882481034413398637710896967279804321042713467179649224766014504*x^3
     -372.0410892732714350882976144039320224571815795418942963903600904713547050597529616695864768904889408*x^4};
     \addplot [lime,domain=0:0.1,dash pattern=on 3pt off 1pt] {0.3796628587501034616094920614718360257603178667293620488696476738937120349688174722116960880927288187
     +2*x*0.3796628587501034616094920614718360257603178667293620488696476738937120349688174722116960880927288187
     -4.367638555796405907668892892349321686325629017277959319603305534781703926169081640936457034108387146*x^2
     +37.22529357787421243436491861519803356882481034413398637710896967279804321042713467179649224766014504*x^3
     -372.0410892732714350882976144039320224571815795418942963903600904713547050597529616695864768904889408*x^4
     +4132.973342158350856260687141114000886043251039894932038174254304102708843814149132659261616807558827*x^5};
      \addplot [cyan,domain=0:0.1,dash pattern=on 3pt off 1pt] {0.3796628587501034616094920614718360257603178667293620488696476738937120349688174722116960880927288187
     +2*x*0.3796628587501034616094920614718360257603178667293620488696476738937120349688174722116960880927288187
     -4.367638555796405907668892892349321686325629017277959319603305534781703926169081640936457034108387146*x^2
     +37.22529357787421243436491861519803356882481034413398637710896967279804321042713467179649224766014504*x^3
     -372.0410892732714350882976144039320224571815795418942963903600904713547050597529616695864768904889408*x^4
     +4132.973342158350856260687141114000886043251039894932038174254304102708843814149132659261616807558827*x^5
     -49510.01767154216559884579057179029976570463102920032677479704234391552432446303632711658604591957489*x^6};
     \addplot [purple,domain=0:0.1,dash pattern=on 3pt off 1pt] {0.3796628587501034616094920614718360257603178667293620488696476738937120349688174722116960880927288187
     +2*x*0.3796628587501034616094920614718360257603178667293620488696476738937120349688174722116960880927288187
     -4.367638555796405907668892892349321686325629017277959319603305534781703926169081640936457034108387146*x^2
     +37.22529357787421243436491861519803356882481034413398637710896967279804321042713467179649224766014504*x^3
     -372.0410892732714350882976144039320224571815795418942963903600904713547050597529616695864768904889408*x^4
     +4132.973342158350856260687141114000886043251039894932038174254304102708843814149132659261616807558827*x^5
     -49510.01767154216559884579057179029976570463102920032677479704234391552432446303632711658604591957489*x^6
     +625284.5652418212199321851011643367256356274655215483909816826427052877954928577128981685190825024459*x^7};
     \addlegendentry{numeric}
     \addlegendentry{0-loop}
     \addlegendentry{1-loop}
     \addlegendentry{2-loop}
     \addlegendentry{3-loop}
     \addlegendentry{4-loop}
     \addlegendentry{5-loop}
     \addlegendentry{6-loop}
     \addlegendentry{7-loop}
  \end{axis}
\end{tikzpicture}
 \caption{Numeric results and weak-coupling approximation at various loop orders
 for the Hagedorn temperature as a function of $g^2$.
 }
\label{fig: weak coupling numerics}
\end{figure}
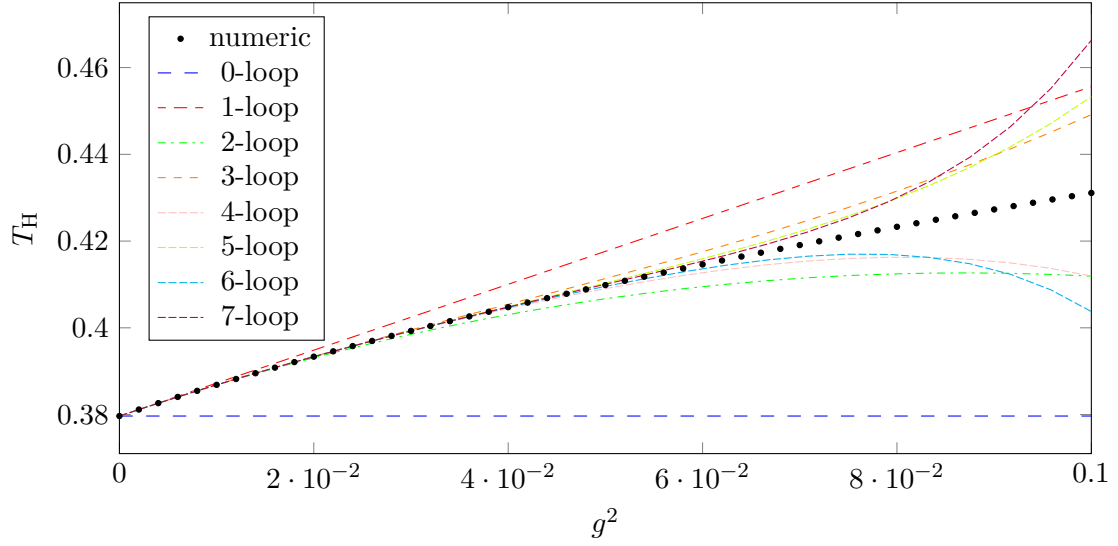

We used our numerical procedure to capture two different regimes of the coupling $g$. We previously reported these results in Ref.\ \cite{Harmark:2018red}. In Fig.\ \ref{fig: weak coupling numerics} we display the results of a series of numerical estimations of the Hagedorn temperature $T_{\rm H}$ as function of $g^2$. We computed 51 values of $T_{\rm H}$, evenly spaced in $g^2$, ending on $g^2=1/10$. This captures the weak-coupling regime, and we compare this successfully to the 7-loop results of Subsec.\ \ref{subsec: perturbative solution}. For this computation we used $K=10$, $N=18$, $U = 31$ and with 20 points in $I_P$. For the last point with $g^2=1/10$, which is the one with the least precision, we computed the Hagedorn temperature $T_{\rm H}=0.43109293576791$ with an estimated accuracy of 14 digits, i.e.\ the uncertainty is in the last digit.

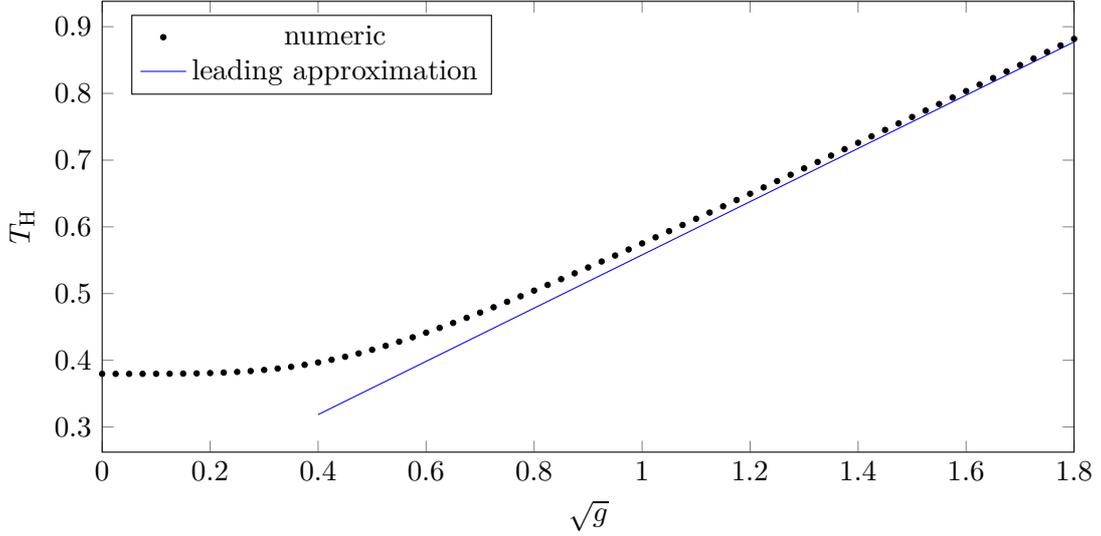
\begin{figure}
\centering
 \begin{tikzpicture}
  \begin{axis}[width=0.95\textwidth,height=0.5\textwidth,
  /pgf/number format/set thousands separator = {},
    xlabel = $\sqrt{g}$,
    ylabel = $\THag$,
    xmin=0,
    xmax=1.8,
    xtick distance=0.2,
    ytick distance=0.1,
    xlabel near ticks,
    ylabel near ticks,
    legend pos = north west,
    ]
    \addplot [only marks, black,mark size=1] table[col sep=comma,x index=0,y index=1,header=false] {datagsqroot.csv};
    \addplot [blue,domain=0.4:1.8]
    {0.3989422804014326779399460599343818684758586311649346576659258296706579258993018385012523339073069364*x
    +0.159032};
     \addlegendentry{numeric}
     \addlegendentry{leading approximation}
  \end{axis}
\end{tikzpicture}
 \caption{Numeric results and leading strong coupling approximation.
 for the Hagedorn temperature as a function of $\sqrt{g}$.
 }
\label{fig: strong coupling numerics}
\end{figure}

In Fig.\ \ref{fig: strong coupling numerics} we display the results of a series of numerical estimations of the Hagedorn temperature $T_{\rm H}$ as function of $\sqrt{g}$. We computed 73 values of $T_{\rm H}$, evenly space in $\sqrt{g}$, ending on $\sqrt{g}=1.8$. This computation required considerably more care and precision to accomplish. For instance, with $\sqrt{g}=1.2625$, we used $K=18$, $N=18$, $U = 31$ and 44 points in $I_P$. This gave an estimated accuracy  of 6 digits with $T_{\rm H}= 0.673348$. 
For $\sqrt{g}=1.75$, we used $K=26$, $N=24$, $U = 51$ and 120 points in $I_P$. This gave an estimated accuracy of 7 digits with $T_{\rm H}= 0.8621292$. 
And for $\sqrt{g}=1.8$ we used $K=26$, $N=30$, $U = 61$ and 160 points 
in $I_P$. This gave an estimated accuracy  of 6 digits with $T_{\rm H}= 0.881729$. For $\sqrt{g} \geq 1.275$ we use a slope estimation with $7$ previous values of $T_{\rm H}$ and $c_{i,n}$ as input to find good initial values for $T_{\rm H}$ and $c_{i,n}$. For $\sqrt{g}=1.8$ we used instead 10 previous values to find good initial data.

We attach the numerical values on which the plots are based in the ancillary files \texttt{datagsquared.csv} and \texttt{datagsqroot.csv}.

The numerical data for $\THag (g)$ in the range $0 \leq \sqrt{g} \leq 1.8$ exhibits an approximately linear behavior 
\begin{equation}
\label{Thag_glarge}
\THag (g) = c_0 \sqrt{g} + c_1 +  \frac{c_2}{\sqrt{g}} +  \frac{c_3}{g}+\CO\left(\frac{1}{\sqrt{g}^3}\right)  \ \ \mbox{for large } \ g  \,,
\end{equation}
with 
\begin{equation}
\label{numc}
c_0 = 0.3989 \spa c_1 = 0.159 \spa c_2=-0.0087 \spa c_3=0.037\,, 
\end{equation}
where the uncertainties in $c_0$ and $c_1$ are in the last digits while the uncertainties in $c_2$ and $c_3$ are $0.0005$ and $0.005$, respectively. The coefficient $c_0$ was previously reported in Ref.\ \cite{Harmark:2018red}.

The approximate leading linear behavior \eqref{Thag_glarge} is quite remarkable, as it agrees well with the Hagedorn temperature of type IIB string theory in flat space. Indeed, for large coupling $g$ we expect that the part of the spectrum of the dual type IIB string theory on AdS$_5\times S^5$ that we probe corresponds to a flat space spectrum \cite{Harmark:2018red}. The Hagedorn temperature of type IIB string theory in flat space is $1/(\sqrt{8}\pi l_s)$ \cite{Sundborg:1984uk}. Using the AdS/CFT dictionary, this becomes 
\begin{equation}
\THag (g) = \sqrt{\frac{g}{2\pi}} + \CO(g^0)  \ \ \mbox{for large } \ g  \,.
\end{equation}
Since $1/\sqrt{2\pi} \simeq  0.39894$ we find agreement with $c_0$ in Eq.\ \eqref{numc}.
Thus, we have found a way to probe flat-space physics of ten-dimensional string theory within $\CN=4$ SYM theory. 

Moreover, since the first appearance of the present paper on the arXiv, the coefficients $c_0$ and $c_1$ have been analytically calculated by considering the inverse Hagedorn temperature as the radius for which a winding mode becomes massless \cite{MaldacenaPrivateCommunication,Urbach:2022xzw}:
\begin{equation}
\label{large_g_TH}
\THag (g)  =  \sqrt{\frac{g}{2\pi}}   + \frac{1}{2\pi}  + \mathcal{O} (1/\sqrt{g})    \ \ \mbox{for large } \ g  \,.
\end{equation}
Numerically $1/(2\pi) \simeq 0.1592$, which thus fits with our numerical data \eqref{numc}.
It would be very interesting to obtain further subleading coefficients as well.

\section{Chemical potentials}
\label{sec: chemical potentials}

In this section, we first discuss in Subsec.\ \ref{subsec: gibbs} the general relation between the Hagedorn temperature and the thermodynamic limit of Gibbs free energy per unit classical scaling dimension when turning on chemical potentials, generalizing Ref.\ \cite{Harmark:2017yrv}. Subsequently, in Subsec.~\ref{subsec: QSC chempots}, we use this to generalize the construction of the Quantum Spectral Curve presented in the previous sections to include non-zero chemical potentials.
Finally, in Subsec.~\ref{subsec: zero_coupling}, we show how our integrability-based approach to the Hagedorn temperature is related to the P\'{o}lya-theory approach for zero 't Hooft coupling.

\subsection{Hagedorn temperature and Gibbs free energy}
\label{subsec: gibbs}

We now generalize the relation found in Ref.\ \cite{Harmark:2017yrv} between the Hagedorn temperature and the Gibbs free energy per unit classical scaling dimension.
This relation forms the basis for applying integrability-based methods to the calculation of the Hagedorn temperature.

We write the full refined partition function of $\CN=4$ SYM theory on $\mathbb{R}\times S^3$ as
\begin{equation}
\label{eq: full refined}
\mathcal{Z}(T,\Omega_i) = \tr \left(\e^{-\beta D + \beta \sum_{i=1}^3 \Omega_i J_i + \beta \sum_{a=1}^2 \Omega_{a+3} S_a} \right) \,.
\end{equation}
Here, $\beta=1/T$ is the inverse temperature and $D$ is the dilatation operator on flat Minkowski space $\mathbb{R}^{1,3}$, which is the image of the Hamiltonian of $\CN=4$ SYM theory on $\mathbb{R}\times S^3$ under a conformal map. $J_1$, $J_2$, $J_3$ are the three $\mathfrak{su}(4)$ R-charges and their chemical potentials are denoted by $\Omega_1$, $\Omega_2$ and $\Omega_3$, respectively. Moreover,  $S_1$ and $S_2$ are the two angular momenta on the $S^3$, with corresponding chemical potentials $\Omega_4$ and $\Omega_5$.

Define the following charges associated to the $\mathfrak{psu}(2,2|4)$ spin chain:
\begin{equation}
q^{(1)} = J_1 - D_0 \spa q^{(2)}=J_2 \spa q^{(3)}=J_3 \spa q^{(4)}=S_1 \spa q^{(5)}=S_2 \,,
\end{equation} 
where
\begin{equation}
D = D_0 + \delta D \,,
\end{equation}
with $D_0$ the classical scaling dimension and $\delta D$ the anomalous scaling dimension.
In terms of these charges, we have
\begin{equation}
\mathcal{Z}(T,\Omega_i) = \tr \left(\e^{-\beta (1-\Omega_1) D_0 - \beta \delta D + \beta \sum_{i=1}^5 \Omega_i q^{(i)}} \right) \,.
\end{equation}

In the planar theory, all states can be written as products of single-trace states, which in turn can be mapped to spin chains. 
We can thus write the single-trace partition function as
\begin{equation}
 Z(T,\Omega_i)=\sum_{m=2}^\infty\e^{-\frac{m}{2}\beta(1-\Omega_1 + F_m(T,\Omega_i))}\eqncom
\end{equation}
where 
\begin{equation}
\label{eq: spin_chain_Z}
 F_m(T,\Omega_i)=-T\frac{2}{m}\log \left( \tr_{\text{spin-chain},D_0=\frac{m}{2}}\left[\e^{-\beta\delta D+ \beta \sum_{i=1}^5 \Omega_i q^{(i)}}\right] \right)    
 \end{equation}
is the spin-chain free energy per unit classical scaling dimension for fixed $D_0=\frac{m}{2}$. 

A multi-trace state in the planar theory is given by a product of single-trace states, in which each bosonic single-trace factor can occur with multiplicity $0,1,2,3,\dots$ and each fermionic single-trace factor can occur with multiplicity $0,1$. Moreover, the energy and other charges of the multi-trace state are given as  sum of the contributions from the individual single-trace factors. 
The multi-trace partition function $\mathcal{Z}(T)$ is thus given by 
\begin{equation}
\label{eq: relation between partition function and free energy}
 \mathcal{Z}(T,\Omega_i)=\exp\sum_{n=1}^\infty\frac{1}{n}\sum_{m=2}^\infty(-1)^{m(n+1)}\e^{-\frac{mn}{2} \beta  (1 - \Omega_1 + F_m(T/n,\Omega_i))}\eqncom
\end{equation}
where the alternating exploits the fact that single-trace states with even $m=2D_0$ are bosons, while those with odd $m=2D_0$ are fermions.

The Hagedorn temperature $\THag$ is the lowest temperature above which the planar partition function diverges. Since $F_m$ is a monotonically decreasing function of the temperature, 
 a divergence first occurs when the $n=1$ contribution to the multi-trace partition diverges. The $n=1$ contribution is
\begin{equation}
\label{eq: n1_contribution}
\sum_{m=2}^\infty(-1)^{m(n+1)}\e^{-\frac{m}{2} \beta  (1 - \Omega_1 + F_m(T,\Omega_i))} \,.
\end{equation}
Define the thermodynamic limit of the Gibbs free energy per unit classical scaling dimension of the $\mathfrak{psu}(2,2|4)$ spin chain:
\begin{equation}
\label{gen_hag_gibbs}
F(T,\Omega_i) = \lim_{m\rightarrow \infty} F_m (T,\Omega_i) \,.
\end{equation}
In terms of this, one sees from Eq.\ \eqref{eq: n1_contribution} using the Cauchy root test that the Hagedorn temperature is determined as \cite{Harmark:2017yrv}
\begin{equation}
\label{eq: general_TH}
F(\THag,\Omega_i) = -1 + \Omega_1 \,,
\end{equation}
since we have $\exp( - \frac{1}{2}\beta (1-\Omega_1 + F(T,\Omega_i))) > 1$ when $T > \THag$.

One can now in principle find $\THag$ for any chemical potentials $\Omega_i$ and any coupling $g$ by computing $F(T,\Omega_i)$. As described in App.\ \ref{app: TBA equations}, this can be done by solving the TBA equations \eqref{newfullTBA1}--\eqref{newfullTBA5} with boundary conditions \eqref{Ysystembcs} to obtain the Y-functions $\CY_{a,s}(u)$ and from this computing $F(T,\Omega_i)$ by Eq.\ \eqref{gibbsfree2}. To simplify this, one can reformulate the TBA equations as Y-system equations, see App.\ \ref{app: Ysystem equations} as used in Ref.\ \cite{Harmark:2017yrv} for the case of zero chemical potentials. Finally, one can alternatively read off $F(T,\Omega_i)$ from Eq.\ \eqref{eq: free_energy_Y}, using the asymptotic behavior of the Y-functions $\CY_{a,s}(u)$. 
However, as is clear from Sec.\ \ref{sec: QSC} and \ref{sec: solving}, a considerably more powerful and efficient approach to find $F(T,\Omega_i)$ is through a set of QSC equations.

\subsection{Quantum Spectral Curve}
\label{subsec: QSC chempots}

We now consider the generalization of the QSC for the case of non-zero chemical potentials. A main part of the QSC does in fact remain unchanged, namely the general structure of the QSC equations and their branch cuts as described in Subsec.\ \ref{subsec: QSC generalities} and \ref{subsec: branch cuts}. 

The main difference compared to the case without chemical potentials lies in the asymptotics of the $\Pb_a$ and $\Qb_i$ functions. To find these asymptotics, we can consider the asymptotic (and thus constant) solution to the T-system. As discussed in App.\ \ref{app: general asymptotic T-system}, it is given by the $\mathfrak{psu}(2,2|4)$ character solution of Ref.\ \cite{Gromov:2010vb} upon identifying the parameters $x_a$ and $y_i$ of Ref.~\cite{Gromov:2010vb} with $\THag$ and the chemical potentials $\Omega_i$ as
\begin{equation}
\label{Hagedorn_xy}
\begin{aligned}
&x_1 = - \e^{\frac{-1 + \Omega_4 + \Omega_5}{2\THag}}
\,, &&
x_2 = - \e^{\frac{-1 - \Omega_4 - \Omega_5}{2\THag}} 
\,, &&
x_3 = - \e^{\frac{1 + \Omega_4 - \Omega_5}{2\THag}}
\,, &&
x_4 = - \e^{\frac{1 - \Omega_4 + \Omega_5}{2\THag}} \,,
\\
&y_1 = \e^{\frac{\Omega_1 + \Omega_2 - \Omega_3}{2\THag}} \,, &&
y_2 = \e^{\frac{\Omega_1 - \Omega_2 + \Omega_3}{2\THag}} \,, &&
y_3 = \e^{\frac{-\Omega_1 + \Omega_2 + \Omega_3}{2\THag}} \,, && 
y_4 = \e^{\frac{-\Omega_1 - \Omega_2 - \Omega_3}{2\THag}} \,.
\end{aligned}
\end{equation}

Fortunately for us, the asymptotics of the $\Pb_a$ and $\Qb_i$ functions that reproduce the $\mathfrak{psu}(2,2|4)$ character solution of Ref.\ \cite{Gromov:2010vb} have already been identified in the context of the twisted spectral problem \cite{Kazakov:2015efa}. However, we have to  make the interchange $\Pb_a \leftrightarrow \Qb_i$ together with $x_a \leftrightarrow y_i$ and $A_a \leftrightarrow B_i$ compared to Ref.\ \cite{Kazakov:2015efa}. The transformation  $\Pb_a \leftrightarrow \Qb_i$, $x_a \leftrightarrow y_i$ and $A_a \leftrightarrow B_i$ is due to the fact that we consider the direct physical theory rather than the mirror theory, as discussed in the Introduction (Sec.\ \ref{sec: Introduction}) as well as in Subsec.\ \ref{subsec: branch cuts} and App.~\ref{app: TBA equations}.%
\footnote{Note that we have also set $\lambda_a=\nu_i=0$ in the more general solution of Ref.\ \cite{Kazakov:2015efa}; cf.\ the discussion in Subsec.\ \ref{subsec: asymptotic solution}.}
This gives
\begin{equation}
\begin{aligned}
\label{PQansatz}
    \Pb_a&\simeq A_a x_a^{-i u}
    u^{\sum_{b< a} \delta_{x_ax_b} -\sum_{i< a}\delta_{x_ay_i}}\,,&\hspace{.5cm}
    \Pb^a&\simeq A^a x_a^{i u}
    u^{\sum_{b> a} \delta_{x_ax_b} -\sum_{i> a}\delta_{x_ay_i}}\\
    \Qb_i&\simeq B_i y_i^{i u}
    u^{-\sum_{a< i}\delta_{x_ay_i}+\sum_{j< i} \delta_{y_iy_j}}\,,&
    \Qb^i&\simeq B^i y_i^{-i u}
    u^{-\sum_{a> i}\delta_{x_ay_i}+\sum_{j> i} \delta_{y_iy_j}}\,,
\end{aligned}
\end{equation}
where
\begin{align}
\label{Asqr}
    A_aA^a&=-\frac 1 {x_a} \frac{\prod_{1\le i \le 4}z_{a,i}}{\prod_{b\ne
    a}z_{b,a}}\qquad (\text{no sum over }a)\,,\\
    \label{Bsqr}
  B_iB^i&=-\frac 1 {y_i} \frac{\prod_{1\le a \le 4}z_{a,i}}{\prod_{j\ne
    i}z_{j,i}}\qquad (\text{no sum over }i)\,,
\end{align}
with
\begin{equation}
  \begin{aligned}
    z_{ab}=-z_{ba}&=
  \begin{cases}x_b-x_a&\textrm{ if }x_a\neq x_b\\
    i x_a(-\sum_{a< c< b}\delta_{x_cx_a}+\sum_{a< i< b}\delta_{x_ay_i}-1)&\textrm{ if }x_a=x_b\textrm{ and }a< b\,,
  \end{cases}\\
z_{ij}=-z_{ji}&=
  \begin{cases}y_i-y_j&\textrm{ if }y_i\neq y_j\\
i y_i(-\sum_{i< k< j}\delta_{y_iy_k}+\sum_{i< a< j} \delta_{x_ay_i}-1
)&\textrm{ if }y_i= y_j\textrm{ and }i< j\,,
\end{cases}\\
z_{ai}=-z_{ia}&=
  \begin{cases}y_i-x_a&\textrm{ if }x_a\neq y_i\\
i x_a (-\sum_{a< b< i}\delta_{x_ax_b}+\sum_{a< j< i}\delta_{x_ay_j})&\textrm{ if }x_a= y_i\textrm{ and }a< i\,,\\
i x_a (\sum_{i< b< a}\delta_{x_ax_b}-\sum_{i< j< a}\delta_{x_ay_j})&\textrm{ if }x_a= y_i\textrm{ and }i< a\,.\\
  \end{cases}
\end{aligned}
\end{equation}

The asymptotics above change discontinuously when certain chemical potentials vanish or become equal to each other, resulting in certain $x_a$ and $y_i$ becoming equal. In the generic case of all chemical potentials being nonvanishing and unequal, the equivalence between the asymptotic T-system and the asymptotic above is illustrated in App.\ \ref{app: asymptotic Q}. Left-right symmetry occurs in the case $\Omega_3=\Omega_5=0$.
Finally, it is easy to see that the asymptotics above reproduce Eqs.\ \eqref{eq: asymptotic Q-system: Ps after H}--\eqref{eq: A and B} in the case where all chemical potentials vanish.

We leave the determination of the gluing conditions, which close the equations, for future work.
We now turn to the special case of the free theory, where the branch cuts vanish and the gluing conditions are thus not required.%
\footnote{Note that the zero-coupling solution is given by exponential factors times polynomials; no (generalized) $\eta$ functions occur, in contrast to the spectral problem.}

\subsection{Zero-coupling limit and single-particle partition functions}
\label{subsec: zero_coupling}

Previously in the literature, the Hagedorn temperature for $\CN=4$ SYM theory with non-zero chemical potentials has only been computed via P\'{o}lya theory \cite{Yamada:2006rx,Harmark:2006di,Harmark:2006ie,Harmark:2007px,Harmark:2014mpa,Suzuki:2017ipd}
based on the methods introduced for the case of vanishing chemical potentials in Refs.\ \cite{Sundborg:1999ue,Aharony:2003sx} at tree level and in Ref.\ \cite{Spradlin:2004pp} at one-loop order. Below we check our tree-level results that emerge from the TBA analysis with the previously derived results of Refs.\ \cite{Yamada:2006rx,Harmark:2006di,Harmark:2007px} using P\'{o}lya theory. This is an important consistency check on our methods.

At zero coupling $g=0$, the full refined partition function \eqref{eq: full refined} can be computed using P\'{o}lya theory \cite{Sundborg:1999ue,Aharony:2003sx}:
\begin{equation}
\label{eq: polya partition function}
\log \mathcal{Z}(T,\Omega_i) = - \sum_{k=1}^\infty \log \left[ 1-\eta_B \left(\frac{T}{k},\Omega_i\right) + (-1)^k \eta_F \left(\frac{T}{k},\Omega_i \right) \right] \,,
\end{equation}
where $\eta_B(T,\Omega_i)$ and $\eta_F(T,\Omega_i)$ are the single-particle partition functions for the bosonic and fermionic modes on the three-sphere, computed for $\CN=4$ SYM theory with chemical potentials in Refs.\ \cite{Yamada:2006rx,Harmark:2006di,Harmark:2007px}.
Defining the total single-particle partition function as 
\begin{equation}
\eta (T,\Omega_i) = \eta_B (T,\Omega_i) + \eta_F (T,\Omega_i) \,,
\end{equation}
we see from Eq.\ \eqref{eq: polya partition function} that the Hagedorn temperature $\THag$ for finite chemical potentials can be determined from
\begin{equation}
\label{eq: SPPF method}
\eta (\THag,\Omega_i)=1 \,.
\end{equation}
Thus, at zero coupling $g=0$ we have two rather explicit sets of methods to compute $\THag$. Either we determine it from the single-particle partition function via Eq.~\eqref{eq: SPPF method}. Or, we use the integrability methods laid out in this paper. Since one cannot solve Eq.~\eqref{eq: SPPF method} explicitly for general chemical potentials $\Omega_i$, a pertinent question is: can we reproduce Eq.~\eqref{eq: SPPF method} from the methods of this paper? 
The answer is affirmative, and a simple route to this is to use the T-system instead of the QSC. At zero coupling $g=0$, the T-system is known to be constant \cite{Harmark:2017yrv}, and hence the general asymptotic T-system reviewed in App.\ \ref{app: T-system} and \ref{app: general asymptotic T-system} should hold for all $u$, i.e.\ $T_{a,s}(u) = T^\infty_{a,s}$. As shown in App.\ \ref{app: zeroth order}, one can derive from the requirement of a constant Y-system that
\begin{equation}
T_{1,0} = 1 \,,
\end{equation}
for $g=0$. In App.\ \ref{app: zeroth order}, we compute $T_{1,0}$ and find that we can identify
\begin{equation}
T_{1,0}(\THag,\Omega_i) = \eta (\THag,\Omega_i) \,.
\end{equation}
Hence, the two methods are equivalent for $g=0$.

\section{Deformations}
\label{sec: deformations}

The maximally supersymmetric Yang-Mills theory admits several deformations that preserve integrability but break some or all of supersymmetry and or Poincar\'{e} symmetry. The best-understood class of these deformations contains so-called diagonal twists and was studied at the level of the Bethe equations in Ref.\ \cite{Beisert:2005if}.
This class encompasses the $\mathcal{N}=1$ supersymmetric one-parameter real-$\beta$ deformation, which is a special case of the  $\mathcal{N}=1$ supersymmetric Leigh-Strassler deformations \cite{Leigh:1995ep}, as well as the non-supersymmetric three-parameter $\gamma_i$ deformation \cite{Frolov:2005dj}.

These deformations change the interaction vertices of the theory -- leaving the tree-level partition function and tree-level Hagedorn temperature trivially the same.
In Ref.~\cite{Fokken:2014moa}, the one-loop corrections to these quantities were calculated in the real-$\beta$ as well as $\gamma_i$ deformation. It was found that while the one-loop partition function depends on the deformation parameters, the one-loop Hagedorn temperature is the same as in $\mathcal{N}=4$ SYM theory.\footnote{The non-conformality \cite{Fokken:2013aea} of the $\gamma_i$ deformation plays no role in the context of the Hagedorn temperature because -- although it is affecting the planar spectrum via finite-size effects \cite{Fokken:2013mza,Fokken:2014soa} -- it only affects a finite number of (single-trace) states, which is irrelevant for the Hagedorn singularity that arises from summing an infinite number of states.}

We will now argue that the Hagedorn temperature of these deformations agrees with that of $\mathcal{N}=4$ SYM theory for any value of the coupling. 
At the heuristic level, we mentioned already in the Introduction that the deformations are encoded in twisted boundary conditions along the direction that becomes infinite in the thermodynamic limit, and thus they do not affect the final result. 

This can equally be seen at the technical level. Deriving TBA equations for the Hagedorn temperature of the deformed theories with diagonal twists following Ref.\ \cite{Harmark:2017yrv}, the starting point are the asymptotic Bethe equations of Ref.\ \cite{Beisert:2005if}. These contain the deformation parameters as a separate factor. Upon taking the logarithm and then the derivative, this factor drops out completely. This shows that the Hagedorn temperature in the deformations with diagonal twist, and in  particular in the real-$\beta$ and $\gamma_i$ deformation, is the same as in $\mathcal{N}=4$ SYM theory at all orders in the 't Hooft coupling.

A related deformation of $\mathcal{N}=4$ SYM theory is the integrable, conformal planar fishnet theory \cite{Gurdogan:2015csr,Sieg:2016vap,Grabner:2017pgm}. It arises by taking a double-scaling limit of the $\gamma_i$ deformation, taking $\gamma_1\to i\infty$, $\lambda\to0$ with $\hat{g}=\e^{-i\gamma_1}\lambda$ fixed.
In this limit, all fields except two complex scalars decouple. 
Naively taking the same double-scaling limit of our result for the Hagedorn temperature would result in $\THag(\lambda=0)$, since the Hagedorn temperature of the $\gamma_i$ deformation is independent of the deformation parameters, as discussed above.
However, the Hagedorn temperature depends on all fields in the theory, whether they decouple or not, and the conformal fishnet theory is conventionally defined without the decoupled fields. 
Using the single-particle partition function of the conformal fishnet theory, $\eta(x)=4x(1-x^2)/(1-x)^4$, the tree-level Hagedorn temperature is given as the solution to the equation $\eta(e^{-1/\THag^{\text{tree,fishnet}}})=1$, $\THag^{\text{tree,fishnet}}=0.508028\dots$;
it clearly differs from the result in $\mathcal{N}=4$ SYM theory.
It would be very interesting to apply our method to calculate the Hagedorn temperature of the conformal fishnet theory at any value of the respective coupling $\hat{g}$, in analogy to what we have done in the present paper for $\mathcal{N}=4$ SYM theory.%
\footnote{The TBA for the conformal fishnet theory is currently known only for a subclass of operators \cite{Basso:2018agi,Basso:2019xay}.
}

\section{Conclusion and outlook}
\label{sec: conclusion}

In this paper, we have derived a Quantum Spectral Curve for the Hagedorn temperature, providing several details deferred in our letters \cite{Harmark:2017yrv,Harmark:2018red}.

The Hagedorn QSC can be efficiently solved perturbatively at weak coupling as well as numerically at finite coupling.
We have extended our previous perturbative results \cite{Harmark:2017yrv,Harmark:2018red} up to and including seven-loop order; we have attached these results in the ancillary file \texttt{PerturbativeResults.m}.
Our perturbative results show interesting number-theoretic properties, namely being expressible in terms of single-valued harmonic polylogarithms (SVHPLs). This is similar to the situation for the spectrum of local composite operators, which is expressible in terms of single-valued multiple zeta values (SVMZVs) \cite{Leurent:2013mr,Marboe:2014gma,Marboe:2018ugv}, which are SVHPLs evaluated at $1$.
 In our case, however, the SVHPLs are evaluated at $\e^{-\frac{1}{\THag(0)}}=\frac{1}{(2+\sqrt{3})^2}$.

 At the technical level, the Hagedorn temperature enters the QSC as a twists. We expect the tools used here, as well as the number-theoretic observations, to be useful also for the systematic solution of the QSC in other cases with twists, such as deformations of $\mathcal{N}=4$ SYM theory \cite{Kazakov:2015efa,Klabbers:2017vtw,Gromov:2017cja,Marboe:2019wyc}, cusped Wilson loops \cite{Gromov:2015dfa,Cavaglia:2018lxi,Grabner:2020nis,Gromov:2021ahm} and color-twisted operators in $\mathcal{N}=4$ SYM theory \cite{Cavaglia:2020hdb}, which were recently investigated in the context of structure constants.
 
Finally, we generalized the Hagedorn QSC to include also chemical potentials, as well as to a class of integrable deformations of $\mathcal{N}=4$ SYM theory. In the letter case, we found that the Hagedorn temperature is identical to the one in $\mathcal{N}=4$ SYM theory for any value of the coupling.

Let us end discussing a number of interesting future directions.
From our numeric solution, we have extracted the leading and subleading behavior of the Hagedorn temperature at large $\lambda$. It would be interesting to extract also further subleading orders, or to develop a systematic approach to solving the Hagedorn QSC perturbatively at strong coupling.
It would also be interesting to solve the Hagedorn QSC in the presence of chemical potentials.
 
 The integrability-based approach to the Hagedorn temperature should also be applicable to further integrable theories. 
 These include deformations of $\mathcal{N}=4$ SYM theory that are not given by diagonal twists, such as the $\eta$ deformation, for which a TBA and QSC has been developed in Refs.\ \cite{Arutynov:2014ota,Klabbers:2017vtw}.
 Further examples of integrable theories to which our approach should be applicable are the ABJM and ABJ theory in the context of AdS$_4$/CFT$_3$ \cite{Aharony:2008ug,Aharony:2008gk}, for which a QSC has equally been studied \cite{Cavaglia:2014exa,Bombardelli:2017vhk,Anselmetti:2015mda,Lee:2017mhh,Lee:2019oml,Lee:2018jvn}.
 Additional examples occur in the context of AdS$_3$/CFT$_2$, for which a QSC has recently been proposed \cite{Cavaglia:2021eqr,Ekhammar:2021pys}. 
 In particular, the theory considered in Ref.\ \cite{Dei:2018jyj} is free of wrapping corrections, making it an interesting starting point to investigate an integrability-based approach to the full partition function.
 
 Moreover, it has been pointed out in Ref.\ \cite{Jiang:2019xdz} that four-point functions of determinant operators exhibit a critical behavior that bears resemblance to Hagedorn behavior 
 and it would be interesting to calculate the critical configuration via a similar integrability based approach as applied here for the Hagedorn temperature. 
 
 Similar techniques to the ones employed for the Hagedorn temperature could also be used to calculate critical exponents, which describe how exactly the partition function diverges when approaching the Hagedorn temperature
 
 In this paper, we have calculated the Hagedorn temperature, which plays the role of a limiting temperature, signaling a phase transition. The low-energy phase ceases to exist at the Hagedorn temperature.
 The confinement-deconfinement phase transition, dual to the Hawking-Page transition, is expected to occur at a lower temperature \cite{Aharony:2003sx}.
 It has recently been shown that the Hagedorn behavior at infinite $N$ is replaced by Lee-Yang behavior at large but finite $N$ \cite{Kristensson:2020nly}. 
 It would be extremely interesting to develop an integrability-based approach also for the confinement-deconfinement temperature.

\begin{acknowledgments}
We are grateful to
Simon Caron-Huot,
Marius de Leeuw,
Nikolay Gromov,
Sebastien Leurent,
Fedor Levkovich-Maslyuk,
Christian Marboe,
Andrew McLeod,
Ryo Suzuki and 
Stijn van Tongeren
for very useful discussions. We thank Juan Maldacena for sharing his calculation \cite{MaldacenaPrivateCommunication} of the first subleading correction at strong coupling with us, for illuminating and interesting correspondences about it, as well as for making us aware of the subsequent work \cite{Urbach:2022xzw}.
T.H.\ acknowledges support from FNU grant number DFF-6108-00340 and the Marie-Curie-CIG grant number 618284.
M.W.\ was supported in part by FNU through grants number DFF-4002-00037, by the ERC advance grant 291092, by the ERC starting grant 757978 and by the research grants 00015369 and 00025445 from Villum Fonden.

\end{acknowledgments}

\appendix

\section{Generalized \texorpdfstring{$\eta$}{eta} functions}
\label{app: eta functions}

In this appendix, we discuss two important properties of the generalized $\eta$ functions \eqref{eq: eta definition}:
\begin{equation}
\label{eq: eta definition appendix}
 \eta^{z_1,\dots,z_k}_{s_1,\dots,s_k}(u)\equiv \sum_{n_1>n_2>\dots >n_k\geq 0}\frac{z_1^{n_1}\dots z_k^{n_k}}{(u+i n_1)^{s_1}\dots(u+i n_k)^{s_k}}
 \,.
\end{equation}

Many properties of the generalized $\eta$ functions were already discussed in Ref.\ \cite{Gromov:2015dfa}, such as their (shuffle) product and their behavior under shifts; we refer the reader to App.~F of Ref.\ \cite{Gromov:2015dfa} for details. Moreover, these relations are conveniently implemented in the \textsc{Mathematica} package \texttt{TwistTools.m} accompanying Ref.\ \cite{Gromov:2015dfa}. We will mainly need two additional properties.

\subsection*{Relations for \texorpdfstring{$\eta$}{eta} with vanishing arguments}

Generalized $\eta$ functions can be further simplified in the case that one of their indices vanishes, $s_i=0$. This simplifications requires us to distinguish whether the corresponding  $z_i\neq 1$ or 
$z_i= 1$; the case $z_i\neq 1$ was already worked out in the \textsc{Mathematica} package \texttt{TwistTools.m} accompanying Ref.\ \cite{Gromov:2015dfa}.

Let us first consider $z_i\neq 1$. We have
\begin{equation}
 \sum_{n_{i-1}>n_i>n_{i+1}}z_i^{n_i}=\frac{z_i^{n_{i-1}}-z_i^{n_{i+1}+1}}{z_i-1}\,,
\end{equation}
from which it follows that
\begin{equation}
 \begin{aligned}
\eta^{z_1,\dots,z_{i-1},z_i,z_{i+1},\dots,z_k}_{s_1,\dots,s_{i-1},0,s_{i+1},\dots,s_k}(u)&= \frac{1}{z_i-1} \eta^{z_1,\dots,z_{i-1} z_i,z_{i+1},\dots,z_k}_{s_1,\dots,s_{i-1},s_{i+1},\dots,s_k}(u)-\frac{z_i}{z_i-1} \eta^{z_1,\dots,z_{i-1},z_i z_{i+1},\dots,z_k}_{s_1,\dots,s_{i-1},s_{i+1},\dots,s_k}(u)\,,\\
 \eta^{z_1,\dots,z_{k-1},z_k}_{s_1,\dots,s_{k-1},0}(u)&=\frac{1}{z_k-1} \eta^{z_1,\dots,z_{k-1} z_k}_{s_1,\dots,s_{k-1}}(u)-\frac{1}{z_k-1} \eta^{z_1,\dots,z_{k-1}}_{s_1,\dots,s_{k-1}}(u)\,. 
 \end{aligned}
\end{equation}

In the case $z_i=1$, however, we have 
\begin{equation}
 \sum_{n_{i-1}>n_i>n_{i+1}}z_i^{n_i}=n_{i-1}-n_{i+1}-1\,.
\end{equation}
Using partial fractioning, we find
\begin{equation}
\frac{n_{i-1}z_{i-1}^{n_{i-1}}}{(u+i n_{i-1})^{s_{i-1}}} = iu \frac{z_{i-1}^{n_{i-1}}}{(u+i n_{i-1})^{s_{i-1}}}-i \frac{z_{i-1}^{n_{i-1}}}{(u+i n_{i-1})^{s_{i-1}-1}}\,,
\end{equation}
and a similar expression for the index $i+1$.
Thus,
\begin{equation}
 \begin{aligned}
\eta^{z_1,\dots,z_{i-1},1,z_{i+1},\dots,z_k}_{s_1,\dots,s_{i-1},0,s_{i+1},\dots,s_k}(u)
&= - i \eta^{z_1,\dots,z_{i-1},z_{i+1},\dots,z_k}_{s_1,\dots,s_{i-1}-1,s_{i+1},\dots,s_k}(u)
+i \eta^{z_1,\dots,z_{i-1}, z_{i+1},\dots,z_k}_{s_1,\dots,s_{i-1},s_{i+1}-1,\dots,s_k}(u)
\\&\phaneq-\eta^{z_1,\dots,z_{i-1},z_{i+1},\dots,z_k}_{s_1,\dots,s_{i-1},s_{i+1},\dots,s_k}(u)
\,,\\
 \eta^{z_1,\dots,z_{k-1},1}_{s_1,\dots,s_{k-1},0}(u)&=iu \eta^{z_1,\dots,z_{k-1}}_{s_1,\dots,s_{k-1}}(u)-i\eta^{z_1,\dots,z_{k-1}}_{s_1,\dots,s_{k-1}-1}(u)\,. 
 \end{aligned}
\end{equation}

An important special case is $s_1=\dots=s_k=0$.
In particular,
\begin{equation}
 \eta^z_0(u)=\frac{\Li_0(z)}{z}=\frac{1}{1-z}\,.
\end{equation}
Note that the sum in $\eta^z_0$ is divergent for $|z|>1$; the right hand side is the appropriate analytic continuation.

\subsection*{Expansion at infinity}

In the perturbative algorithm, it is required to expand the $\eta$ functions around $u=\infty$.
Consider an individual term in Eq.\ \eqref{eq: eta definition}.
We have 
\begin{equation}
\frac{z_i^{n_i}}{(i n_i+i+u)^{s_i}}=\sum_{{j_1}=0}^{\infty} \frac{1}{u^{s_i+j_i}}(-i)^{j_i} \binom{s_i+j_i-1}{j_i} z_i^{n_i}( n_i+1)^{j_i}\,.
\end{equation}
Thus,
\begin{equation}
 \eta^{z_1,\dots,z_k}_{s_1,\dots,s_k}(u+i)=\sum_{j=0}^{\infty}\frac{(-i)^j}{u^{j+\sum_{i=1}^ks_i}}\sum_{\substack{j_1,\dots,j_k\geq0\\j_1+\dots+j_k=j}}\frac{\binom{s_1+j_1-1}{j_1}\dots\binom{s_k+j_k-1}{j_k}}{z_1\dots z_k}\Li_{-j_1,\dots,-j_k}(z_1,\dots,z_k)\,.
\end{equation}
In this expansion, we in particular encounter the formally divergent quantities $\Li_{-n}(1)$ for $n\in \mathbb{N}_0$, which we regularize as $\Li_{-n}(1)=\zeta_{-n}$.

\section{TBA equations and Y-system}
\label{app: TBA and Y}

In this appendix, we provide further details on the TBA and Y-system equations for the Hagedorn temperature, which we deferred in our letter \cite{Harmark:2017yrv}.

\subsection{TBA equations}
\label{app: TBA equations}

In the following, we review the TBA equations for the $\mathfrak{psu}(2,2|4)$ spin chain of $\CN=4$ SYM theory at finite temperature and in the presence of chemical potentials. These are obtained in complete analogy with the TBA equations of the spectral problem for $\CN=4$ SYM theory
\cite{Arutyunov:2009zu,Bombardelli:2009ns,Gromov:2009bc,Arutyunov:2009ur,Gromov:2009tv,Cavaglia:2010nm}. The only subtle difference is that in our case we use the direct physical theory, based on the direct association between temperature of $\CN=4$ SYM and the temperature of the spin chain, whereas one uses instead the so-called mirror theory in the spectral problem, since in that case one should perform a double Wick rotation as explained in the Introduction. 

\begin{figure}
\centering
 \begin{tikzpicture}[scale=0.8]
    \foreach \n in {1, 2, ..., 7} {
        \draw[fill=blue] (0, \n ) circle (0.25);
	\draw[fill=blue] (1, \n ) circle (0.25);
	\draw[fill=blue] (-1, \n ) circle (0.25);
	}
    \foreach \n in {3, 4, ..., 7} {
        \draw (2, \n ) circle (0.25);
	\draw (-2, \n ) circle (0.25);
	}
        \draw[densely dotted] (0, 8 ) circle (0.25);
	\draw[densely dotted] (1, 8 ) circle (0.25);
	\draw[densely dotted] (-1, 8 ) circle (0.25);
	\draw[densely dotted] (2, 8 ) circle (0.25);
	\draw[densely dotted] (-2, 8 ) circle (0.25);
    \foreach \n in {1, 2, ..., 5} {
        \draw[fill=blue] (\n + 1,1) circle (0.25);
	\draw[fill=blue] (-\n - 1,1) circle (0.25);
        }
    \foreach \n in {3, 4, ..., 6} {
        \draw (\n, 2) circle (0.25);
	\draw (-\n, 2) circle (0.25);
	}
    \foreach \n in {-6,-5, ..., 6} {
        \draw (\n, 0) circle (0.25);
	}	
        \draw[densely dotted] (6 + 1,1) circle (0.25);
	\draw[densely dotted] (-6 - 1,1) circle (0.25);
        \draw[densely dotted] (6 + 1,0) circle (0.25);
	\draw[densely dotted] (-6 - 1,0) circle (0.25);
        \draw[densely dotted] (6 + 1,2) circle (0.25);
	\draw[densely dotted] (-6 - 1,2) circle (0.25);
	\draw[fill=blue] (2,2) circle (0.25);
	\draw[fill=blue] (-2,2) circle (0.25);
 	\draw[->] (0,9) -- (0,9.5) node[above] {$a$};
 	\draw[->] (8,0) -- (8.5,0) node[right] {$s$};
\end{tikzpicture}
\caption{There exists one Y-function $\CY_{a,s}(u)$ for each $(a,s)\in M$ (defined in Eq.\ \eqref{eq: definition M} and shown in blue) and one T-function $T_{a,s}$ for each $(a,s)\in \hat{M}$ (defined in Eq.\ \eqref{eq: definition Mhat} and shown as union of blue and white).}
\label{fig: Y and T hook}
\end{figure}
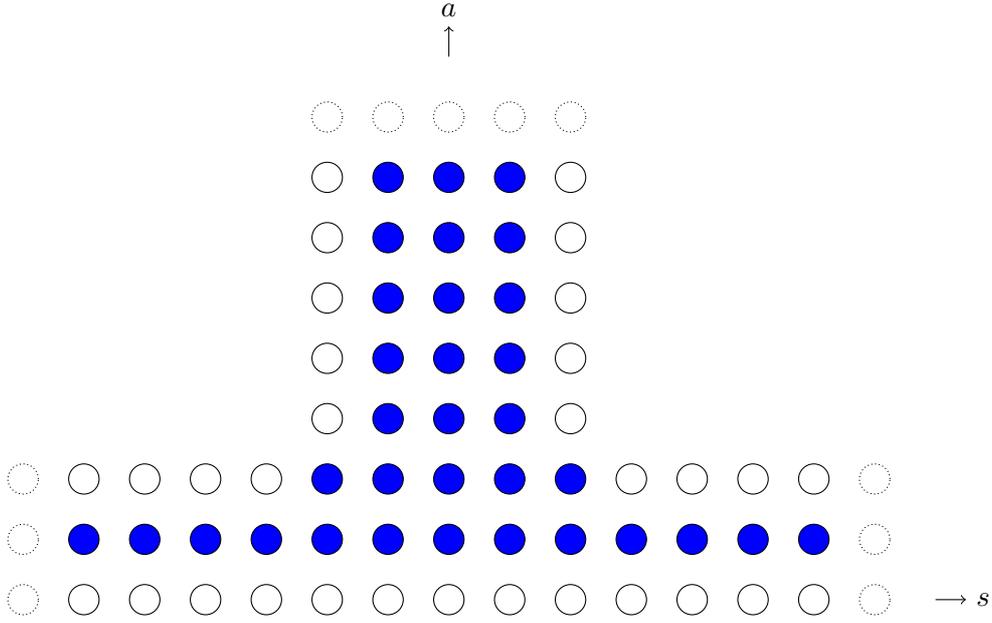

To obtain TBA equations for the $\mathfrak{psu}(2,2|4)$ spin chain of $\CN=4$ SYM theory, one starts with the asymptotic Bethe ansatz of Refs.\ \cite{Beisert:2004hm,Beisert:2006ez}. 
One then assumes the string hypothesis, which is a specific assumption to what constitutes the complete set of solutions to the asymptotic Bethe ansatz equations. In particular, these solutions are organized in terms of so-called strings.
One considers an ensemble of configurations of strings of various types and lengths in the continuum limit by sending the length of the spin chains to infinity. This is described by densities of values of the centers of the strings realized by the particular configuration of strings in the ensemble as well as the density of center values not realized.  From this one can define the entropy per unit classical scaling dimension $s$, the  energy per unit classical scaling dimension $e$, and the charges per unit classical scaling dimension $\tilde{q}^{(i)}$ with $i=1,...,5$.
Imposing the first law of thermodynamics $\delta e = T \delta s + \sum_i \Omega_i \delta \tilde{q}^{(i)}$ at temperature $T$ and for chemical potentials $\Omega_i$, one obtains the TBA equations in terms of Y-functions and an associated free energy per unit classical scaling dimension. The Y-functions are defined as the ratio of the density of values not realized  over the density of values that are realized by the configurations in the ensemble. 
Following the notation of Ref.\ \cite{Harmark:2017yrv}, one can define Y-functions $\CY_{a,s}(u)$ with $(a,s)$ in the set $M$,
\begin{equation}
\label{eq: definition M}
(a,s) \in M=\{(a,s)\in\mathbb{N}_{\geq0}\times\mathbb{Z} \,|\,a=1 \vee |s|\leq 2 \vee \pm s=a=2\} \,.
\end{equation}
They form a so-called hook, illustrated in Fig.\ \ref{fig: Y and T hook}.

In terms of the Y-functions $\CY_{a,s}(u)$, the TBA equations for the direct theory take the following form:
\begin{eqnarray}
\label{newfullTBA1}
\log \CY_{n,0} &=& -\frac{1}{T} \epsilon_n - \sum_{m=1}^\infty \log (1+\CY_{m,0} ) \star ( K^{m,n}+ \Sigma^{m,n} )  - \sum_{a=\pm 1} \sum_{m=1}^\infty     \log (1+\CY_{m+1,a} ) \star  \Theta^{m,n}_{a,0}  \nn \\ && - \sum_{a=\pm 1} \left\{ \log (1+\CY_{1,a} ) \checkstar  \Theta^{1,n}_{3a,0} + \log (1+\CY^{-1}_{2,2a} ) \checkstar  \Theta^{1,n}_{4a,0} \right\}  - \frac{n}{T} ( \Omega_{1}-\Omega_{2} )  \,,
\end{eqnarray}
\begin{eqnarray}
\label{newfullTBA2}
\log \CY_{n+1,\pm 1}(u) &=& -\sum_{m=1}^\infty     \log (1+\CY_{m+1,\pm 1}  (v) ) \star  K^{m,n}(v,u) -  \log \frac{1+\CY_{1,\pm 1}(v)}{1+\CY_{2,\pm 2} ^{-1} (v) } \checkstar a_n (v-u)\nn \\ &&-  \sum_{m=1}^\infty     \log (1+\CY_{m,0} (v) ) \star  \Theta^{m,n}_{0,\pm 1}(v,u)   - \frac{n}{T} ( \Omega_{2}\pm \Omega_{3} )  \,,
\end{eqnarray}
\begin{eqnarray}
\label{newfullTBA3}
\log \CY_{1,\pm (n+1)} (u)&=& \sum_{m=1}^\infty     \log (1+\CY_{1,\pm (m+1)}^{-1}(v) ) \star  K^{m,n}(v,u) +  \log \frac{1+\CY_{1,\pm 1}(v)}{1+\CY_{2,\pm 2} ^{-1} (v)} \checkstar a_n (v-u) \nn \\ && + \frac{n}{T} (\Omega_4 \pm \Omega_5) \,,
\end{eqnarray}
\begin{eqnarray}
\label{newfullTBA4}
\log \CY_{1,\pm 1} (u) &=& -\sum_{m=1}^\infty \log (1+\CY_{m,0} (v) ) \star \Theta_{0,\pm 3}^{m,1} (v,u) - \sum_{m=1}^\infty  \log \frac{ 1+ \CY_{m+1,\pm 1}(v) }{   1+ \CY_{1,\pm (m+1)}^{-1}(v)  } \star a_m(v-u)\nn \\ &&
- \frac{1}{2T} ( \Omega_2 \pm \Omega_3 - \Omega_4 \mp \Omega_5) \,,
\end{eqnarray}
\begin{eqnarray}
\label{newfullTBA5}
\log \CY_{2,\pm 2} (u) &=& \sum_{m=1}^\infty \log (1+\CY_{m,0} (v) ) \star \Theta_{0,\pm 4}^{m,1} (v,u) + \sum_{m=1}^\infty  \log \frac{ 1+ \CY_{m+1,\pm 1}(v) }{   1+ \CY_{1,\pm (m+1)}^{-1}(v)  }\star a_m(v-u)\nn \\ &&
+ \frac{1}{2T} ( \Omega_2 \pm \Omega_3 - \Omega_4 \mp \Omega_5) \,,
\end{eqnarray}
where $n  \geq 1$ and $u \in \R \setminus ( - 2g,2g)$ using the kernels $\epsilon$, $K$, $\Sigma$, $\Theta$ and $a$ defined in App.\ \ref{sec:kernels}. We define here the convolutions
\begin{align}
\label{starconv}
(f \star g) (u) &= \int_{-\infty}^\infty \de v f(v) g(v,u)\,,\\
\label{checkstarconv}
(f \checkstar g) (u) &= \left( \int_{-\infty}^{-2g} \de v + \int_{2g}^\infty \de v \right)  f(v) g(v,u)\,,\\
\label{hatstarconv}
(f \hatstar g) (u) &= \int_{-2g}^{2g} \de v f(v) g(v,u)\,,
\end{align}
for the functions $f(u)$ and $g(v,u)$. 
The above TBA equations for the direct theory were previously considered in Refs.\ \cite{Cavaglia:2010nm,Arutynov:2014ota} but in different thermodynamic limits.

Note that we have assumed 
\begin{equation}
\label{assume_omega}
\Omega_1 \geq \Omega_2 \geq \Omega_3 \geq 0 \spa \Omega_4 \geq \Omega_5 \geq 0 \,.
\end{equation}
Concretely, this is important to get the right asymptotics for large $n$.%
\footnote{Note that one can always reparametrize the given charges to obtain this.}
One finds the large $n$ asymptotics
\begin{equation}
\label{Ysystembcs}
\begin{array}{c} \ds
\lim_{n\rightarrow \infty} \frac{\log \CY_{n,0}}{n} = - \frac{\Omega_1-\Omega_2}{T}
\spa  \lim_{n\rightarrow \infty} \frac{\log \CY_{n,\pm 1}}{n} = - \frac{\Omega_2 \pm \Omega_3}{T}
 \\[4mm] \ds \lim_{n\rightarrow \infty} \frac{\log \CY_{1,\pm n}}{n} = \frac{\Omega_4 \pm \Omega_5}{T}  \,.\end{array}
\end{equation}

The Gibbs free energy per unit classical scaling dimension is given by
\begin{equation}
\label{gibbsfree2}
F (T,\Omega_{i})= - T \sum_{n=1}^\infty  \int_{-\infty}^\infty du \, \theta_n (u) \log ( 1 + \CY_{n,0} (u) ) \,,
\end{equation}
with $\theta_n(u)$ defined in Eq.\ \eqref{theta_n_def}. The TBA equations \eqref{newfullTBA1}--\eqref{newfullTBA5} determine the Y-functions $\CY_{a,s}(u)$ at a given temperature $T$ and chemical potentials $\Omega_i$. Inserting this into Eq.\ \eqref{gibbsfree2}, one finds the Gibbs free energy of the $\mathfrak{psu}(2,2|4)$ spin chain at the temperature $T$ and chemical potentials $\Omega_i$. 
From the Gibbs free energy \eqref{gibbsfree2}, one can now determine the Hagedorn temperature $\THag$ from Eq.~\eqref{eq: general_TH}.

\subsection{Y-system}
\label{app: Ysystem equations}

While the TBA equations in principle solve the problem of determining the Gibbs free energy per unit classical scaling dimension $F (T,\Omega_{i})$ -- and thus the Hagedorn temperature -- in practice they are quite difficult to work with. A first step towards a simplification is to recast them  in terms of so-called Y-system equations, as done for the case of the spectral problem in Refs.~\cite{Gromov:2009tv,Bombardelli:2009ns,Gromov:2009bc,Arutyunov:2009ur,Cavaglia:2010nm}.
In case of zero chemical potentials, we used this in Ref.\ \cite{Harmark:2017yrv} to determine the Hagedorn temperature up to and including order $g^4$ (two loops).

Analytically extending the Y-functions $\CY_{a,s}(u)$, one finds from the analytic properties of the TBA equations \eqref{newfullTBA1}--\eqref{newfullTBA5} that they are analytic in the strip with $\mbox{Im}(u)< \frac{1}{2} |a-|s||$.
One can now derive the Y-system equations
\begin{equation}
\label{eq: general Y-system}
 \log\CY_{a,s}=\log\frac{(1+\CY_{a,s-1})(1+\CY_{a,s+1})}{(1+\CY_{a-1,s}^{-1})(1+\CY_{a+1,s}^{-1})}\star s\eqncom
\end{equation}
valid on $\mathbb{R}$, where
\begin{equation}
s(u)=(2 \cosh \pi u)^{-1} \,.
\end{equation}
The convolution with $\CY_{1, \pm 1}$ and $\CY_{2,\pm 2}$ in Eq.\ \eqref{eq: general Y-system} for $(a,s)=(2,\pm 1),(1,\pm 2)$ is understood to be $\checkstar$.
The Y-system equations \eqref{eq: general Y-system} holds for all $(a,s)\in M$ except when $(a,s)=(1,0)$, $(a,s)=(1,\pm 1)$ or $(a,s)=(2,\pm 2)$.
For the exceptions $(a,s)=(1,\pm 1)$ and $(a,s)=(2,\pm 2)$, we have the non-local equations
\begin{equation}
\label{eq: Y11_times_Y22}
\log \CY_{1,\pm 1}\CY_{2,\pm 2} (u) = \sum_{m=1}^\infty \log (1+\CY_{m,0} (v) ) \star ( \Theta_{0,\pm 4}^{m,1}- \Theta_{0,\pm 3}^{m,1} ) (v,u)  \,,
\end{equation}
and
\begin{equation}
\log \frac{\CY_{2,\pm 2}}{ \CY_{1,\pm 1}}  =  \sum_{m=1}^\infty  \log \frac{ (1+ \CY_{m+1,\pm 1})^2 }{  ( 1+ \CY_{1,\pm (m+1)}^{-1})^2 (1+\CY_{m,0})  } \star a_m 
+ \frac{1}{T} ( \Omega_2 \pm \Omega_3 - \Omega_4 \mp \Omega_5) \,,
\end{equation}
with $a_n(u)$ defined in Eq.\ \eqref{an_def}. 
For the exception $(a,s)=(1,0)$, we find
\begin{equation}
\label{eq: CY_10}
 \log \CY_{1,0}  = - \rho\hatstar s  + \log (1+\CY_{1,-1}) (1+\CY_{1,1})\checkstar s-\log(1+\CY_{2,0}^{-1})\star s\,,
\end{equation}
with the source term $\rho(u)$ given by
\begin{equation}
 \begin{aligned}
\rho &=  \frac{\epsilon_0}{T}    +  \log (1+\CY_{1,-1})(1+\CY_{1,1}) (1+\CY_{2,-2}^{-1})(1+\CY_{2,2}^{-1})  \checkstar  H_0   \\ &\phaneq  +  \sum_{m=1}^\infty \log (1+\CY_{m+1,-1})(1+\CY_{m+1,1})\star 
\left(H_m +H_{-m}\right)   \\&\phaneq+ \sum_{m=1}^\infty \log (1+\CY_{m,0} ) \star \Sigma^{m} \,.
 \end{aligned}
\label{eq: rho}
\end{equation}
See App.\ \ref{sec:kernels} for the definitions of $\epsilon_0$, $H_m$ and $\Sigma_m$. The above equations provide an equivalent yet more tractable version of the TBA equations \eqref{newfullTBA1}--\eqref{newfullTBA5}.

\subsection{Hagedorn temperature from asymptotics of Y-functions}

For large $u$, one can infer from the TBA equations \eqref{newfullTBA1}--\eqref{newfullTBA5} that the Y-functions $\CY_{a,s}(u)$ asymptote to finite values
\begin{equation}
\label{Y_infty}
\CY_{a,s}^\infty = \lim_{u\rightarrow \infty} \CY_{a,s} (u) \,.
\end{equation}
This is important primarily with respect to reformulating the TBA equations in terms of a QSC \cite{Harmark:2018red} but it is also useful for solving the Y-system at zero coupling $g=0$ \cite{Harmark:2017yrv}. 
As we describe in App.\ \ref{app: asymptotic T}, the asymptotic Y-system $\CY_{a,s}^\infty$ can be deduced from a constant T-system, and this in turn provides the seed for all-important boundary conditions for the asymptotic behavior of the $\Pb_a(u)$ and $\Qb_i(u)$ in the QSC, as described in Subsec.\ \ref{subsec: asymptotic solution} and App.\ \ref{app: asymptotic Q}. 
A crucial piece in this is how we can infer the Gibbs free energy per unit classical scaling dimension $F(T,\Omega_i)$ from the asymptotic behavior of Y-functions in the Y-system, since this in turns makes it possible to find it from the asymptotics of the T-functions, and therefore also from the asymptotics of the $\Pb_a(u)$ and $\Qb_i(u)$ functions in the QSC. 
Below we provide this piece, by showing that one can infer the Gibbs free energy per unit classical scaling dimension $F(T,\Omega_i)$ directly from the asymptotic Y-system $\CY_{a,s}^\infty$. Indeed, this provides the argument behind Eq.\ \eqref{eq: zth}, which we used in Ref.~\cite{Harmark:2018red}.

From Eq.\ \eqref{theta034}, we have
\begin{equation}
\label{theta_diff}
(\Theta_{0,\pm 4}^{n,1}- \Theta_{0, \pm 3}^{n,1}) (v,u) = - \frac{i}{2\pi} \partial_v \log \frac{x(v+\frac{i}{2}n) - \frac{g^2}{x(u)}}{x(v-\frac{i}{2}n) - \frac{g^2}{x(u)}} + \frac{i}{2\pi} \partial_v \log \frac{x(v+\frac{i}{2}n) - x(u)}{x(v-\frac{i}{2}n) - x(u)} \,.
\end{equation}
We consider $u \in \R \setminus ( - 2g,2g)$. For large $u$, we have $x(u) = u + \CO( 1/u)$. Considering the first term, one observes that since $g^2/x(u)$ goes to zero for $u\rightarrow \infty$; thus, one obtains $- \theta_n(v)$ as one can infer from the definition \eqref{theta_n_def}. The second term instead goes to zero if $v$ is kept fixed for $u\rightarrow \infty$. However, if $v-u$ is finite, it is non-zero for $u\rightarrow \infty$. Since this requires $v\rightarrow \infty$, one has $x(v+\frac{i}{2}n)\simeq v$; hence, the second term gives $a_n(v-u)$, as one can infer from the definition \eqref{an_def}. Thus, we have derived
 \begin{equation}
\label{theta_diff2}
\lim_{u \rightarrow \infty} (\Theta_{0,\pm 4}^{n,1}- \Theta_{0, \pm 3}^{n,1}) (v,u) = - \theta_n (v) + a_n (v-u) \,.
\end{equation}
Employing Eq.\ \eqref{eq: Y11_times_Y22} with Eq.\ \eqref{Y_infty}, we find
\begin{equation}
\label{asymptY11Y22a}
\log \CY_{1,\pm 1}^\infty \CY_{2,\pm 2}^\infty = - \sum_{n=1}^\infty \left[ \theta_n\star  \log (1+\CY_{n,0})\right](0) + \lim _{u\rightarrow \infty} \sum_{n=1}^\infty \left[\log (1+\CY_{n,0})\star a_n \right](u)  \,.
\end{equation}
For the second term, one observes that one picks up only contributions for large $v$ in the integral since $u\rightarrow \infty$. Hence,
\begin{equation}
\label{asymptY11Y22}
\log \CY_{1,\pm 1}^\infty \CY_{2,\pm 2}^\infty =  - \sum_{n=1}^\infty \left[ \theta_n\star  \log (1+\CY_{n,0})\right](0) + \sum_{n=1}^\infty \log (1+\CY_{n,0}^\infty) \,.
\end{equation}
Note that one can deduce $\CY_{1,1}^\infty \CY_{2,2}^\infty=\CY_{1,-1}^\infty \CY_{2,-2}^\infty$ from this.
A slight rewriting of the Gibbs free energy per unit classical scaling dimension \eqref{gibbsfree2} reveals
\begin{equation}
\label{gibbsfree3}
F(T,\Omega_i) = - T \sum_{n=1}^\infty [ \theta_n \star  \log (1+\CY_{n,0} ) ](0) \,.
\end{equation}
Therefore, combining this with Eq.\ \eqref{asymptY11Y22} one finds that
\begin{equation}
\label{eq: free_energy_Y}
F(T,\Omega_i) = - T \left\{    \sum_{n=1}^\infty \log (1+\CY^{\infty}_{n,0}) -   \log \CY_{1,1}^\infty \CY_{2,2}^\infty     \right\}  \,.
\end{equation}
Thus, we can obtain the Gibbs free energy per unit classical scaling dimension $F(T,\Omega_i)$ directly from the asymptotic values of the Y-functions $\CY^\infty_{a,s}$ defined by Eq.\ \eqref{Y_infty}. By Eq.~\eqref{eq: general_TH}, this means in turn that one can obtain the Hagedorn temperature in terms of $\CY^\infty_{a,s}$.

\subsection{Definitions of functions and kernels}
\label{sec:kernels}

In this appendix, we have collected the various definitions for families of functions and kernels that are needed for formulating the TBA equations \eqref{newfullTBA1}--\eqref{newfullTBA5}. 

We define for a positive integer $n$ and $u\in \C$ the functions
\begin{align}
\label{an_def}
a_n (u) & = \frac{i}{2\pi} \frac{d}{du} \log \frac{u+\frac{in}{2}}{u-\frac{in}{2}} = \frac{i}{2\pi} \left\{ \frac{1}{u+\frac{i}{2}n} - \frac{1}{u-\frac{i}{2}n} \right\} = \frac{n}{2\pi \left( u^2 + \frac{n^2}{4} \right)}\,,\\
\label{theta_n_def}
\theta_n(u)& = \frac{i}{2\pi} \frac{d}{du} \log \frac{x(u+\frac{in}{2})}{x(u-\frac{in}{2})}\,,\\
\label{epsilon_n_def}
\epsilon_n(u)& =  g^2 \left( \frac{i}{x( u + \frac{i}{2}n )} -  \frac{i}{x( u - \frac{i}{2}n )} \right)\,,\qquad \text{ for }n>0\,.
\end{align}
We extend these three families of functions to $n=0$ by taking the limit $n \rightarrow 0^+$:
\begin{align}
\label{a0_def}
a_0 (u)  &=  \delta (u) \,,\\
\label{theta0}
\theta_0 (u) &= \left\{ \begin{array}{l} 0 \ \ \mbox{for} \ \ | u | > 2 g \,, \\[2mm] \frac{1}{\pi\sqrt{4g^2-u^2}} \ \ \mbox{for} \ \  |u| < 2g \,, \end{array} \right. \\
\label{epsilon0}
\epsilon_0 (u) &= \begin{cases}
                  0 &\mbox{for }  | u | \geq 2 g\,,\\
                  2\sqrt{4g^2-u^2} & \mbox{for }  |u| < 2g\,.
                 \end{cases}
\end{align}

We turn now to the kernels.
For $n \neq m$, we define the kernel $K^{mn}(v,u)$ via
\begin{equation}
\label{thetmn1}
K^{mn} (v,u) = a_{|n-m|}(v-u) + 2 a_{|n-m|+2}(v-u) + 2 a_{n+m-2}(v-u) + a_{n+m}(v-u)   \,.
\end{equation}
For $n=m$, we define
\begin{equation}
\label{thetmn2}
K^{nn} (v,u) = a_{2n}(v-u) + 2 \sum_{j=1}^{n-1} a_{2n-2j}(v-u)   \,.
\end{equation}
Moreover, we define the kernels
\begin{equation}
\Sigma^{m,n} ( v,u ) =  \frac{i}{2\pi}  \sum_{k=1}^m \sum_{j=1}^n \partial_v \log  \sigma^2 \left( v + \frac{i}{2} ( m+1-2k) , u + \frac{i}{2} ( n+1-2j) \right)\,,
\end{equation}
\begin{eqnarray}
\Theta^{m,n}_{\pm 1,0} (v,u) &=& \frac{i}{2\pi} \sum_{k=1}^{m+1}  \partial_v \log \frac{c(u+\frac{i}{2} n, v + \frac{i}{2} (m+2-2k))}{c(u-\frac{i}{2} n, v + \frac{i}{2} (m+2-2k))} \nn \\ && + \frac{i}{2\pi} \sum_{k=1}^{m-1}  \partial_v \log \frac{d(u+\frac{i}{2} n, v + \frac{i}{2} (m-2k))}{d(u-\frac{i}{2} n, v + \frac{i}{2} (m-2k))}\,,
\end{eqnarray}
\begin{eqnarray}
\Theta^{m,n}_{0,\pm 1} (v,u) &=& \frac{i}{2\pi} \sum_{k=1}^{n+1}  \partial_v \log \frac{a(v-\frac{i}{2} m, u + \frac{i}{2} (n+2-2k))}{a(v+\frac{i}{2} m, u + \frac{i}{2} (n+2-2k))} \nn \\ && + \frac{i}{2\pi} \sum_{k=1}^{n-1}  \partial_v \log \frac{b(v-\frac{i}{2} m, u + \frac{i}{2} (n-2k))}{b(v+\frac{i}{2} m, u + \frac{i}{2} (n-2k))}\,,
\end{eqnarray}
\begin{equation}
\Theta_{\pm 3,0}^{1,n} (v,u) = \frac{i}{2\pi} \partial_v \log \frac{c(u+\frac{i}{2} n,v)}{c(u-\frac{i}{2} n,v)}
\spa
\Theta_{\pm 4,0}^{1,n} (v,u) = \frac{i}{2\pi} \partial_v \log \frac{d(u-\frac{i}{2} n,v)}{d(u+\frac{i}{2} n,v)}\,,
\end{equation}
\begin{equation}
\label{theta034}
\Theta_{0,\pm 3}^{m,1} (v,u) = \frac{i}{2\pi} \partial_v \log \frac{a(v-\frac{i}{2} m,u)}{a(v+\frac{i}{2} m,u)}
\spa
\Theta_{0,\pm 4}^{m,1} (v,u) = \frac{i}{2\pi} \partial_v \log \frac{b(v-\frac{i}{2} m,u)}{b(v+\frac{i}{2} m,u)}\,,
\end{equation}
where we defined
\begin{equation}
a(v,u) = \frac{x(v)-x(u)}{\sqrt{x(v)}} \spa b(v,u) = \frac{x(v)- \frac{g^2}{2x(u)}}{\sqrt{x(v)}}\,,
\end{equation}
\begin{equation}
c(u,v) = x(u)-x(v) \spa d(u,v) = x(u) - \frac{g^2}{2x(v)}\,.
\end{equation}
We also define the kernels
\begin{equation}
 H_m(v,u) =\frac{i}{2\pi}\partial_v\log \frac{x(u-i0)-\frac{g^2}{x(v+\frac{i}{2}m)}}{x(u+i0)-\frac{g^2}{ x(v+\frac{i}{2}m)}}\,,
\end{equation}
and
\begin{equation}
\begin{aligned}
\Sigma^{m} (v,u) =& \frac{i}{2\pi} \partial_v  \left( \log \frac{R^2(x(v+ \frac{im}{2}),x(u+i0))}{R^2(x(v+ \frac{im}{2}),x(u-i0))} 
+ 
\log \frac{R^2(x(v- \frac{im}{2}),x(u-i0))}{R^2(x(v- \frac{im}{2}),x(u+i0))} \right) \,,
\end{aligned}
\end{equation}
which is defined in terms of the dressing factor \cite{Beisert:2006ez}
\begin{equation}
\begin{aligned}
\sigma^2 (u,v) =&  \frac{R^2(x^+(u),x^+(v))R^2(x^-(u),x^-(v))}{R^2(x^+(u),x^-(v))R^2(x^-(u),x^+(v))}      \,,
\end{aligned}
\end{equation}
with $x^\pm(u) = x\left(u\pm \frac{i}{2}\right)$.

\section{Asymptotic T-system}
\label{app: asymptotic T}

In this appendix, we determine the asymptotic values of the T-system functions, from which one can infer the the asymptotic values of the Y-functions $\CY^\infty_{a,s}$ defined by Eq.\ \eqref{Y_infty}. The purpose is to generalize the asymptotic T-system found previously in Ref.\ \cite{Harmark:2017yrv} for zero chemical potentials to the case of non-zero chemical potentials. This is crucial for identifying the correct asymptotic behavior of the $\Pb_a(u)$ and $\Qb_i(u)$ functions in the QSC with generic chemical potentials in App.\ \ref{app: asymptotic Q} and furthermore to the case of general chemical potentials presented in Subsec.\ \ref{subsec: QSC chempots}. 

\subsection{T-system}
\label{app: T-system}

To set up the framework for our analysis of the  asymptotic values of the T-system functions, we review first very briefly what a T-system is.

One can translate the Y-system equations of App.\ \ref{app: Ysystem equations} into equations for a T-system. 
To this end, one introduces the T-functions $T_{a,s}(u)$  with $(a,s) \in \hat{M}$ where $\hat{M}$ is the so-called T-hook set
\begin{equation}
\label{eq: definition Mhat}
\hat{M} = \{ (a,s) \in \mathbb{Z}_{\geq 0} \times \mathbb{Z}\, |\, \mbox{min}(a,|s|) \leq 2 \}  \,;
\end{equation}
cf.\ Fig.\ \ref{fig: Y and T hook}.
The T-functions are set to zero outside the T-hook $\hat{M}$.
The T-functions are related to the Y-functions as follows:
\begin{equation}
\label{eq: Tsystem}
\CY_{a,s} = \frac{T_{a,s+1} T_{a,s-1}}{T_{a+1,s}T_{a-1,s}} \,.
\end{equation}

The T-functions should obey the Hirota equations
\begin{equation}
\label{hirota_eqs}
T_{a,s}^+ T_{a,s}^- = T_{a+1,s} T_{a-1,s} + T_{a,s+1} T_{a,s-1} \,.
\end{equation}
In addition to this, the T-functions should obey certain analyticity properties listed in Ref.~\cite{Gromov:2011cx}. 
Note that there are certain gauge freedoms of the T-system functions $T_{a,s}(u)$ relating different T-systems that correspond to the same Y-system; see for instance Ref.~\cite{Gromov:2011cx}.
We impose the following gauge conditions on the T-functions:
\begin{equation}
\label{eq: T_gauge}
 T_{2,n}=T_{n,2} \ \mbox{and}\ T_{2,-n}=T_{n,-2} \ \mbox{for}\  n \geq 2 \,.
\end{equation}

\subsection{General asymptotic T-system}
\label{app: general asymptotic T-system}

We turn now to the asymptotic values of the T-functions $T_{a,s}(u)$. We define these as
\begin{equation}
T_{a,s}^\infty = \lim_{u \rightarrow \infty} T_{a,s} (u) \,.
\end{equation}
The asymptotic values of the Y-functions $\CY^\infty_{a,s}$ defined in Eq.\ \eqref{Y_infty} are connected to asymptotic values of the T-functions $T_{a,s}^\infty$ as
\begin{equation}
\label{eq: Tsystem infinity}
\CY_{a,s}^\infty = \frac{T_{a,s+1}^\infty T_{a,s-1}^\infty}{T_{a+1,s}^\infty T_{a-1,s}^\infty} \,.
\end{equation}
It follows from the Hirota equations \eqref{hirota_eqs} that 
\begin{equation}
\label{eq: const_Hirota}
(T_{a,s}^\infty )^2 = T^\infty_{a+1,s} T^\infty_{a-1,s} + T^\infty_{a,s+1} T^\infty_{a,s-1} \,,
\end{equation}
for $(a,s)\in \hat{M}$. We inherit the gauge choice \eqref{eq: T_gauge} and impose the additional gauge condition $T^\infty_{0,s} = 1$ for $s\in \mathbb{Z}$. This is possible to impose since it involves only a constant transformation of the T-system, which is consistent with Eq.\ \eqref{eq: T_gauge}. 
Considering now the formula \eqref{eq: free_energy_Y} for the Gibbs free energy per unit classical scaling dimension in terms of $\CY_{a,s}^\infty$, we can translate this into a relation between the free energy and the asymptotic T-system:
\begin{equation}
\label{eq: FT_asymptT}
F(T,\Omega_i) = T \lim_{n\rightarrow \infty} \log \frac{T^\infty_{n+1,0}}{T^\infty_{n,0}} \,.
\end{equation}
Using Eq.\ \eqref{eq: general_TH}, this reveals a direct connection between the Hagedorn temperature $\THag$ and the asymptotic T-system.

The general solution to the constant Hirota equations \eqref{eq: const_Hirota} is the $\mathfrak{psu}(2,2|4)$ character solution of Ref.\ \cite{Gromov:2010vb}. The solution is presented in terms of the eight variables $x_1$, $x_2$, $x_3$, $x_4$, $y_1$, $y_2$, $y_3$ and $y_4$, where one can think of $x_1$, $x_2$, $x_3$, $x_4$ as associated with the $\mathfrak{su}(2,2)$ subalgebra and $y_1$, $y_2$, $y_3$, $y_4$, as associated with the $\mathfrak{su}(4)$ subalgebra. 
For $a \geq |s|$, the solution is
\begin{equation}
\label{constT1}
T^\infty_{a,s} = (-1)^{a+as} \left( \frac{x_3 x_4}{y_1y_2y_3y_4} \right)^{s-a} \frac{\det \left( S_i ^{\theta_{j,s+2}} y_i^{j-4-(a+2)\theta_{j,s+2}} \right)_{1\leq i,j \leq 4}}{\det \left( S_i ^{\theta_{j,2}} y_i^{j-4-2\theta_{j,s+2}} \right)_{1\leq i,j \leq 4}} \,,
\end{equation}
with 
\begin{equation}
\label{constT4}
S_i = \frac{(y_i-x_3)(y_i-x_4)}{(y_i-x_1)(y_i-x_2)} \spa\theta_{j,s} = \left\{ \begin{array}{l} 1 \ , \ j > s \\ 0 \ , \ j \leq s \end{array} \right. \,.
\end{equation}
For $s \geq a \geq 0$, the solutions is
\begin{equation}
\label{constT2}
T^\infty_{a,s} = \frac{\det \left( Z_i ^{1-\theta_{j,a}} x_i^{2-j+(s-2)(1-\theta_{j,a})} \right)_{1\leq i,j \leq 2}}{\det \left( Z_i ^{1-\theta_{j,0}} x_i^{2-j-2(1-\theta_{j,0})} \right)_{1\leq i,j \leq 2}} \,,
\end{equation}
with 
\begin{equation}
\label{constT3}
 Z_i = \frac{(x_i-y_1)(x_i-y_2)(x_i-y_3)(x_i-y_4)}{(x_i-x_3)(x_i-x_4)} \,.
\end{equation}
From the above, one finds $T^\infty_{a,s}$ for $s \geq 0$. 
For negative $s$, one can find $T^\infty_{a,s}$ via the general relation
\begin{equation}
\label{Tneg_s}
T^\infty_{a,s} (x_1,x_2,x_3,x_4|y_1,y_2,y_3,y_4) = \left( \frac{y_1y_2y_3y_4}{x_1x_2x_3x_4} \right)^a T^\infty_{a,-s} \left( \frac{1}{x_4},\frac{1}{x_3},\frac{1}{x_2},\frac{1}{x_1}\big|\frac{1}{y_4},\frac{1}{y_3},\frac{1}{y_2},\frac{1}{y_1} \right) \,.
\end{equation}

Note that one has manifestly $T^\infty_{0,s}=1$ in the above solution. 
The gauge choice \eqref{eq: T_gauge} corresponds to 
\begin{equation}
\label{xy_con0}
x_1 x_2 x_3 x_4 = y_1 y_2 y_3 y_4  \,.
\end{equation}
One can check that making the rescaling $x_a \rightarrow \alpha x_a$ and $y_i \rightarrow \alpha y_i$ results in $T^\infty_{a,s} \rightarrow \alpha^{as} T^{\infty}_{a,s}$ which corresponds to a gauge transformation of the asymptotic T-system. Thus, one can freely make a rescaling of this kind to impose $y_1y_2y_3y_4=1$, thus 
\begin{equation}
\label{xy_con}
x_1 x_2 x_3 x_4 = y_1 y_2 y_3 y_4 = 1 \,.
\end{equation}

We impose the following ordering in the case of non-zero chemical potentials:
\begin{equation}
\label{order_xy}
-x_1 > -x_2 > 0  \spa -x_3 > -x_4  > 0  \spa y_1 > y_2 > y_3 > y_4 > 0 \,.
\end{equation}
The $x_i$'s are negative due to the boundary conditions of the fermions in the partition function. Indeed, from \eqref{Q10} we see that negativity of the $x_i$'s is in accordance with the signs in the $(-e^{-1/(2T_{\rm H})})^{\pm iu}$ factors in Eq.~\eqref{eq: asymptotic Q-system: Ps after H}.
The inequalities \eqref{order_xy} assume
\begin{equation}
\label{assume_omega_strict}
\Omega_1 > \Omega_2 > \Omega_3 \geq 0 \spa \Omega_4 > \Omega_5 \geq 0 \,.
\end{equation}
This means the above solution is not fully general as it does not work for the full range of chemical potentials constrained by the bounds \eqref{assume_omega} that one can always assume without loss of generality. However, one can describe the cases in which some or all of the more general bounds \eqref{assume_omega} are saturated by taking limits of the above solution. For instance, as we review in App.\ \ref{app: asympt Tsystem zero chem}, one finds the case of zero chemical potentials by a limit of the above \cite{Harmark:2017yrv}. Thus, in this sense, we can think of the above solution for the asymptotic T-system as fully general.

Combining the above solution for $T^\infty_{a,s}$ with the Y-system asymptotics \eqref{Ysystembcs}, we find
\begin{equation}
\label{xy_chempots}
\begin{array}{c}\ds
\e^{(\Omega_2-\Omega_3)/T} = \frac{y_1}{y_2} \spa
\e^{(\Omega_1-\Omega_2)/T} = \frac{y_2}{y_3}\spa
\e^{(\Omega_2+\Omega_3)/T} = \frac{y_3}{y_4}\,,
\\[4mm] \ds
\e^{(\Omega_4+\Omega_5)/T} = \frac{x_1}{x_2} \spa
\e^{(\Omega_4-\Omega_5)/T} = \frac{x_3}{x_4}\,.
\end{array}
\end{equation}
From this, we find the $y_i$'s in terms of the chemical potentials and temperature:
\begin{equation}
\label{the_ys}
y_1 = \e^{\frac{\Omega_1 + \Omega_2 - \Omega_3}{2T}} \spa
y_2 = \e^{\frac{\Omega_1 - \Omega_2 + \Omega_3}{2T}} \spa 
y_3 = \e^{\frac{-\Omega_1 + \Omega_2 + \Omega_3}{2T}} \spa 
y_4 = \e^{\frac{-\Omega_1 - \Omega_2 - \Omega_3}{2T}} \,.
\end{equation}
Using the asymptotics of the solution \eqref{constT1}--\eqref{constT4} with Eq.\ \eqref{eq: FT_asymptT},  we find
\begin{equation}
\exp \left( \frac{F(T,\Omega_i)}{T} \right)= \frac{y_1 y_2}{x_3 x_4} \,.
\end{equation}
Thus, we deduce 
\begin{equation}
\label{the_xs}
\begin{aligned}
&x_1 = - \e^{\frac{F(T,\Omega_i) -\Omega_1 + \Omega_4 + \Omega_5}{2T}}
\,,&&
x_2 = - \e^{\frac{F(T,\Omega_i) -\Omega_1 - \Omega_4 - \Omega_5}{2T}} \,,
\\
&x_3 = - \e^{\frac{-F(T,\Omega_i) +\Omega_1 + \Omega_4 - \Omega_5}{2T}}
\,,&&
x_4 = - \e^{\frac{-F(T,\Omega_i) +\Omega_1 - \Omega_4 + \Omega_5}{2T}} \,.
\end{aligned}
\end{equation}
Inserting now the relation to the Hagedorn temperature \eqref{eq: general_TH}, one finds Eq.\ \eqref{Hagedorn_xy}.
Once inserted into the above solution for $T^\infty_{a,s}$, this provides the asymptotic T-system for a Hagedorn temperature $\THag$ with chemical potentials obeying the bounds \eqref{assume_omega_strict}. From this, one can approach the cases in which some or all of the more general bounds \eqref{assume_omega} are saturated by taking limits of this solution.

\subsection{Explicit form of the asymptotic T-system}
\label{app: new_form_asy_T}

For connecting to the Q-system, and hence the QSC, it is useful to rewrite the general solution \eqref{constT1}--\eqref{Tneg_s} for the asymptotic T-system $T^\infty_{a,s}$ in a more explicit form.
We explicitly present the generic case \eqref{assume_omega_strict} here. The cases in which some of the inequalities \eqref{assume_omega} are saturated can be obtained by taking limits of the expressions below, as illustrated for the case of vanishing chemical potentials in the subsequent subsection.

We make the following definitions:\footnote{Note that the constants $\CA_a$ and $\CB_i$ are related to the coefficients $A_a$ and $B_i$ from Eqs.\ \eqref{Asqr} and \eqref{Bsqr} in the main text as $A_a A^a = \CA_a
$ and $B_i B^i = \CB_i$.}
\begin{equation}
\label{CA_and_CB}
\CA_a = \frac{\prod_{i=1}^4 ( x_a - y_i ) }{x_a \prod_{b\neq a} ( x_b - x_a )} \spa \qquad \CB_i = -\frac{\prod_{a=1}^4 ( x_a - y_i ) }{y_i \prod_{j\neq i} ( y_j - y_i )} \,.
\end{equation}
For the right band $s \geq a \geq 0$, we record
\begin{equation}
\label{Tinfty_right}
T^\infty_{0,s} = 1 \spa \quad T^\infty_{1,s} = - \CA_1 x_1^s - \CA_2 x_2^s \spa \quad T^\infty_{2,s} = - \frac{(x_1-x_2)^2}{x_1x_2} \CA_1 \CA_2 (x_1x_2)^s \,.
\end{equation}
For the left band $s \leq -a \leq 0$, we record
\begin{equation}
\label{Tinfty_left}
T^\infty_{0,s} = 1 \spa \quad T^\infty_{1,s} = \CA_3 x_3^s + \CA_4 x_4^s \spa \quad T^\infty_{2,s} = - \frac{(x_3-x_4)^2}{x_3x_4} \CA_3 \CA_4 (x_3x_4)^s \,.
\end{equation}
Finally, for the upper band $a \geq |s|$, we record
\begin{equation}
\label{Tinfty_Ta1}
T^\infty_{a,1} =  \left( \frac{y_1y_2y_3y_4}{x_1x_2x_3x_4} \right)^{a-1} \CA_1 \CA_2 \sum_{i=1}^4 \CB_i \frac{(x_1-x_2)^2 y_i^2}{(x_1-y_i)^2 (x_2-y_i)^2}   \left( - \frac{x_1x_2}{y_i} \right)^a \,,
\end{equation}
\begin{equation}
\label{Tinfty_Tam1}
T^\infty_{a,-1} = - \CA_3 \CA_4 \sum_{i=1}^4 \CB_i \frac{(x_3-x_4)^2 y_i^2}{(x_3-y_i)^2 (x_4-y_i)^2}   \left( - \frac{y_i}{x_3x_4} \right)^a \,,
\end{equation}
\begin{equation}
\label{Tinfty_Ta2}
T^\infty_{a,2} = - \frac{(x_1-x_2)^2}{x_1x_2} \CA_1 \CA_2 (x_1x_2)^a \spa
T^\infty_{a,-2} = - \frac{(x_3-x_4)^2}{x_3x_4} \CA_3 \CA_4 (x_3x_4)^{-a} \,.
\end{equation}
In addition, one finds $T^\infty_{a,0}$ from the general relation 
\begin{equation}
\label{Tinfty_Ta0}
T^\infty_{a,0} = \frac{(T^\infty_{a,1})^2 - T^\infty_{a+1,1} T^\infty_{a-1,1}}{T^\infty_{a,2}} \,.
\end{equation}
One can check that the above formulas give the same result as Eqs.\ \eqref{constT1}--\eqref{Tneg_s}.

The above explicit form \eqref{Tinfty_right}--\eqref{Tinfty_Ta2} for $T^\infty_{a,s}$ is useful in that one can directly infer the dependence on $a$ and $s$. This dependence is instead hidden in the character solution \eqref{constT1}--\eqref{Tneg_s}.
We use this explicit form below in App.\ \ref{app: asymptotic Q} to connect to the large-$u$ behavior of the QSC.

\subsection{Asymptotic T-system for zero chemical potentials}
\label{app: asympt Tsystem zero chem}

One can find the asymptotic T-system for zero chemical potentials by first setting $\Omega_3=\Omega_5=0$ and subsequently taking the limits $\Omega_1\rightarrow 0$, $\Omega_2\rightarrow 0$ and $\Omega_4\rightarrow 0$ of the solution \eqref{Tinfty_right}--\eqref{Tinfty_Ta2} of App.\ \ref{app: new_form_asy_T}. 
One finds 
\begin{equation}
\label{consTbfgen1}
\begin{aligned}
T^\infty_{a,0} &=  \left(e^{\frac{F(T)}{2T}}\right)^{2a} \frac{a -2\tanh \tfrac{F(T)}{4T}}{12 \tanh^4 \frac{F(T)}{4T}}  \big( a^3   -6a^2\tanh \tfrac{F(T)}{4T}  \\&\hspace{0.35\linewidth}+ \left( 12\tanh^2 \tfrac{F(T)}{4T}-1\right) a  -6\tanh^3 \tfrac{F(T)}{4T} \big) \,,
\\
T^\infty_{a,\pm 1} &=  (-1)^{a} \left(e^{\frac{F(T)}{2T}}\right)^{2a} \frac{a-3\tanh \frac{F(T)}{4T}}{6\tanh^4 \frac{F(T)}{4T}}  \left(a^2-3a\tanh \tfrac{F(T)}{4T} + 3\tanh^2 \tfrac{F(T)}{4T}-1\right) \,,
\\
T^\infty_{a,\pm 2} &= \frac{1}{\tanh^4 \frac{F(T)}{4T}}  \left(e^{\frac{F(T)}{2T}}\right)^{2a}\,,
\end{aligned}
\end{equation}
for $a \geq |s|$, and
\begin{equation}
\label{consTbfgen2}
\begin{aligned}
T^\infty_{0,s}&=1\,,\\
T^\infty_{1,s} &= \frac{(-1)^s}{\tanh^2 \frac{F(T)}{4T}} \left[ |s| - \frac{1-3\tanh^2 \frac{F(T)}{4T}}{2\tanh \frac{F(T)}{4T}} \right] \left(e^{\frac{F(T)}{2T}}\right)^{|s|}  \,,\\
T^\infty_{2,s} &= \frac{1}{\tanh^4 \frac{F(T)}{4T}}  \left(e^{\frac{F(T)}{2T}}\right)^{2|s|}\,,
\end{aligned}
\end{equation}
for $|s| \geq a$. 
This is the T-system given in our letter \cite{Harmark:2017yrv}, where it was written in terms of the parameter 
\begin{equation}
\label{eq: zth}
 z = -\tanh \frac{F(T)}{4T}\,.
\end{equation}
Imposing the Hagedorn temperature condition \eqref{eq: general_TH} with zero chemical potentials further fixes $F(\THag)=-1$.

\section{Zeroth order T-system and the Hagedorn temperature}
\label{app: zeroth order}

In this appendix, we provide details on the relation between the integrability-based method and the P\'{o}lya-theory method, which we mentioned in Subsec.\ \ref{subsec: zero_coupling}.

For vanishing 't Hooft coupling $\lambda=0$, the Y-system is constant. Thus, it is equal to the asymptotic Y-system $\CY_{a,s}|_{\lambda=0} = \CY^\infty_{a,s}|_{\THag=\THag(\lambda=0)}$. Similarly, for the T-system
\begin{equation}
T_{a,s}|_{\lambda=0} = T^\infty_{a,s}|_{\THag=\THag(\lambda=0)} \,.
\end{equation}
For vanishing 't Hooft coupling, one can see from Eq.\ \eqref{eq: Y11_times_Y22} along with the definition of the kernels in App.\ \ref{sec:kernels} that a constant Y-system implies $\CY_{1,1} \CY_{2,2} = \CY_{1,-1} \CY_{2,-2} = 1$. Using Eq.\ \eqref{eq: Tsystem} together with the gauge \eqref{eq: T_gauge} as well as $T_{0,1}^\infty=1$, we derive the condition
\begin{equation}
\label{T10isone}
T^\infty_{1,0}|_{\lambda=0} = 1 \,.
\end{equation}
Imposing Eq.\ \eqref{xy_con}, we compute from the general asymptotic T-system \eqref{constT1}--\eqref{constT4} 
\begin{equation}
T^\infty_{1,0} = \eta(x,y)  \,,
\end{equation}
where we defined
\begin{equation}
\eta(x,y) = \eta_s (x,y) + \eta_v (x,y) + \eta_{f1} (x,y)+ \eta_{f2} (x,y)\,,
\end{equation}
with
\begin{align}
\eta_s (x,y) &= \frac{(x_3 x_4 - x_1 x_2)(y_1y_2 + y_3 y_4 + y_1 y_3 + y_2 y_4 + y_2 y_3 + y_1 y_4)}{(x_4-x_1)(x_3-x_1)(x_4-x_2)(x_3-x_2)}\,,\\
\eta_v (x,y) &= \frac{2+ 2x_1^2x_2^2 - 2x_1 x_2 (x_1+x_2)(x_3+x_4)+(x_1 x_3+x_2 x_4)(x_1 x_4+x_2 x_3)}{(x_4-x_1)(x_3-x_1)(x_4-x_2)(x_3-x_2)}\,,\\
\eta_{f1} (x,y) &=  \sum_{i=1}^4 y_i^{-1}  \frac{x_1 + x_2 - x_3 - x_4}{(x_4-x_1)(x_3-x_1)(x_4-x_2)(x_3-x_2)}\,,\\
\eta_{f2} (x,y) &=  \sum_{i=1}^4 y_i  \frac{(x_3+x_4)x_1 x_2- (x_1+x_2)x_3 x_4    }{(x_4-x_1)(x_3-x_1)(x_4-x_2)(x_3-x_2)} \,.
\end{align} 
Noting that Eq.\ \eqref{Hagedorn_xy} implies
\begin{equation}
\begin{array}{c}\ds
e^{-\frac{1}{\THag}} = x_1 x_2 \spa \e^{\frac{\Omega_4}{\THag}} = x_1 x_3 \spa \e^{\frac{\Omega_5}{\THag}} = x_1 x_4  \,,
\\[2mm]\ds
e^{\frac{\Omega_1}{\THag}} = y_1 y_2 \spa e^{\frac{\Omega_2}{\THag}} = y_1 y_3 \spa e^{\frac{\Omega_3}{\THag}} = y_2 y_3  \,,
\end{array} 
\end{equation}
one can easily check that $\eta(x,y) = \eta(\THag,\Omega_i)$ is the single-particle partition function for $\CN=4$ SYM theory at zero 't Hooft coupling, with $\eta_s$ originating from the scalar fields, $\eta_v$ from the gauge boson and $\eta_{f1}$ as well as $\eta_{f2}$ from the fermions. As derived in Refs.\ \cite{Sundborg:1999ue,Aharony:2003sx}, the condition for the Hagedorn temperature $\THag$ at zero 't Hooft coupling is $\eta(\THag,\Omega_i)=1$. This corresponds precisely to the requirement \eqref{T10isone}. Therefore, we have shown that the Hagedorn temperature at zero 't Hooft coupling for any value of the chemical potentials correspond to the one obtained using the single-particle partition function for $\CN=4$ SYM theory in Refs.\ \cite{Sundborg:1999ue,Aharony:2003sx}.

\section{Asymptotic Q-system with generic chemical potentials}
\label{app: asymptotic Q}

In this appendix, we use the results for the asymptotic T-system $T^\infty_{a,s}$ recorded in App.\ 
\ref{app: general asymptotic T-system} and \ref{app: new_form_asy_T} to obtain the large-$u$ asymptotic behavior of the Q-system for the case of generic chemical potentials obeying the bounds \eqref{assume_omega_strict}. From this, one can infer the asymptotic behavior of $\Qb_i(u)$ and $\Pb_a(u)$. In Subsec.\ \ref{subsec: QSC chempots} this is extended to the fully general case.
Note that this treatment is a special case of the one considered in Ref.\ \cite{Kazakov:2015efa}, as discussed in Subsec.~\ref{subsec: QSC chempots}. 

Consider the asymptotic T-system $T^\infty_{a,s}$ appropriate for obtaining the Hagedorn temperature. This is given by Eqs.\ \eqref{Tinfty_right}--\eqref{Tinfty_Ta0} with $x_i$ and $y_i$ given by Eq.\ \eqref{Hagedorn_xy}.
This T-system is associated with a Q-system through the relation~(4.11) of Ref.\ \cite{Gromov:2010km}. For the left and right bands, we have
\begin{equation}
\begin{array}{l}
T^\infty_{0,s}=Q_{1234|1234}^{[-s]}Q_{\varnothing|\varnothing}^{[s]}\quad \text{for} \quad s\leq0 \,,
\\[3mm]
T^\infty_{1,s}=Q_{123|1234}^{[-s]}Q_{4|\varnothing}^{[s]}-Q_{124|1234}^{[-s]}Q_{3|\varnothing}^{[s]}\quad \text{for} \quad s\leq-1 \,,
\\[3mm]
T^\infty_{2,s}=Q_{12|1234}^{[-s]}Q_{34|\varnothing}^{[s]}\quad \text{for} \quad s\leq-2 \,,
\end{array}
\end{equation}
and
\begin{equation}
\begin{array}{l}
T^\infty_{0,s}=Q_{\varnothing|\varnothing}^{[s]}Q_{1234|1234}^{[-s]} \quad \text{for} \quad s\geq0 \,,
\\[3mm]
T^\infty_{1,s}=Q_{1|\varnothing}^{[s]}Q_{234|1234}^{[-s]}-Q_{2|\varnothing}^{[s]}Q_{134|1234}^{[-s]}\quad \text{for} \quad s\geq1 \,,
\\[3mm]
T^\infty_{2,s}= Q_{12|\varnothing}^{[s]}Q_{34|1234}^{[-s]}   \quad \text{for} \quad s\geq2 \,,
\end{array}
\end{equation}
while for the upper band
\begin{equation}
\begin{array}{l}
T^\infty_{a,-2}=Q_{12|1234}^{[a]}Q_{34|\varnothing}^{[-a]}\mbox{   for  } a\geq2 \,,
\\[3mm]
T^\infty_{a,-1}=(-1)^a\big(Q_{12|123}^{[a]}Q_{34|4}^{[-a]}-Q_{12|124}^{[a]}Q_{34|3}^{[-a]}+Q_{12|134}^{[a]}Q_{34|2}^{[-a]}-Q_{12|234}^{[a]}Q_{34|1}^{[-a]} \big)\mbox{   for  } a\geq1 \,,
\\[3mm]
T^\infty_{a,0}=Q_{12|12}^{[a]}Q_{34|34}^{[-a]}-Q_{12|13}^{[a]}Q_{34|24}^{[-a]}+Q_{12|14}^{[a]}Q_{34|23}^{[-a]}+Q_{12|34}^{[a]}Q_{34|12}^{[-a]}
\\[1mm] \hphantom{T_{a,0} = }
-Q_{12|24}^{[a]}Q_{34|13}^{[-a]}+Q_{12|23}^{[a]}Q_{34|14}^{[-a]}\mbox{   for  } a\geq0\,,
\\[3mm]
T^\infty_{a,1}=(-1)^a \big(Q_{12|1}^{[a]}Q_{34|234}^{[-a]}-Q_{12|2}^{[a]}Q_{34|134}^{[-a]}+Q_{12|3}^{[a]}Q_{34|124}^{[-a]} -Q_{12|4}^{[a]}Q_{34|123}^{[-a]} \big)\mbox{   for  } a\geq1\,,
\\[3mm]
T^\infty_{a,2}=Q_{12|\varnothing}^{[a]}Q_{34|1234}^{[-a]}\mbox{   for  } a\geq2  \,.
\end{array}
\end{equation}

One can reproduce the general asymptotic T-system \eqref{Tinfty_right}--\eqref{Tinfty_Ta0} from the Q-system as follows. We choose the gauge
\begin{equation}
\label{Qee}
Q_{\varnothing|\varnothing} = 1  \,.
\end{equation}
Furthermore, one can show that it is possible to impose \cite{Kazakov:2015efa}
\begin{equation}
\label{Qee2}
Q_{1234|1234} = 1  \,.
\end{equation}
One uses now the ansatz
\begin{equation}
\label{Q10}
\begin{array}{c}
Q_{a|\varnothing}= \Pb_a(u) =  A_a x_a^{-iu} \spa Q_{\varnothing|j} = \Qb_j(u) =  B_j y_j^{iu} \,, 
\\[2mm]
Q^{a|\varnothing}=\Pb^a(u) =  A^a x_a^{iu} \spa Q^{\varnothing|j} =\Qb^j(u) =  B^j y_j^{-iu} \,,
\end{array}
\end{equation}
for $a,j=1,2,3,4$. For the constants $A_a$, $A^a$, $B_j$ and $B^j$ we impose the constraints
\begin{equation}
\label{ABsqr}
A_a A^a = \CA_a
\spa
B_i B^i = \CB_i 
\end{equation}
where $\CA_a$ and $\CB_i$ are defined by Eq.\ \eqref{CA_and_CB} and where there are no sums over $a$ and $i$ on the left-hand sides of the two expressions. 
As mentioned in Subsec.\ \ref{subsec: QSC chempots}, this ansatz is related to a special case of the large-$u$ asymptotics of the twisted QSC in Ref.\ \cite{Kazakov:2015efa}.
From Eqs.\ \eqref{Q10} and \eqref{ABsqr}, the Q-system can now be inferred from the QQ-relations \eqref{eq: Q-system general 1}--\eqref{eq: Q-system general 3} as well as the definitions of the Hodge dual Q-functions \eqref{eq: Hodge dual Q}. In particular, we record
\begin{align}
\label{Q11}
Q_{a|j}&=x_a^{-iu} y_j^{iu} \frac{\sqrt{x_a y_j} }{x_a - y_j} A_a  B_j
\,,\\
Q^{a|j}&=-x_a^{iu} y_j^{-iu} \frac{\sqrt{x_a y_j} }{x_a - y_j} A^a  B^j\,,\\ 
\label{Q20}
Q_{ab|\varnothing}&= (x_a x_b)^{- \frac{1}{2}-iu} A_a A_b (x_a-x_b) 
\,,\\
Q^{ab|\varnothing}  &=(x_a x_b)^{- \frac{1}{2}+iu} A^a A^b (x_b-x_a)
\,,
\end{align}
for $a,j=1,2,3,4$.
For the left and right bands, it is obvious from Eqs.\ \eqref{Qee} and \eqref{Qee2} that $T_{0,s}=1$. We record furthermore the relations
\begin{align}
T^\infty_{1,s} &= \left\{ \begin{array}{l} 
 - \Pb_1 ^{[s]} (\Pb^1)^{[-s]} - \Pb_2 ^{[s]} (\Pb^2)^{[-s]} 
\quad \text{for} \quad s\geq1 \,, \\[2mm]  \Pb_4 ^{[s]} (\Pb^4)^{[-s]} + \Pb_3 ^{[s]} (\Pb^3)^{[-s]} 
\quad \text{for} \quad s\leq-1 \,, \end{array} \right. \\
T^\infty_{2,s} &= \left\{ \begin{array}{l} 
 Q_{12|\varnothing}^{[s]}(Q^{12|\varnothing})^{[-s]} \quad \text{for} \quad s\geq2 \,,
\\[2mm]  
Q_{34|\varnothing}^{[s]}(Q^{34|\varnothing})^{[-s]}   \quad \text{for} \quad s\leq -2 \,.
\end{array} \right. 
\end{align}
Combining these relations with Eq.\ \eqref{Q10}, we reproduce the general asymptotic T-system \eqref{Tinfty_left} and \eqref{Tinfty_right} for the left and right bands. 

For the upper band, we record
\begin{equation}
T^\infty_{a,2} = Q_{12|\varnothing}^{[a]}(Q^{12|\varnothing})^{[-a]} \spa T^\infty_{a,-2} =Q_{34|\varnothing}^{[-a]}(Q^{34|\varnothing})^{[a]} \,,
\end{equation}
for $a\geq 2$, and
\begin{equation}
T^\infty_{a,1}=-(-1)^a \sum_{j=1}^4 Q_{12|j}^{[a]} (Q^{12|j})^{[-a]}
\spa T^\infty_{a,-1}=(-1)^a \sum_{j=1}^4 Q_{34|j}^{[a]} (Q^{34|j})^{[-a]} \,,
\end{equation}
for $a \geq 1$. Combining these relations with Eq.\ \eqref{Q20},
 we reproduce the general asymptotic T-system \eqref{Tinfty_Ta1}--\eqref{Tinfty_Ta2} for the upper band. 
It is not necessary to check that $T^\infty_{a,0}$ matches, since that is determined by the relation \eqref{Tinfty_Ta0}.

Using Eqs.\ \eqref{ABsqr} and \eqref{xy_con0}, one derives the following identities
\begin{equation}
\label{id_Asqr}
\sum_{a=1}^4 A_a A^a =  
\sum_{i=1}^4 B_i B^i =  0\,,
\qquad
\sum_{a=1}^4 A_a A^a \frac{x_a}{x_a - y_i} =
\sum_{i=1}^4 B_i B^i \frac{x_a}{x_a - y_i}  = -1\,,
\end{equation}
\begin{equation}
\label{id_Asqr_adv2}
B_i B^j \sum_{a=1}^4  \frac{A_a A^a x_a}{(x_a - y_i)(x_a - y_j)} = \frac{1}{y_i} \delta_{ij}
\spa
A_a A^b \sum_{i=1}^4  \frac{B_i B^i y_i}{(x_a - y_i)(x_b - y_i)} = \frac{1}{x_a} \delta_{ab}\,.
\end{equation}
From Eqs.\ \eqref{PQansatz} and \eqref{Asqr}--\eqref{Bsqr}, one finds the Q-functions \eqref{Q10} and \eqref{Q20}
which reproduce the asymptotic T-system \eqref{Tinfty_right}--\eqref{Tinfty_Ta0} for generic chemical potentials (satisfying the bounds \eqref{assume_omega_strict}). This confirms the large-$u$ asymptotics \eqref{Q10}--\eqref{ABsqr} of the QSC.

Employing the first identities \eqref{id_Asqr} together with \eqref{Q10} and \eqref{xy_con}, one finds
\begin{equation}
\Pb_a \Pb^a =  0 \spa \Qb_i \Qb^i =  0 \,.
\end{equation}
Moreover, the second identities \eqref{id_Asqr} together with Eqs.\ \eqref{Q10}, \eqref{Q11} and \eqref{xy_con} give the relations \eqref{eq: QSC equation 2}. Finally, the identities \eqref{id_Asqr_adv2} together with Eq.\ \eqref{Q11} give Eq.\ \eqref{eq: QSC orthonormality}. 

Finally, we note that for $\Omega_3=\Omega_5=0$ one has left-right symmetry of the QSC, and the $\Pb_a$ and $\Qb_i$ functions should satisfy Eqs.\ \eqref{eq: QSC upper indices}--\eqref{eq: chi matrix}. 
Indeed, using $\Omega_3=\Omega_5=0$ one finds from Eq.\ \eqref{Hagedorn_xy} that $x_4 = 1/x_1$, $x_3=1/x_2$, $y_4=1/y_1$ and $y_3=1/y_2$. Using this, $\Pb^a = \chi^{ab} \Pb_b$ and $\Qb^i = \chi^{ij} \Qb_j$ provided $A^a = \chi^{ab} A_b$ and $B^i = \chi^{ij} B_j$. These relations require $A_1 A^1 = - A_4 A^4$, $A_2 A^2 = - A_3 A^3$, $B_1 B^1 = - B_4 B^4$ and $B_2 B^2 = - B_3 B^3$ which can be derived from Eqs.\ \eqref{Asqr} and \eqref{Bsqr}. Finally, it is easily checked that $Q^{a|i} = \chi^{ab} \chi^{ij} Q_{b|j}$ from Eq.\ \eqref{Q11}.

\section{Asymptotic \texorpdfstring{$Q_{a|i}$}{Qai} for vanishing chemical potentials}
\label{app: asymptotic Qai}

In this appendix, we give the functions $Q_{a|i}$ that correspond to the asymptotic T-system in the case of vanishing chemical potentials; we deferred these in Subsec.\ \ref{subsec: asymptotic solution}.
In particular, these $Q_{a|i}$ functions give $Q_{a|i}^{(0)}$ when setting $\THag=\THag^{(0)}$, thus providing the remaining part of the starting point for the perturbative solution to the QSC.

We have
\begin{equation}
\label{eq: assymptotic Qai 1}
\begin{aligned}
Q_{1|1}(u)&=A_1 B_1  \left(-\e^{-\frac{1}{2\THag}}\right)^{-i u}
\frac{1}{2}(-i) \sech \left(\tfrac{1}{4 \THag}\right),
\\
Q_{1|2}(u)&=A_1 B_2  \left(-\e^{-\frac{1}{2\THag}}\right)^{-i u}
\frac{1}{4} \sech\left(\tfrac{1}{4 \THag}\right) \left[\tanh \left(\tfrac{1}{4 \THag}\right)-2 i u\right],
\\
Q_{1|3}(u)&=A_1 B_3  \left(-\e^{-\frac{1}{2\THag}}\right)^{-i u}
\frac{1}{8}(- i) \sech\left(\tfrac{1}{4 \THag}\right) \big[4 i u \tanh \left(\tfrac{1}{4 \THag}\right)\\&\hspace{0.5\textwidth}+2 \sech^2\left(\tfrac{1}{4 \THag}\right)+4 u^2-1\big],
\\
Q_{1|4}(u)&=A_1 B_4  \left(-\e^{-\frac{1}{2\THag}}\right)^{-i u}
\frac{1}{32} \sech^3\left(\tfrac{1}{4 \THag}\right) \big[3 \left(4 u^2+1\right) \sinh \left(\tfrac{1}{2 \THag}\right)\\&\hspace{0.2\textwidth}-2 i \left(4 u^3+u\right) \cosh \left(\tfrac{1}{2 \THag}\right)-12 \tanh \left(\tfrac{1}{4 \THag}\right)-8 i u^3+22 i u\big],
\\
Q_{2|1}(u)&=A_2 B_1  \left(-\e^{-\frac{1}{2\THag}}\right)^{-i u}
\frac{1}{8} \csch\left(\tfrac{1}{4 \THag}\right) \sech^2\left(\tfrac{1}{4 \THag}\right) \left[1-2 i u \sinh \left(\tfrac{1}{2 \THag}\right)\right],
\\
Q_{2|2}(u)&=A_2 B_2  \left(-\e^{-\frac{1}{2\THag}}\right)^{-i u}
\frac{1}{16} \big[2 u \csch\left(\tfrac{1}{4 \THag}\right)\\&\hspace{0.2\textwidth}+\sech\left(\tfrac{1}{4 \THag}\right) \left(2 u \tanh \left(\tfrac{1}{4 \THag}\right)-i \left(\sech^2\left(\tfrac{1}{4 \THag}\right)+8 u^2\right)\right)\big],
\\
Q_{2|3}(u)&=A_2 B_3  \left(-\e^{-\frac{1}{2\THag}}\right)^{-i u}
\frac{1}{64} \csch\left(\tfrac{1}{4 \THag}\right) \sech^4\left(\tfrac{1}{4 \THag}\right) \big[\left(4 u^2+7\right) \cosh \left(\tfrac{1}{2 \THag}\right)
\\&+u \left(4 u \cosh \left(\tfrac{1}{\THag}\right)-i \left(\left(4 u^2-1\right) \sinh \left(\tfrac{1}{\THag}\right)+2 \left(4 u^2+5\right) \sinh \left(\tfrac{1}{2 \THag}\right)\right)\right)-5\big],
 \\
 Q_{2|4}(u)&=A_2 B_4  \left(-\e^{-\frac{1}{2\THag}}\right)^{-i u}
 \frac{1}{64} \big[2 u \left(4 u^2-5\right) \csch\left(\tfrac{1}{4 \THag}\right)\\&+\sech\left(\tfrac{1}{4 \THag}\right) \big\{(2 u \tanh \left(\tfrac{1}{4 \THag}\right) \left(18 \sech^2\left(\tfrac{1}{4 \THag}\right)+20 u^2+11\right)\\&-i \left(-3 \left(12 u^2+7\right) \sech^2\left(\tfrac{1}{4 \THag}\right)+18 \sech^4\left(\tfrac{1}{4 \THag}\right)+8 \left(4 u^4+u^2\right)\right)\big\}\big],
\end{aligned}
\end{equation}
and
\begin{equation}
\label{eq: assymptotic Qai 2}
 Q_{a+2|i}(u)=\frac{A_{a+2}}{A_a}Q_{a+2|i}(-u)\quad \text{for }a=1,2\,.
\end{equation}

\bibliographystyle{utphys2}
\bibliography{mybib}

\providecommand{\href}[2]{#2}\begingroup\raggedright\begin{thebibliography}{10}

\bibitem{Harmark:2017yrv}
T.~Harmark and M.~Wilhelm, ``{The Hagedorn temperature of AdS$_5$/CFT$_4$ via
  integrability},''
  \href{http://dx.doi.org/10.1103/PhysRevLett.120.071605}{{\em Phys. Rev.
  Lett.} {\bfseries 120} no.~7, (2018) 071605},
\href{http://arxiv.org/abs/1706.03074}{{\ttfamily arXiv:1706.03074 [hep-th]}}.

\bibitem{Harmark:2018red}
T.~Harmark and M.~Wilhelm, ``{The Hagedorn temperature of AdS$_5$/CFT$_4$ at
  finite coupling via the Quantum Spectral Curve},''
  \href{http://dx.doi.org/10.1016/j.physletb.2018.09.033}{{\em Phys. Lett. B}
  {\bfseries 786} (2018) 53--58},
  \href{http://arxiv.org/abs/1803.04416}{{\ttfamily arXiv:1803.04416
  [hep-th]}}.

\bibitem{Beisert:2010jr}
N.~Beisert {\em et~al.}, ``{Review of AdS/CFT Integrability: An Overview},''
  \href{http://dx.doi.org/10.1007/s11005-011-0529-2}{{\em Lett. Math. Phys.}
  {\bfseries 99} (2012) 3--32},
\href{http://arxiv.org/abs/1012.3982}{{\ttfamily arXiv:1012.3982 [hep-th]}}.

\bibitem{Bombardelli:2016rwb}
D.~Bombardelli, A.~Cagnazzo, R.~Frassek, F.~Levkovich-Maslyuk, F.~Loebbert,
  S.~Negro, I.~M. Szécsényi, A.~Sfondrini, S.~J. van Tongeren, and
  A.~Torrielli, ``{An integrability primer for the gauge-gravity
  correspondence: An introduction},''
  \href{http://dx.doi.org/10.1088/1751-8113/49/32/320301}{{\em J. Phys.}
  {\bfseries A49} no.~32, (2016) 320301},
\href{http://arxiv.org/abs/1606.02945}{{\ttfamily arXiv:1606.02945 [hep-th]}}.

\bibitem{Arutyunov:2009zu}
G.~Arutyunov and S.~Frolov, ``{String hypothesis for the AdS$_5\times S^5$
  mirror},'' \href{http://dx.doi.org/10.1088/1126-6708/2009/03/152}{{\em JHEP}
  {\bfseries 03} (2009) 152},
\href{http://arxiv.org/abs/0901.1417}{{\ttfamily arXiv:0901.1417 [hep-th]}}.

\bibitem{Bombardelli:2009ns}
D.~Bombardelli, D.~Fioravanti, and R.~Tateo, ``{Thermodynamic Bethe Ansatz for
  planar AdS/CFT: A Proposal},''
  \href{http://dx.doi.org/10.1088/1751-8113/42/37/375401}{{\em J. Phys.}
  {\bfseries A42} (2009) 375401},
\href{http://arxiv.org/abs/0902.3930}{{\ttfamily arXiv:0902.3930 [hep-th]}}.

\bibitem{Gromov:2009bc}
N.~Gromov, V.~Kazakov, A.~Kozak, and P.~Vieira, ``{Exact Spectrum of Anomalous
  Dimensions of Planar $\mathcal{N} = 4$ Supersymmetric Yang-Mills Theory: TBA
  and excited states},''
  \href{http://dx.doi.org/10.1007/s11005-010-0374-8}{{\em Lett. Math. Phys.}
  {\bfseries 91} (2010) 265--287},
\href{http://arxiv.org/abs/0902.4458}{{\ttfamily arXiv:0902.4458 [hep-th]}}.

\bibitem{Arutyunov:2009ur}
G.~Arutyunov and S.~Frolov, ``{Thermodynamic Bethe Ansatz for the AdS$_5 \times
  S^5$ Mirror Model},''
  \href{http://dx.doi.org/10.1088/1126-6708/2009/05/068}{{\em JHEP} {\bfseries
  05} (2009) 068},
\href{http://arxiv.org/abs/0903.0141}{{\ttfamily arXiv:0903.0141 [hep-th]}}.

\bibitem{Gromov:2009tv}
N.~Gromov, V.~Kazakov, and P.~Vieira, ``{Exact Spectrum of Anomalous Dimensions
  of Planar $\mathcal{N}=4$ Supersymmetric Yang-Mills Theory},''
  \href{http://dx.doi.org/10.1103/PhysRevLett.103.131601}{{\em Phys. Rev.
  Lett.} {\bfseries 103} (2009) 131601},
\href{http://arxiv.org/abs/0901.3753}{{\ttfamily arXiv:0901.3753 [hep-th]}}.

\bibitem{Cavaglia:2010nm}
A.~Cavaglia, D.~Fioravanti, and R.~Tateo, ``{Extended Y-system for the
  $AdS_5/CFT_4$ correspondence},''
  \href{http://dx.doi.org/10.1016/j.nuclphysb.2010.09.015}{{\em Nucl. Phys.}
  {\bfseries B843} (2011) 302--343},
\href{http://arxiv.org/abs/1005.3016}{{\ttfamily arXiv:1005.3016 [hep-th]}}.

\bibitem{Gromov:2013pga}
N.~Gromov, V.~Kazakov, S.~Leurent, and D.~Volin, ``{Quantum Spectral Curve for
  Planar $\mathcal{N} =4$ Super-Yang-Mills Theory},''
  \href{http://dx.doi.org/10.1103/PhysRevLett.112.011602}{{\em Phys. Rev.
  Lett.} {\bfseries 112} no.~1, (2014) 011602},
\href{http://arxiv.org/abs/1305.1939}{{\ttfamily arXiv:1305.1939 [hep-th]}}.

\bibitem{Gromov:2014caa}
N.~Gromov, V.~Kazakov, S.~Leurent, and D.~Volin, ``{Quantum spectral curve for
  arbitrary state/operator in AdS$_{5}$/CFT$_{4}$},''
  \href{http://dx.doi.org/10.1007/JHEP09(2015)187}{{\em JHEP} {\bfseries 09}
  (2015) 187},
\href{http://arxiv.org/abs/1405.4857}{{\ttfamily arXiv:1405.4857 [hep-th]}}.

\bibitem{Marboe:2014gma}
C.~Marboe and D.~Volin, ``{Quantum spectral curve as a tool for a perturbative
  quantum field theory},''
  \href{http://dx.doi.org/10.1016/j.nuclphysb.2015.08.021}{{\em Nucl. Phys.}
  {\bfseries B899} (2015) 810--847},
\href{http://arxiv.org/abs/1411.4758}{{\ttfamily arXiv:1411.4758 [hep-th]}}.

\bibitem{Marboe:2017dmb}
C.~Marboe and D.~Volin, ``{The full spectrum of AdS5/CFT4 I: Representation
  theory and one-loop Q-system},''
  \href{http://dx.doi.org/10.1088/1751-8121/aab34a}{{\em J. Phys.} {\bfseries
  A51} no.~16, (2018) 165401},
\href{http://arxiv.org/abs/1701.03704}{{\ttfamily arXiv:1701.03704 [hep-th]}}.

\bibitem{Marboe:2018ugv}
C.~Marboe and D.~Volin, ``{The full spectrum of AdS$_5$/CFT$_4$ II: Weak
  coupling expansion via the quantum spectral curve},''
\href{http://arxiv.org/abs/1812.09238}{{\ttfamily arXiv:1812.09238 [hep-th]}}.

\bibitem{Gromov:2015wca}
N.~Gromov, F.~Levkovich-Maslyuk, and G.~Sizov, ``{Quantum Spectral Curve and
  the Numerical Solution of the Spectral Problem in AdS5/CFT4},''
  \href{http://dx.doi.org/10.1007/JHEP06(2016)036}{{\em JHEP} {\bfseries 06}
  (2016) 036},
\href{http://arxiv.org/abs/1504.06640}{{\ttfamily arXiv:1504.06640 [hep-th]}}.

\bibitem{Hegedus:2016eop}
A.~Heged\"{u}s and J.~Konczer, ``{Strong coupling results in the
  AdS$_{5}$/CFT$_{4}$ correspondence from the numerical solution of the quantum
  spectral curve},'' \href{http://dx.doi.org/10.1007/JHEP08(2016)061}{{\em
  JHEP} {\bfseries 08} (2016) 061},
\href{http://arxiv.org/abs/1604.02346}{{\ttfamily arXiv:1604.02346 [hep-th]}}.

\bibitem{Alfimov:2014bwa}
M.~Alfimov, N.~Gromov, and V.~Kazakov, ``{QCD Pomeron from AdS/CFT Quantum
  Spectral Curve},'' \href{http://dx.doi.org/10.1007/JHEP07(2015)164}{{\em
  JHEP} {\bfseries 07} (2015) 164},
\href{http://arxiv.org/abs/1408.2530}{{\ttfamily arXiv:1408.2530 [hep-th]}}.

\bibitem{Gromov:2015vua}
N.~Gromov, F.~Levkovich-Maslyuk, and G.~Sizov, ``{Pomeron Eigenvalue at Three
  Loops in $\mathcal N=$ 4 Supersymmetric Yang-Mills Theory},''
  \href{http://dx.doi.org/10.1103/PhysRevLett.115.251601}{{\em Phys. Rev.
  Lett.} {\bfseries 115} no.~25, (2015) 251601},
\href{http://arxiv.org/abs/1507.04010}{{\ttfamily arXiv:1507.04010 [hep-th]}}.

\bibitem{Alfimov:2020obh}
M.~Alfimov, N.~Gromov, and V.~Kazakov, ``{$\mathcal{N}=4$ SYM Quantum Spectral
  Curve in BFKL regime},'' \href{http://arxiv.org/abs/2003.03536}{{\ttfamily
  arXiv:2003.03536 [hep-th]}}.

\bibitem{Gromov:2015dfa}
N.~Gromov and F.~Levkovich-Maslyuk, ``{Quantum Spectral Curve for a cusped
  Wilson line in $ \mathcal{N}=4 $ SYM},''
  \href{http://dx.doi.org/10.1007/JHEP04(2016)134}{{\em JHEP} {\bfseries 04}
  (2016) 134},
\href{http://arxiv.org/abs/1510.02098}{{\ttfamily arXiv:1510.02098 [hep-th]}}.

\bibitem{Cavaglia:2018lxi}
A.~Cavaglià, N.~Gromov, and F.~Levkovich-Maslyuk, ``{Quantum spectral curve
  and structure constants in $ \mathcal{N}=4 $ SYM: cusps in the ladder
  limit},'' \href{http://dx.doi.org/10.1007/JHEP10(2018)060}{{\em JHEP}
  {\bfseries 10} (2018) 060}, \href{http://arxiv.org/abs/1802.04237}{{\ttfamily
  arXiv:1802.04237 [hep-th]}}.

\bibitem{Grabner:2020nis}
D.~Grabner, N.~Gromov, and J.~Julius, ``{Excited States of One-Dimensional
  Defect CFTs from the Quantum Spectral Curve},''
  \href{http://dx.doi.org/10.1007/JHEP07(2020)042}{{\em JHEP} {\bfseries 07}
  (2020) 042}, \href{http://arxiv.org/abs/2001.11039}{{\ttfamily
  arXiv:2001.11039 [hep-th]}}.

\bibitem{Gromov:2021ahm}
N.~Gromov, J.~Julius, and N.~Primi, ``{Open Fishchain in $\mathcal{N}=4$
  Supersymmetric Yang-Mills Theory},''
  \href{http://arxiv.org/abs/2101.01232}{{\ttfamily arXiv:2101.01232
  [hep-th]}}.

\bibitem{Cavaglia:2021bnz}
A.~Cavagli\`a, N.~Gromov, J.~Julius, and M.~Preti, ``{Integrability and
  Conformal Bootstrap: One Dimensional Defect CFT},''
  \href{http://arxiv.org/abs/2107.08510}{{\ttfamily arXiv:2107.08510
  [hep-th]}}.

\bibitem{Gromov:2016rrp}
N.~Gromov and F.~Levkovich-Maslyuk, ``{Quark-anti-quark potential in $
  \mathcal{N} = 4$ SYM},''
  \href{http://dx.doi.org/10.1007/JHEP12(2016)122}{{\em JHEP} {\bfseries 12}
  (2016) 122},
\href{http://arxiv.org/abs/1601.05679}{{\ttfamily arXiv:1601.05679 [hep-th]}}.

\bibitem{Cavaglia:2020hdb}
A.~Cavaglia, D.~Grabner, N.~Gromov, and A.~Sever, ``{Colour-twist operators.
  Part I. Spectrum and wave functions},''
  \href{http://dx.doi.org/10.1007/JHEP06(2020)092}{{\em JHEP} {\bfseries 06}
  (2020) 092}, \href{http://arxiv.org/abs/2001.07259}{{\ttfamily
  arXiv:2001.07259 [hep-th]}}.

\bibitem{Kazakov:2015efa}
V.~Kazakov, S.~Leurent, and D.~Volin, ``{T-system on T-hook: Grassmannian
  Solution and Twisted Quantum Spectral Curve},''
  \href{http://dx.doi.org/10.1007/JHEP12(2016)044}{{\em JHEP} {\bfseries 12}
  (2016) 044},
\href{http://arxiv.org/abs/1510.02100}{{\ttfamily arXiv:1510.02100 [hep-th]}}.

\bibitem{Klabbers:2017vtw}
R.~Klabbers and S.~J. van Tongeren, ``{Quantum Spectral Curve for the
  eta-deformed AdS$_5\times S^5$ superstring},''
  \href{http://dx.doi.org/10.1016/j.nuclphysb.2017.10.005}{{\em Nucl. Phys.}
  {\bfseries B925} (2017) 252--318},
\href{http://arxiv.org/abs/1708.02894}{{\ttfamily arXiv:1708.02894 [hep-th]}}.

\bibitem{Gromov:2017cja}
N.~Gromov, V.~Kazakov, G.~Korchemsky, S.~Negro, and G.~Sizov, ``{Integrability
  of Conformal Fishnet Theory},''
  \href{http://dx.doi.org/10.1007/JHEP01(2018)095}{{\em JHEP} {\bfseries 01}
  (2018) 095},
\href{http://arxiv.org/abs/1706.04167}{{\ttfamily arXiv:1706.04167 [hep-th]}}.

\bibitem{Marboe:2019wyc}
C.~Marboe and E.~Widén, ``{The fate of the Konishi multiplet in the
  $\beta$-deformed Quantum Spectral Curve},''
  \href{http://dx.doi.org/10.1007/JHEP01(2020)026}{{\em JHEP} {\bfseries 01}
  (2020) 026}, \href{http://arxiv.org/abs/1902.01248}{{\ttfamily
  arXiv:1902.01248 [hep-th]}}.

\bibitem{Levkovich-Maslyuk:2020rlp}
F.~Levkovich-Maslyuk and M.~Preti, ``{Exploring the ground state spectrum of
  $\gamma$-deformed $\mathcal{N}=4$ SYM},''
  \href{http://arxiv.org/abs/2003.05811}{{\ttfamily arXiv:2003.05811
  [hep-th]}}.

\bibitem{Sundborg:1984uk}
B.~Sundborg, ``{Thermodynamics of Superstrings at High-energy Densities},''
\href{http://dx.doi.org/10.1016/0550-3213(85)90235-4}{{\em Nucl. Phys.}
  {\bfseries B254} (1985) 583--592}.

\bibitem{Sundborg:1999ue}
B.~Sundborg, ``{The Hagedorn transition, deconfinement and $\mathcal{N}=4$ SYM
  theory},'' \href{http://dx.doi.org/10.1016/S0550-3213(00)00044-4}{{\em Nucl.
  Phys.} {\bfseries B573} (2000) 349--363},
\href{http://arxiv.org/abs/hep-th/9908001}{{\ttfamily arXiv:hep-th/9908001
  [hep-th]}}.

\bibitem{Spradlin:2004pp}
M.~Spradlin and A.~Volovich, ``{A Pendant for Polya: The One-loop partition
  function of $\mathcal{N}=4$ SYM on $\mathbb{R}\times S^3$},''
  \href{http://dx.doi.org/10.1016/j.nuclphysb.2005.01.007}{{\em Nucl. Phys.}
  {\bfseries B711} (2005) 199--230},
\href{http://arxiv.org/abs/hep-th/0408178}{{\ttfamily arXiv:hep-th/0408178
  [hep-th]}}.

\bibitem{Atick:1988si}
J.~J. Atick and E.~Witten, ``{The Hagedorn Transition and the Number of Degrees
  of Freedom of String Theory},''
\href{http://dx.doi.org/10.1016/0550-3213(88)90151-4}{{\em Nucl. Phys.}
  {\bfseries B310} (1988) 291--334}.

\bibitem{Witten:1998zw}
E.~Witten, ``{Anti-de Sitter space, thermal phase transition, and confinement
  in gauge theories},'' {\em Adv. Theor. Math. Phys.} {\bfseries 2} (1998)
  505--532,
\href{http://arxiv.org/abs/hep-th/9803131}{{\ttfamily arXiv:hep-th/9803131
  [hep-th]}}.

\bibitem{Aharony:2003sx}
O.~Aharony, J.~Marsano, S.~Minwalla, K.~Papadodimas, and M.~Van~Raamsdonk,
  ``{The Hagedorn - deconfinement phase transition in weakly coupled large N
  gauge theories},'' \href{http://dx.doi.org/10.4310/ATMP.2004.v8.n4.a1}{{\em
  Adv. Theor. Math. Phys.} {\bfseries 8} (2004) 603--696},
\href{http://arxiv.org/abs/hep-th/0310285}{{\ttfamily arXiv:hep-th/0310285
  [hep-th]}}.

\bibitem{Brown:2004ugm}
F.~C.~S. Brown, ``{Polylogarithmes multiples uniformes en une variable},''
\href{http://dx.doi.org/10.1016/j.crma.2004.02.001}{{\em Compt. Rend. Math.}
  {\bfseries 338} no.~7, (2004) 527--532}.

\bibitem{Leurent:2013mr}
S.~Leurent and D.~Volin, ``{Multiple zeta functions and double wrapping in
  planar $\mathcal{N}=4$ SYM},''
  \href{http://dx.doi.org/10.1016/j.nuclphysb.2013.07.020}{{\em Nucl. Phys.}
  {\bfseries B875} (2013) 757--789},
\href{http://arxiv.org/abs/1302.1135}{{\ttfamily arXiv:1302.1135 [hep-th]}}.

\bibitem{MaldacenaPrivateCommunication}
J.~Maldacena, ``Correction to the Hagedorn temperature in $AdS_5\times S^5$.''
  Unpublished note.

\bibitem{Urbach:2022xzw}
E.~Y. Urbach, ``{String Stars in Anti de Sitter Space},''
  \href{http://arxiv.org/abs/2202.06966}{{\ttfamily arXiv:2202.06966
  [hep-th]}}.

\bibitem{Fokken:2014moa}
J.~Fokken and M.~Wilhelm, ``{One-Loop Partition Functions in Deformed
  $\mathcal{N}=4$ SYM Theory},''
  \href{http://dx.doi.org/10.1007/JHEP03(2015)018}{{\em JHEP} {\bfseries 03}
  (2015) 018},
\href{http://arxiv.org/abs/1411.7695}{{\ttfamily arXiv:1411.7695 [hep-th]}}.

\bibitem{Yamada:2006rx}
D.~Yamada and L.~G. Yaffe, ``{Phase diagram of $\mathcal{N}=4$ super-Yang-Mills
  theory with R-symmetry chemical potentials},''
  \href{http://dx.doi.org/10.1088/1126-6708/2006/09/027}{{\em JHEP} {\bfseries
  09} (2006) 027},
\href{http://arxiv.org/abs/hep-th/0602074}{{\ttfamily arXiv:hep-th/0602074
  [hep-th]}}.

\bibitem{Harmark:2006di}
T.~Harmark and M.~Orselli, ``{Quantum mechanical sectors in thermal
  $\mathcal{N}=4$ super Yang-Mills on $\mathbb{R}\times S^3$},''
  \href{http://dx.doi.org/10.1016/j.nuclphysb.2006.08.022}{{\em Nucl. Phys.}
  {\bfseries B757} (2006) 117--145},
\href{http://arxiv.org/abs/hep-th/0605234}{{\ttfamily arXiv:hep-th/0605234
  [hep-th]}}.

\bibitem{Harmark:2006ie}
T.~Harmark, K.~R. Kristjansson, and M.~Orselli, ``{Magnetic
  Heisenberg-chain/pp-wave correspondence},''
  \href{http://dx.doi.org/10.1088/1126-6708/2007/02/085}{{\em JHEP} {\bfseries
  02} (2007) 085}, \href{http://arxiv.org/abs/hep-th/0611242}{{\ttfamily
  arXiv:hep-th/0611242}}.

\bibitem{Harmark:2007px}
T.~Harmark, K.~R. Kristjansson, and M.~Orselli, ``{Decoupling limits of
  $\mathcal{N}=4$ super Yang-Mills on $\mathbb{R}\times S^3$},''
  \href{http://dx.doi.org/10.1088/1126-6708/2007/09/115}{{\em JHEP} {\bfseries
  09} (2007) 115},
\href{http://arxiv.org/abs/0707.1621}{{\ttfamily arXiv:0707.1621 [hep-th]}}.

\bibitem{Harmark:2014mpa}
T.~Harmark and M.~Orselli, ``{Spin Matrix Theory: A quantum mechanical model of
  the AdS/CFT correspondence},''
  \href{http://dx.doi.org/10.1007/JHEP11(2014)134}{{\em JHEP} {\bfseries 11}
  (2014) 134},
\href{http://arxiv.org/abs/1409.4417}{{\ttfamily arXiv:1409.4417 [hep-th]}}.

\bibitem{Suzuki:2017ipd}
R.~Suzuki, ``{Refined Counting of Necklaces in One-loop $\mathcal{N}=4$ SYM},''
  \href{http://dx.doi.org/10.1007/JHEP06(2017)055}{{\em JHEP} {\bfseries 06}
  (2017) 055},
\href{http://arxiv.org/abs/1703.05798}{{\ttfamily arXiv:1703.05798 [hep-th]}}.

\bibitem{Arutyunov:2010gu}
G.~Arutyunov, M.~de~Leeuw, and S.~J. van Tongeren, ``{Twisting the Mirror
  TBA},'' \href{http://dx.doi.org/10.1007/JHEP02(2011)025}{{\em JHEP}
  {\bfseries 02} (2011) 025}, \href{http://arxiv.org/abs/1009.4118}{{\ttfamily
  arXiv:1009.4118 [hep-th]}}.

\bibitem{Gromov:2017blm}
N.~Gromov, ``{Introduction to the Spectrum of $\mathcal{N}=4$ SYM and the
  Quantum Spectral Curve},''
\href{http://arxiv.org/abs/1708.03648}{{\ttfamily arXiv:1708.03648 [hep-th]}}.

\bibitem{Kazakov:2018ugh}
V.~Kazakov, {\em {Quantum Spectral Curve of $\gamma$-twisted ${\cal N}=4$ SYM
  theory and fishnet CFT}}, vol.~30,
  \href{http://dx.doi.org/10.1142/9789813233867\_0016}{p.~1840010}.
\newblock 2018.
\newblock \href{http://arxiv.org/abs/1802.02160}{{\ttfamily arXiv:1802.02160
  [hep-th]}}.

\bibitem{Levkovich-Maslyuk:2019awk}
F.~Levkovich-Maslyuk, ``{A review of the AdS/CFT Quantum Spectral Curve},''
  \href{http://dx.doi.org/10.1088/1751-8121/ab7137}{{\em J. Phys. A} {\bfseries
  53} no.~28, (2020) 283004}, \href{http://arxiv.org/abs/1911.13065}{{\ttfamily
  arXiv:1911.13065 [hep-th]}}.

\bibitem{Duhr:2014woa}
C.~Duhr, \href{http://dx.doi.org/10.1142/9789814678766_0010}{``{Mathematical
  aspects of scattering amplitudes},''} in {\em {Proceedings, Theoretical
  Advanced Study Institute in Elementary Particle Physics: Journeys Through the
  Precision Frontier: Amplitudes for Colliders (TASI 2014): Boulder, Colorado,
  June 2-27, 2014}}, pp.~419--476.
\newblock 2015.
\newblock
\href{http://arxiv.org/abs/1411.7538}{{\ttfamily arXiv:1411.7538 [hep-ph]}}.
\newblock

\bibitem{Vollinga:2004sn}
J.~Vollinga and S.~Weinzierl, ``{Numerical evaluation of multiple
  polylogarithms},'' \href{http://dx.doi.org/10.1016/j.cpc.2004.12.009}{{\em
  Comput. Phys. Commun.} {\bfseries 167} (2005) 177},
\href{http://arxiv.org/abs/hep-ph/0410259}{{\ttfamily arXiv:hep-ph/0410259
  [hep-ph]}}.

\bibitem{Goncharov:2010jf}
A.~B. Goncharov, M.~Spradlin, C.~Vergu, and A.~Volovich, ``{Classical
  Polylogarithms for Amplitudes and Wilson Loops},''
  \href{http://dx.doi.org/10.1103/PhysRevLett.105.151605}{{\em Phys. Rev.
  Lett.} {\bfseries 105} (2010) 151605},
  \href{http://arxiv.org/abs/1006.5703}{{\ttfamily arXiv:1006.5703 [hep-th]}}.

\bibitem{Remiddi:1999ew}
E.~Remiddi and J.~A.~M. Vermaseren, ``{Harmonic polylogarithms},''
  \href{http://dx.doi.org/10.1142/S0217751X00000367}{{\em Int. J. Mod. Phys.}
  {\bfseries A15} (2000) 725--754},
\href{http://arxiv.org/abs/hep-ph/9905237}{{\ttfamily arXiv:hep-ph/9905237
  [hep-ph]}}.

\bibitem{Chen}
K.-T. Chen, ``{Iterated Path Integrals},''
  \href{http://dx.doi.org/10.1090/S0002-9904-1977-14320-6}{{\em Bull. Amer.
  Math. Soc.} {\bfseries 83} no.~5, (1977) 831--879}.

\bibitem{G91b}
A.~B. Goncharov, ``{Geometry of Configurations, Polylogarithms, and Motivic
  Cohomology},''
  \href{http://dx.doi.org/http://dx.doi.org/10.1006/aima.1995.1045}{{\em Adv.
  Math.} {\bfseries 114} no.~2, (1995) 197--318}.

\bibitem{Goncharov:1998kja}
A.~B. Goncharov, ``{Multiple Polylogarithms, Cyclotomy and Modular
  Complexes},'' \href{http://dx.doi.org/10.4310/MRL.1998.v5.n4.a7}{{\em Math.
  Res. Lett.} {\bfseries 5} (1998) 497--516},
\href{http://arxiv.org/abs/1105.2076}{{\ttfamily arXiv:1105.2076 [math.AG]}}.

\bibitem{Panzer:2014caa}
E.~Panzer, ``{Algorithms for the symbolic integration of hyperlogarithms with
  applications to Feynman integrals},''
  \href{http://dx.doi.org/10.1016/j.cpc.2014.10.019}{{\em Comput. Phys.
  Commun.} {\bfseries 188} (2015) 148--166},
\href{http://arxiv.org/abs/1403.3385}{{\ttfamily arXiv:1403.3385 [hep-th]}}.

\bibitem{Dixon:2012yy}
L.~J. Dixon, C.~Duhr, and J.~Pennington, ``{Single-valued harmonic
  polylogarithms and the multi-Regge limit},''
  \href{http://dx.doi.org/10.1007/JHEP10(2012)074}{{\em JHEP} {\bfseries 10}
  (2012) 074}, \href{http://arxiv.org/abs/1207.0186}{{\ttfamily arXiv:1207.0186
  [hep-th]}}.

\bibitem{hyperlogprocedures}
O.~Schnetz, ``HyperLogProcedures.''
\newblock \url{https://www.math.fau.de/person/oliver-schnetz/}. Maple program.

\bibitem{Alfimov:2018cms}
M.~Alfimov, N.~Gromov, and G.~Sizov, ``{BFKL spectrum of $ \mathcal{N} = 4$:
  non-zero conformal spin},''
  \href{http://dx.doi.org/10.1007/JHEP07(2018)181}{{\em JHEP} {\bfseries 07}
  (2018) 181}, \href{http://arxiv.org/abs/1802.06908}{{\ttfamily
  arXiv:1802.06908 [hep-th]}}.

\bibitem{Gromov:2010vb}
N.~Gromov, V.~Kazakov, and Z.~Tsuboi, ``{PSU$(2,2|4)$ Character of
  Quasiclassical AdS/CFT},''
  \href{http://dx.doi.org/10.1007/JHEP07(2010)097}{{\em JHEP} {\bfseries 07}
  (2010) 097},
\href{http://arxiv.org/abs/1002.3981}{{\ttfamily arXiv:1002.3981 [hep-th]}}.

\bibitem{Beisert:2005if}
N.~Beisert and R.~Roiban, ``{Beauty and the twist: The Bethe ansatz for twisted
  $\mathcal{N}=4$ SYM},''
  \href{http://dx.doi.org/10.1088/1126-6708/2005/08/039}{{\em JHEP} {\bfseries
  08} (2005) 039},
\href{http://arxiv.org/abs/hep-th/0505187}{{\ttfamily arXiv:hep-th/0505187
  [hep-th]}}.

\bibitem{Leigh:1995ep}
R.~G. Leigh and M.~J. Strassler, ``{Exactly marginal operators and duality in
  four-dimensional $\mathcal{N}=1$ supersymmetric gauge theory},''
  \href{http://dx.doi.org/10.1016/0550-3213(95)00261-P}{{\em Nucl. Phys.}
  {\bfseries B447} (1995) 95--136},
\href{http://arxiv.org/abs/hep-th/9503121}{{\ttfamily arXiv:hep-th/9503121
  [hep-th]}}.

\bibitem{Frolov:2005dj}
S.~Frolov, ``{Lax pair for strings in Lunin-Maldacena background},''
  \href{http://dx.doi.org/10.1088/1126-6708/2005/05/069}{{\em JHEP} {\bfseries
  05} (2005) 069},
\href{http://arxiv.org/abs/hep-th/0503201}{{\ttfamily arXiv:hep-th/0503201
  [hep-th]}}.

\bibitem{Fokken:2013aea}
J.~Fokken, C.~Sieg, and M.~Wilhelm, ``{Non-conformality of ${{\gamma
  }_{i}}$-deformed $\mathcal{N} = 4$ SYM theory},''
  \href{http://dx.doi.org/10.1088/1751-8113/47/45/455401}{{\em J. Phys.}
  {\bfseries A47} (2014) 455401},
\href{http://arxiv.org/abs/1308.4420}{{\ttfamily arXiv:1308.4420 [hep-th]}}.

\bibitem{Fokken:2013mza}
J.~Fokken, C.~Sieg, and M.~Wilhelm, ``{The complete one-loop dilatation
  operator of planar real $\beta$-deformed $ \mathcal{N} $ = 4 SYM theory},''
  \href{http://dx.doi.org/10.1007/JHEP07(2014)150}{{\em JHEP} {\bfseries 07}
  (2014) 150},
\href{http://arxiv.org/abs/1312.2959}{{\ttfamily arXiv:1312.2959 [hep-th]}}.

\bibitem{Fokken:2014soa}
J.~Fokken, C.~Sieg, and M.~Wilhelm, ``{A piece of cake: the ground-state
  energies in $\gamma_{i}$-deformed $ \mathcal{N} $ = 4 SYM theory at leading
  wrapping order},'' \href{http://dx.doi.org/10.1007/JHEP09(2014)078}{{\em
  JHEP} {\bfseries 09} (2014) 078},
\href{http://arxiv.org/abs/1405.6712}{{\ttfamily arXiv:1405.6712 [hep-th]}}.

\bibitem{Gurdogan:2015csr}
O.~G\"urdo\u{g}an and V.~Kazakov, ``{New Integrable 4D Quantum Field Theories
  from Strongly Deformed Planar $\mathcal N = 4 $ Supersymmetric Yang-Mills
  Theory},'' \href{http://dx.doi.org/10.1103/PhysRevLett.117.201602}{{\em Phys.
  Rev. Lett.} {\bfseries 117} no.~20, (2016) 201602},
  \href{http://arxiv.org/abs/1512.06704}{{\ttfamily arXiv:1512.06704
  [hep-th]}}. [Addendum: Phys.Rev.Lett. 117, 259903 (2016)].

\bibitem{Sieg:2016vap}
C.~Sieg and M.~Wilhelm, ``{On a CFT limit of planar $\gamma_i$-deformed
  $\mathcal{N}=4$ SYM theory},''
  \href{http://dx.doi.org/10.1016/j.physletb.2016.03.004}{{\em Phys. Lett. B}
  {\bfseries 756} (2016) 118--120},
  \href{http://arxiv.org/abs/1602.05817}{{\ttfamily arXiv:1602.05817
  [hep-th]}}.

\bibitem{Grabner:2017pgm}
D.~Grabner, N.~Gromov, V.~Kazakov, and G.~Korchemsky, ``{Strongly
  $\gamma$-Deformed $\mathcal{N}=4$ Supersymmetric Yang-Mills Theory as an
  Integrable Conformal Field Theory},''
  \href{http://dx.doi.org/10.1103/PhysRevLett.120.111601}{{\em Phys. Rev.
  Lett.} {\bfseries 120} no.~11, (2018) 111601},
  \href{http://arxiv.org/abs/1711.04786}{{\ttfamily arXiv:1711.04786
  [hep-th]}}.

\bibitem{Basso:2018agi}
B.~Basso and D.-l. Zhong, ``{Continuum limit of fishnet graphs and AdS sigma
  model},'' \href{http://dx.doi.org/10.1007/JHEP01(2019)002}{{\em JHEP}
  {\bfseries 01} (2019) 002}, \href{http://arxiv.org/abs/1806.04105}{{\ttfamily
  arXiv:1806.04105 [hep-th]}}.

\bibitem{Basso:2019xay}
B.~Basso, G.~Ferrando, V.~Kazakov, and D.-l. Zhong, ``{Thermodynamic Bethe
  Ansatz for Biscalar Conformal Field Theories in any Dimension},''
  \href{http://dx.doi.org/10.1103/PhysRevLett.125.091601}{{\em Phys. Rev.
  Lett.} {\bfseries 125} no.~9, (2020) 091601},
  \href{http://arxiv.org/abs/1911.10213}{{\ttfamily arXiv:1911.10213
  [hep-th]}}.

\bibitem{Arutynov:2014ota}
G.~Arutyunov, M.~de~Leeuw, and S.~van Tongeren, ``{The exact spectrum and
  mirror duality of the $(\text{AdS}_5{\times}S^5)_\eta$ superstring},''
  \href{http://dx.doi.org/10.1007/s11232-015-0243-9}{{\em Theor. Math. Phys.}
  {\bfseries 182} no.~1, (2015) 23--51},
  \href{http://arxiv.org/abs/1403.6104}{{\ttfamily arXiv:1403.6104 [hep-th]}}.
[Teor. Mat. Fiz.182,no.1,28(2014)].

\bibitem{Aharony:2008ug}
O.~Aharony, O.~Bergman, D.~L. Jafferis, and J.~Maldacena, ``{$\mathcal{N}=6$
  superconformal Chern-Simons-matter theories, M2-branes and their gravity
  duals},'' \href{http://dx.doi.org/10.1088/1126-6708/2008/10/091}{{\em JHEP}
  {\bfseries 10} (2008) 091}, \href{http://arxiv.org/abs/0806.1218}{{\ttfamily
  arXiv:0806.1218 [hep-th]}}.

\bibitem{Aharony:2008gk}
O.~Aharony, O.~Bergman, and D.~L. Jafferis, ``{Fractional M2-branes},''
  \href{http://dx.doi.org/10.1088/1126-6708/2008/11/043}{{\em JHEP} {\bfseries
  11} (2008) 043}, \href{http://arxiv.org/abs/0807.4924}{{\ttfamily
  arXiv:0807.4924 [hep-th]}}.

\bibitem{Cavaglia:2014exa}
A.~Cavaglià, D.~Fioravanti, N.~Gromov, and R.~Tateo, ``{Quantum Spectral Curve
  of the $\mathcal N=$ 6 Supersymmetric Chern-Simons Theory},''
  \href{http://dx.doi.org/10.1103/PhysRevLett.113.021601}{{\em Phys. Rev.
  Lett.} {\bfseries 113} no.~2, (2014) 021601},
  \href{http://arxiv.org/abs/1403.1859}{{\ttfamily arXiv:1403.1859 [hep-th]}}.

\bibitem{Bombardelli:2017vhk}
D.~Bombardelli, A.~Cavaglià, D.~Fioravanti, N.~Gromov, and R.~Tateo, ``{The
  full Quantum Spectral Curve for $AdS_4/CFT_3$},''
  \href{http://dx.doi.org/10.1007/JHEP09(2017)140}{{\em JHEP} {\bfseries 09}
  (2017) 140},
\href{http://arxiv.org/abs/1701.00473}{{\ttfamily arXiv:1701.00473 [hep-th]}}.

\bibitem{Anselmetti:2015mda}
L.~Anselmetti, D.~Bombardelli, A.~Cavagli\`a, and R.~Tateo, ``{12 loops and
  triple wrapping in ABJM theory from integrability},''
  \href{http://dx.doi.org/10.1007/JHEP10(2015)117}{{\em JHEP} {\bfseries 10}
  (2015) 117}, \href{http://arxiv.org/abs/1506.09089}{{\ttfamily
  arXiv:1506.09089 [hep-th]}}.

\bibitem{Lee:2017mhh}
R.~N. Lee and A.~I. Onishchenko, ``{ABJM quantum spectral curve and Mellin
  transform},'' \href{http://dx.doi.org/10.1007/JHEP05(2018)179}{{\em JHEP}
  {\bfseries 05} (2018) 179}, \href{http://arxiv.org/abs/1712.00412}{{\ttfamily
  arXiv:1712.00412 [hep-th]}}.

\bibitem{Lee:2019oml}
R.~N. Lee and A.~I. Onishchenka, ``{ABJM quantum spectral curve at twist 1:
  algorithmic perturbative solution},''
  \href{http://dx.doi.org/10.1007/JHEP11(2019)018}{{\em JHEP} {\bfseries 11}
  (2019) 018}, \href{http://arxiv.org/abs/1905.03116}{{\ttfamily
  arXiv:1905.03116 [hep-th]}}.

\bibitem{Lee:2018jvn}
R.~Lee and A.~Onishchenko, ``{Toward an analytic perturbative solution for the
  ABJM quantum spectral curve},''
  \href{http://dx.doi.org/10.1134/S0040577919020077}{{\em Teor. Mat. Fiz.}
  {\bfseries 198} no.~2, (2019) 292--308},
  \href{http://arxiv.org/abs/1807.06267}{{\ttfamily arXiv:1807.06267
  [hep-th]}}.

\bibitem{Cavaglia:2021eqr}
A.~Cavagli\`a, N.~Gromov, B.~Stefa\'nski, Jr., and A.~Torrielli, ``{Quantum
  Spectral Curve for AdS$_3$/CFT$_2$: a proposal},''
  \href{http://arxiv.org/abs/2109.05500}{{\ttfamily arXiv:2109.05500
  [hep-th]}}.

\bibitem{Ekhammar:2021pys}
S.~Ekhammar and D.~Volin, ``{Mondromy Bootstrap for SU(2|2) Quantum Spectral
  Curves: From Hubbard model to AdS3/CFT2},''
  \href{http://arxiv.org/abs/2109.06164}{{\ttfamily arXiv:2109.06164
  [math-ph]}}.

\bibitem{Dei:2018jyj}
A.~Dei and A.~Sfondrini, ``{Integrable S matrix, mirror TBA and spectrum for
  the stringy AdS$_{3}\times S^{3}\times S^{3}\times S^{1}$ WZW model},''
  \href{http://dx.doi.org/10.1007/JHEP02(2019)072}{{\em JHEP} {\bfseries 02}
  (2019) 072}, \href{http://arxiv.org/abs/1812.08195}{{\ttfamily
  arXiv:1812.08195 [hep-th]}}.

\bibitem{Jiang:2019xdz}
Y.~Jiang, S.~Komatsu, and E.~Vescovi, ``{Structure constants in $ \mathcal{N} =
  4$ SYM at finite coupling as worldsheet g-function},''
  \href{http://dx.doi.org/10.1007/JHEP07(2020)037}{{\em JHEP} {\bfseries 07}
  no.~07, (2020) 037}, \href{http://arxiv.org/abs/1906.07733}{{\ttfamily
  arXiv:1906.07733 [hep-th]}}.

\bibitem{Kristensson:2020nly}
A.~T. Kristensson and M.~Wilhelm, ``{From Hagedorn to Lee-Yang: partition
  functions of $ \mathcal{N} $ = 4 SYM theory at finite N},''
  \href{http://dx.doi.org/10.1007/JHEP10(2020)006}{{\em JHEP} {\bfseries 10}
  (2020) 006}, \href{http://arxiv.org/abs/2005.06480}{{\ttfamily
  arXiv:2005.06480 [hep-th]}}.

\bibitem{Beisert:2004hm}
N.~Beisert, V.~Dippel, and M.~Staudacher, ``{A Novel long range spin chain and
  planar $\mathcal{N}=4$ super Yang-Mills},''
  \href{http://dx.doi.org/10.1088/1126-6708/2004/07/075}{{\em JHEP} {\bfseries
  07} (2004) 075},
\href{http://arxiv.org/abs/hep-th/0405001}{{\ttfamily arXiv:hep-th/0405001
  [hep-th]}}.

\bibitem{Beisert:2006ez}
N.~Beisert, B.~Eden, and M.~Staudacher, ``{Transcendentality and Crossing},''
  \href{http://dx.doi.org/10.1088/1742-5468/2007/01/P01021}{{\em J. Stat.
  Mech.} {\bfseries 0701} (2007) P01021},
\href{http://arxiv.org/abs/hep-th/0610251}{{\ttfamily arXiv:hep-th/0610251
  [hep-th]}}.

\bibitem{Gromov:2011cx}
N.~Gromov, V.~Kazakov, S.~Leurent, and D.~Volin, ``{Solving the AdS/CFT
  Y-system},'' \href{http://dx.doi.org/10.1007/JHEP07(2012)023}{{\em JHEP}
  {\bfseries 07} (2012) 023},
\href{http://arxiv.org/abs/1110.0562}{{\ttfamily arXiv:1110.0562 [hep-th]}}.

\bibitem{Gromov:2010km}
N.~Gromov, V.~Kazakov, S.~Leurent, and Z.~Tsuboi, ``{Wronskian Solution for
  AdS/CFT Y-system},'' \href{http://dx.doi.org/10.1007/JHEP01(2011)155}{{\em
  JHEP} {\bfseries 01} (2011) 155},
\href{http://arxiv.org/abs/1010.2720}{{\ttfamily arXiv:1010.2720 [hep-th]}}.

\end{thebibliography}\endgroup

\end{document}